\documentclass{jfm}

\usepackage[version=4]{mhchem}
\usepackage{amsmath}
\usepackage{graphicx}
\usepackage{epstopdf, epsfig}
\usepackage{amsfonts}
\usepackage{amssymb}
\usepackage{float}
\usepackage{xcolor}
\usepackage{bm}
\usepackage[acronym]{glossaries}
\usepackage{soul}

\shorttitle{Two-phase lattice Boltzmann model}
\shortauthor{S.A. Hosseini, B.Dorschner and I.V. Karlin}

\title{Towards a consistent lattice Boltzmann model for two-phase fluid}

\author{S.A.~Hosseini\aff{1}
  B.~Dorschner\aff{1}
 \and I.V. Karlin\aff{1}\corresp{\email{ikarlin@ethz.ch}}}

\affiliation{\aff{1}Department of Mechanical and Process Engineering, ETH Zurich, 8092 Zurich, Switzerland.}

\begin{document}
	
\maketitle

\begin{abstract}
We propose a kinetic framework for single-component non-ideal isothermal flows. Starting from a kinetic model for a non-ideal fluid, we show that under conventional scaling the Navier-Stokes equations with a non-ideal equation of state are recovered in the hydrodynamic limit. A scaling based on the smallness of velocity increments is then introduced, which recovers the full Navier-Stokes-Korteweg equations. The proposed model is realized on a standard lattice and validated on a variety of benchmarks.
Through a detailed study of thermodynamic properties including co-existence densities, surface tension, Tolman length and sound speed, we show thermodynamic consistency, well-posedness and convergence of the proposed model.
Furthermore, hydrodynamic consistency is demonstrated by verification of Galilean invariance of the  dissipation rate of shear and normal modes and the study of visco-capillary coupling effects. Finally, the model is validated on dynamic test cases in three dimensions with complex geometries and large density ratios such as drop impact on textured surfaces and mercury drops coalescence.
\end{abstract}

    \section{Introduction}
    Multi-phase flows are omnipresent in science and technology. From micro-droplets coalescing in clouds, to solidification or melting of alloys and diesel droplets evaporation and subsequent combustion, all involve multiple interacting phases and moving interfaces. This ubiquity fueled wide efforts focused on the development of predictive mathematical models and numerical tools for multi-phase flows.
    While significant attention  has been focused on sharp interface methods requiring  efficient tracking of the evolving and deforming interfaces, and imposing jump conditions  \citep{sethian_level_2003,scardovelli_direct_1999,popinet_numerical_2018,prosperetti_computational_2009}, the ever-growing range of temperatures and pressures involved in typical systems of interest is making thermodynamic consistency of the computational models essential. 
    For instance, dramatically different thermodynamic regimes are encountered in diesel engines during the compression phase, in aeoronautical engines during take-off while most rocket engines operate in trans- and super-critical regimes, where the interface thickness becomes comparable to the flow scales.
    Nucleation and cavitation are yet another example, where the sharp interface limit does not hold and modifications to the classical nucleation theory \citep{debenedetti_metastable_1997},
    related to curvature-dependence of the surface tension, are required.
    In such cases, an accurate account of non-ideality of the fluid, including a finite interface thickness, is crucial for predictive numerical simulations of the flow physics.
    At a macorscopic level, a primer example for thermodynamics of non-ideal fluids is the second-gradient theory, first introduced by van der Waals for single-component fluids \citep{van_der_waals_thermodynamische_1894},
    leading to the Navier-Stokes equations supplemented with the Korteweg stress tensor \citep{korteweg_sur_1901}, and is a starting point for numerical methods known as diffuse interface approach \citep{anderson_diffuse-interface_1998}.
    On the other hand, extension of the Boltzmann equation to dense gases within the Enskog hard-sphere collision model \citep{enskog_warmeleitung_1921} and Vlasov mean-field approximation \citep{vlasov_many-particle_1961} provides a kinetic-theory basis for dynamics of non-ideal fluid \citep{chapman_mathematical_1939}.
    
    Since the pioneering work of \cite{shan_lattice_1993}, the lattice Boltzmann method (LBM) gained popularity as a viable numerical tool targeting the hydrodynamic regime of multi-phase flows. 
    Despite their popularity and wide usage, most multi-phase models for LBM, apart from limited studies in \cite{he_thermodynamic_2002}, lack a clear kinetic-theory framework.
    For instance, the so-called pseudo-potential lattice Boltzmann models lack a clear and consistent continuous kinetic model and scaling law recovering the full target macroscopic system. Thorough analyses of the bulk thermodynamic and interface properties of the models (especially near the critical state) are also very scarce. Furthermore, ever since their inception, such models have continuously struggled with larger density ratio simulations achieving at best, via different strategies, ratios of the order of $10^3$ as reflected by a number of recent reviews \citep{chen_critical_2014,li_lattice_2016}.
    
    In this paper, we revisit the construction of the lattice Boltzmann model for isothermal two-phase flows. 
    We propose a flexible kinetic framework for dense fluids with non-ideal equations of states. Using the lattice Boltzmann method discretization strategy, and under proper scaling, the model is shown to recover the full Navier-Stokes-Korteweg system of equations. Through a detailed study of thermodynamic properties, the model is shown to be well-posed and convergent to the capillary fluid thermodynamics. The well-posedness of the model and proper consideration of the proposed scaling is shown to guarantee recovery of the hydrodynamic-scale dynamics both at very large density ratio and near critical point.

    The outline is as follows: We begin in section \ref{sec:vdW} with a summary of the second-gradient theory due to  \cite{van_der_waals_thermodynamische_1894} and \cite{korteweg_sur_1901}. In section \ref{sec:statmech}, following a more microscopic approach, we consider  a class of kinetic models suitable for a non-ideal fluid. 
    We  proceed in section \ref{sec:new_scaling} with  a scaling assumption of small flow velocity increments which leads to a lattice Bhatnagar--Gross-Krook (LBGK) equation with a new realization of the nonlocal force that guarantees consistency with Korteweg's stress. 
    
    Thermodynamics of the LBGK model is validated in section \ref{sec:thermo_properties}. In section \ref{sec:coexistence}, we demonstrate convergence of vapour-liquid coexistence to the principle of corresponding states \citep{guggenheim_principle_1945}, independently of the equation of state and for liquid-vapour density ratio up to at least $\sim10^{11}$. The remainder of the paper is based on the van der Waals equation of state. In section \ref{sec:surface_temperature}, we show that the surface tension in the present LBGK model obeys a temperature scaling in excellent agreement with the theory. 
    After verifying  that the proposed model allows for choosing  surface tension independently of the density ratio in section \ref{sec:tunability}, we show in section \ref{sec:Tolman} that it is also consistent with Gibbs' theory of dividing surfaces \citep{gibbs_equilibrium_1874}. Simulations presented in section \ref{sec:Tolman} reveal a generalized Laplace law and uncover the effect of curvature on the surface tension, in agreement with the theory by \cite{tolman_effect_1949}. 
    Finally, in section \ref{sec:interface_W}, we show that the interface width scales with the temperature in accord with van der Waals theory.
    
    We turn to probing hydrodynamic features of our  model in section \ref{sec:hydro_consistent}. In section \ref{sec:Poiseuille}, we demonstrate that it correctly implements the jump condition for the stresses at the liquid-vapour interface in the simulation of layered Poiseuille flow. In section \ref{sec:acoustics}, Galilean invariance is demonstrated by measuring dissipation of normal modes in a moving reference frame.
    The viscosity-capillarity coupling is probed in section \ref{sec:Rayleigh} by measuring the frequency of higher-order capillary waves, in excellent agreement with Rayleigh's theory \citep{rayleigh_capillary_1879}. We also demonstrate that damping rate of capillary wave agrees with analytical solution. Validation of bulk properties is concluded by measuring the isothermal speed of sound in section \ref{sec:SoS}, where excellent comparison to the theoretical prediction is demonstrated for large density ratios. 
    In section \ref{sec:contact}, the model is extended to the simulation of a fluid-solid interface, and is is validated by demonstrating the Young--Laplace law and a liquid column motion in a channel with non-uniform wettability.
    In section \ref{sec:droplets}, the model is used to simulate water impact on textured superhydrophobic surfaces and mercury droplets coalescence to demonstrate its ability to handle  simulations at extremely high density ratios. Conclusions are drawn in section \ref{sec:conclusion}.
    
    \section{Model for two-phase flows}
    \subsection{Second-gradient theory: Korteweg's stress and capillary fluid equations\label{sec:vdW}}
    In the second-gradient theory as introduced by \citet{van_der_waals_thermodynamische_1894}, free energy per unit volume is expressed as:
    \begin{equation}\label{eq:2nd_grad_energy}
        \mathcal{A}_{\rm vdW} = \mathcal{A} + \frac{1}{2}\kappa {\lvert \bm{\nabla}\rho\lvert}^2,
    \end{equation}
    where $\mathcal{A}$ is the bulk free energy per unit volume, $\rho$ is the density and $\kappa$ is the capillary coefficient. The second term represents the interface energy while the bulk free energy is solely a function of the local density and temperature~\citep{giovangigli_kinetic_2020}. The equilibrium state of the corresponding system is obtained by minimizing free energy in a given volume under the constraint of constant total mass, leading to the stress tensor \citep{anderson_diffuse-interface_1998},
    \begin{equation}
        \bm{T}_{K} =  \bm{\nabla}\otimes\frac{\partial\mathcal{L}}{\partial(\bm{\nabla}\rho)}- \mathcal{L}\bm{I},
    \end{equation}
    where $\bm{I}$ is unit tensor and $\mathcal{L}$ is the Lagrange function,
    \begin{equation}
        \mathcal{L} = \mathcal{A} + \frac{1}{2}\kappa {\lvert \bm{\nabla}\rho\lvert}^2 - \lambda\rho,
    \end{equation}
    and $\lambda$ is the Lagrange multiplier for the mass constraint, or chemical potential,
    \begin{align}
    	\lambda=\frac{\partial\mathcal{A}}{\partial\rho}-\kappa\bm{\nabla}^2\rho, 
    \end{align}
     where $\bm{\nabla}^2$ is the Laplace operator. This in turn leads to the following Korteweg's stress tensor~\citep{korteweg_sur_1901}:
    \begin{equation}
        \bm{T}_{K} = \left(P-\kappa\rho\bm{\nabla}^2\rho - \frac{1}{2}\kappa {\lvert \bm{\nabla}\rho\lvert}^2 \right)\bm{I} + \kappa \bm{\nabla}\rho\otimes\bm{\nabla}\rho,
    \end{equation}
    where 
    \begin{align}
    	P=\rho \frac{\partial\mathcal{A}}{\partial\rho}-\mathcal{A},
    	\label{eq:pressure}
    \end{align}	
    	 is the thermodynamic pressure, or equation of state. From the local balance equations for mass and momentum one obtains the macroscopic governing laws for an iso-thermal capillary fluid:
    \begin{align}\label{eq:macro_mass}
      &  \partial_t\rho + \bm{\nabla}\cdot\rho \bm{u} = 0,\\
      &  \partial_t\rho\bm{u} + \bm{\nabla}\cdot\rho \bm{u}\otimes\bm{u} + \bm{\nabla}\cdot\bm{T}= 0,\label{eq:momentum_balance}
    \end{align}
   where $\bm{u}$ is the fluid velocity and the stress tensor $\bm{T}$ is
    \begin{equation}\label{eq:macro_stress}
        \bm{T} = \bm{T}_K + \bm{T}_{\rm NS}.
    \end{equation}
The Navier--Stokes viscous stress tensor reads,
\begin{equation}\label{eq:NS_stress}
	\bm{T}_{\rm NS}=-\mu\bm{S}-\eta (\bm{\nabla}\cdot\bm{u})\bm{I},
\end{equation}
where $\bm{S}$ is the trace-free rate-of-strain tensor,
\begin{equation}\label{eq:viscous_stress}
	\bm{S}=\bm{\nabla}\bm{u} + {\bm{\nabla}\bm{u}}^{\dagger} -\frac{2}{3}(\bm{\nabla}\cdot\bm{u})\bm{I},
\end{equation}
and $\mu$ and $\eta$ are the dynamic and the bulk viscosity, respectively. 
    
    The momentum balance equation (\ref{eq:momentum_balance}) can  be recast in the following form,
    \begin{equation}\label{eq:momentum_balance_force}
    		\partial_t\rho\bm{u} + \bm{\nabla}\cdot\rho \bm{u}\otimes\bm{u} + \bm{F}_{\rm K} +\bm{\nabla}\cdot\bm{T}_{\rm NS}=0,
    \end{equation}
    where Korteweg's {\it force} $\bm{F}_{\rm K}$ is the divergence of the Korteweg pressure tensor,
    \begin{equation}
        \bm{F}_{\rm K}=\bm{\nabla}\cdot\bm{T}_{\rm K}.
    \end{equation}
    The latter can be written in the following form,
    \begin{align}
    	\label{eq:Kforce}
	    \bm{F}_{\rm K}=\bm{\nabla}P_0+\bm{\nabla}\left(P-P_0\right) {-}\kappa\rho\bm{\nabla}(\bm{\nabla}^2\rho),
    \end{align} 
    where we have introduced a {\it reference pressure} $P_0$. Navier--Stokes momentum equations with Korteweg's force \eqref{eq:Kforce} shall be a target for  reconstruction by a suitable kinetic model.
    \subsection{Kinetic model for non-ideal fluid \label{sec:statmech}}
    \newacronym{bbgky}{BBGKY}{Bogolioubov--Born--Green--Kirkwood--Yvon}
    \newacronym{ret}{RET}{revised Enskog theory}
    In order to introduce a kinetic model for non-ideal fluid, we begin with the first Bogolioubov--Born--Green--Kirkwood--Yvon (BBGKY) equation,
    \begin{equation}\label{eq:Boltzmann_eq}
        \partial_t f + \bm{v}\cdot\bm{\nabla} f = \mathcal{J}=\int\int  \bm{\nabla} V\left(\lvert \bm{r}-\bm{r}_1\rvert\right)\cdot \frac{\partial}{\partial \bm{v}}f_{2}(\bm{r},\bm{v}, \bm{r}_1,\bm{v}_1,t) d\bm{v}_1d\bm{r}_1,
    \end{equation}
    where $f(\bm{r},\bm{v},t)$ and $f_{2}(\bm{r},\bm{v},\bm{r}_1,\bm{v}_1,t)$ are the one- and the two-particle distribution functions, respectively, $\bm{r}$, $\bm{r}_1$ and $\bm{v}$, $\bm{v}_1$ are particles position and velocity, while $V$ is a potential of pair interaction. 
    The local equilibrium state is defined by the Maxwellian $f^{\rm eq}$ at constant temperature $T$, parameterized by the local values of density $\rho$ and flow velocity $\bm{u}$, 
    \begin{equation}\label{eq:LM}
    	f^{\rm eq}=\frac{\rho}{\left(2\pi RT\right)^{3/2}}\exp\left[-\frac{(\bm{v}-\bm{u})^2}{2RT}\right],
    \end{equation}
where $R$ is gas constant.
     Furthermore, let us introduce a projector $\mathcal{K}$ onto local equilibrium at constant temperature, 
    \begin{align}
    	\mathcal{K}\mathcal{J}=
    	\left(\dfrac{\partial f^{\rm eq}}{\partial \rho}-\frac{1}{\rho}\bm{u}\cdot\frac{\partial f^{\rm eq}}{\partial \bm{u}}\right)\int \mathcal{J} d\bm{v}
    	+\frac{1}{\rho}\dfrac{\partial f^{\rm eq}}{\partial \bm{u}}\cdot \int \bm{v}\mathcal{J} d\bm{v}.
    	\label{eq:projector}
    \end{align}
    The projector property, $\mathcal{K}^2=\mathcal{K}$, can be verified by a direct computation. With the projector \eqref{eq:projector}, the interaction term in \eqref{eq:Boltzmann_eq} is split  into two parts by writing an identity,
    \begin{equation}\label{eq:enskog_projection}
        \mathcal{J} = \left(1-\mathcal{K}\right)\mathcal{J} + \mathcal{K} \mathcal{J}.
    \end{equation}
    The first term,
    \begin{align}
	\mathcal{J}_{\rm loc}=\left(1-\mathcal{K}\right)\mathcal{J},
    \end{align}
    satisfies the local conservation of both mass and momentum,
    \begin{align}
	\mathcal{K}\mathcal{J}_{\rm loc}=0.
    \end{align}
    It is conventional to model the locally conserving part of the interaction with a single relaxation time Bhatnagar--Gross--Krook (BGK) approximation,
    \begin{equation}\label{eq:introduction_of_bgk}
        \mathcal{J}_{\rm loc} \to  \mathcal{J}_{\rm BGK} = -\frac{1}{\tau}\left(1-\mathcal{K}\right)f = -\frac{1}{\tau}\left(f- f^{\rm eq}\right),
    \end{equation}
    where the relaxation time $\tau$ is  a  free parameter. 
The second term in the identity \eqref{eq:enskog_projection},
\begin{align}
	\mathcal{J}_{\rm nloc}=\mathcal{K}\mathcal{J},
\end{align}
satisfies the local mass but not the local momentum conservation. 
Indeed, after integration by part in the velocity $\bm{v}$ and neglecting boundary integrals, we arrive at
    \begin{equation}\label{eq:final_non_kinetic_contributions}
    \mathcal{J}_{\rm nloc} = -\frac{1}{\rho}\frac{\partial f^{\rm eq}}{\partial \bm{u}}\cdot\bm{F}_{\rm nloc},
    \end{equation}
    where the force $\bm{F}_{\rm nloc}$ reads,
    \begin{equation}
    	\bm{F}_{\rm nloc}=\int\int\int \bm{\nabla}V\left(\lvert \bm{r}-\bm{r}_1\rvert\right)f_{2}(\bm{r},\bm{v}, \bm{r}_1,\bm{v}_1,t) d\bm{v}_1d\bm{r}_1d\bm{v}.
    	\label{eq:force_general}
    \end{equation}
    Collecting the BGK approximation together with the nonlocal contribution, a generic kinetic  model may be written,
    \begin{equation}\label{eq:final_kinetic_model}
	\partial_t f + \bm{v}\cdot\bm{\nabla} f = -\frac{1}{\tau}\left(f - f^{\rm eq}\right) 
	- \frac{1}{\rho}\frac{\partial f^{\rm eq}}{\partial \bm{u}}\cdot\bm{F}_{\rm nloc}.
    \end{equation}
    Evaluation of the force \eqref{eq:force_general} requires us to specify the particles interaction. It is customary to invoke the Enskog--Vlasov model \citep{enskog_warmeleitung_1921,vlasov_many-particle_1961} where both hard-sphere collisions and a weak long-range attraction potential contribute to a non-local momentum transfer. 
    For the hard-sphere Enskog part, a de-localization of the collision is responsible for a non-vanishing contribution of momentum transfer through the distance between the centres of the spheres upon their impact while the Vlasov approximation contributes non-locally to the momentum transfer from a distributed mean-field force. 
    Evaluation of both the Enskog and Vlasov contributions to the force \eqref{eq:force_general} proceeds along familiar lines \citep{chapman_mathematical_1939,he_thermodynamic_2002} and is reported in Appendices \ref{ap:ShortRange} and \ref{ap:LongRangeMeanField} for completeness,
    \begin{equation}
        \bm{F}_{\rm EV} = \bm{F}_{\rm E}+\bm{F}_{\rm V}.
    \label{eq:EVforce}
    \end{equation}
    The first term is the lowest-order contribution of the collisional momentum transfer from the Enskog hard-sphere model,
    \begin{equation}
    \bm{F}_{\rm E} = 
    \bm{\nabla}b\rho^2\chi RT+ O(\bm{\nabla}^3\rho).
    \label{eq:Eforce}
    \end{equation}
   Here  $b=4v_{\rm HS}$, with $v_{\rm HS}=\mathcal{V}_{\rm HS}/m$ the specific volume of hard-sphere of diameter $d$ and mass $m$, while $\mathcal{V}_{\rm HS}=\pi d^3/6$ is the volume of the sphere. Moreover, $\chi$ is the equilibrium two-particle correlation function, evaluated at the  local density reduced by the specific volume of hard sphere, $\chi=\chi(b\rho(\bm{r},t))$; To the lowest order, $\chi=1+(5/8)b\rho+O((b\rho)^2)$, cf.\ \citet{chapman_mathematical_1939}.
    The second term in \eqref{eq:EVforce} is the contribution of a long-range attraction potential $V$ in the mean field Vlasov approximation. 
    To third order in the gradient of density, 
    \begin{equation}
    	\bm{F}_{\rm V} = 
    	-\bm{\nabla}a\rho^2
    	- \kappa \rho\bm{\nabla}\bm{\nabla}^2\rho 
    	+O(\bm{\nabla}^5\rho),
    	\label{eq:Vforce}
    \end{equation}
    where parameters $a$ and $\kappa$ are,
    \begin{align}
	a &= -2\pi\int_{d}^{\infty} r^2V(r) dr,\\
	\kappa &= -\frac{2\pi}{3}\int_{d}^{\infty} r^4V(r) dr.
	\end{align}
    Thus, with the approximations specified, the non-local force \eqref{eq:EVforce} becomes, 
    \begin{equation}
	\bm{F}_{\rm EV} = 
	\bm{\nabla}\left(P_{\rm EV}-P_0\right)
	- \kappa \rho\bm{\nabla}\bm{\nabla}^2\rho,
	\label{eq:EVforce1}
    \end{equation}
    where the reference ideal gas pressure $P_0$ is provided by the local Maxwellian \eqref{eq:LM},
    \begin{equation}
	P_0=\frac{1}{3}\int \lvert\bm{v}-\bm{u}\rvert^2 f^{\rm eq}d\bm{v}=\rho R T,
    \end{equation}
    while the equation of state of the Enskog--Vlasov gas is of van der Waals type,
    \begin{equation}\label{eq:EVpressure}
  	P_{\rm EV}=\rho RT(1 + b\rho\chi) -a\rho^2.
    \end{equation}
    This allows us to extend the Enskog--Vlasov kinetic model and a phenomenological equation of state $P$ \eqref{eq:pressure} can be used instead of $P_{\rm EV}$ \eqref{eq:EVpressure}.
    Moreover, the reference pressure $P_0$ can be made selective by rescaling the local equilibrium, 
    \begin{equation}
	f^{\rm eq}=\frac{\rho}{\left(2\pi P_0/\rho\right)^{D/2}}\exp\left[-\frac{(\bm{v}-\bm{u})^2}{2P_0/\rho}\right],
	\label{eq:eq_ref}
    \end{equation}
    where $D$ is the space dimension. While the Enskog--Vlasov partition above corresponds to selecting $P_0 = \rho R T$, an alternative is provided by \citet{reyhanian_thermokinetic_2020}, where $P_0 = P$ is chosen. 
Using the rescaled equilibria (\ref{eq:eq_ref}), a family of kinetic models parameterized by the reference pressure reads,
    \begin{equation}\label{eq:gen_kinetic_model_ext}
	\partial_t f + \bm{v}\cdot\bm{\nabla} f = -\frac{1}{\tau}\left(f - f^{\rm eq}\right) - \frac{1}{\rho}\frac{\partial f^{\rm eq}}{\partial \bm{u}}\cdot\left[\bm{\nabla}\left(P-P_0\right)-\kappa \rho\bm{\nabla}\bm{\nabla}^2\rho\right].
    \end{equation}
    The kinetic equation \eqref{eq:gen_kinetic_model_ext} shall be considered as a semi-phenomenological model of nonideal fluid, with the relaxation time $\tau$, the capillarity coefficient $\kappa$, pressure $P$ and reference pressure $P_0$ as phenomenological input parameters, while the Enskog--Vlasov realization will serve as a representative example for estimates of various flow regimes.  

    The analysis of the kinetic model \eqref{eq:gen_kinetic_model_ext} under the conventional scaling of a small deviation from a uniform state \citep{chapman_mathematical_1939}, 
    \begin{equation}\label{eq:scaling_1}
	\bm{\nabla}\to \epsilon\bm{\nabla},\ \partial_t\to\epsilon\partial_t,
    \end{equation} 
    is detailed in Appendix \ref{ap:CE_cont}. To second order in space derivatives, the resulting momentum balance equation reads,
    \begin{equation}\label{eq:momentum_balance_force_weak}
	\partial_t\rho\bm{u} + \epsilon\bm{\nabla}\cdot\rho \bm{u}\otimes\bm{u} + \epsilon\bm{\nabla}P +\epsilon\bm{\nabla}\cdot\epsilon\bm{T}_{\rm NS}+O(\epsilon^3)
	=0,
    \end{equation}
    where the dynamic viscosity $\mu$ and the bulk viscosity $\eta$ in the Navier--Stokes stress tensor (\ref{eq:NS_stress}) are defined by the reference pressure ($D=3$),
    \begin{align}
	\mu&=\tau P_0,\label{eq:visc_gen}\\
	\eta&=\left(\frac{5}{3}-\frac{\partial\ln P_0}{\partial\ln\rho}\right)\tau P_0.\label{eq:bulk_visc_gen}
    \end{align}
    Thus, the momentum balance equation (\ref{eq:momentum_balance_force_weak}) is form-invariant with respect to the choice of reference pressure, provided $P_0$ satisfies a {sub-isentropic condition},
    \begin{equation}\label{eq:subisentropic}
    {P_0}\le C\rho^{5/3},
    \end{equation}
    for some $C>0$. With (\ref{eq:subisentropic}), the bulk viscosity (\ref{eq:bulk_visc_gen}) is positive and vanishes when the reference pressure follows an isentropic process for ideal monatomic gas, $P_0=C\rho^{5/3}$. For example, any polytropic process, $P_0=A\rho^n$, $1\le n\le 5/3$ satisfies the sub-isentropic condition and results in $\eta=(5/3-n)\tau P_0$.
    Special case of isothermal process $n=1$ returns $\eta=(2/3)\tau P_0$, and the viscous stress tensor becomes,
    \begin{equation}
	\bm{T}_{\rm NS}=-\tau P_0\left(\bm{\nabla}\bm{u}+\bm{\nabla}\bm{u}^\dagger\right).
    \end{equation}
    On the other hand, when compared to the two-phase momentum  equation \eqref{eq:momentum_balance_force}, the macroscopic limit recovers only the nonideal gas component thereof while missing Korteweg's capillarity contribution. 
    Indeed, the third-order term, $\sim\epsilon^3\rho\bm{\nabla}\bm{\nabla}^2\rho$  in (\ref{eq:Vforce}) and (\ref{eq:gen_kinetic_model_ext}),
    does {\it not} contribute to the momentum equation (\ref{eq:momentum_balance_force_weak}) under the scaling \eqref{eq:scaling_1}. 
    This is consistent with the well-known results from kinetic theory \citep{chapman_mathematical_1939} and is not surprising: 
    The scaling \eqref{eq:scaling_1} is essentially based on the Knudsen number, which overrides the relative contribution of the capillarity term by two orders, cf. Appendix \ref{ap:CE_cont}. 
    Thus, under weak non-uniformity assumption \eqref{eq:scaling_1}, the capillarity terms are seen as higher-order, Burnett-level contributions, and cannot appear in the main (first and second) orders in the momentum balance equation \eqref{eq:momentum_balance_force_weak}. 
    In fact, condition \eqref{eq:scaling_1} rules out situations at an interface between phases where gradients of density become large over a relatively short distance.
    Therefore, in order for the kinetic model (\ref{eq:gen_kinetic_model_ext}) to recover in-full the momentum balance (\ref{eq:momentum_balance_force}), a different scaling needs to be applied. 
    \subsection{Scaling by velocity increment and lattice Boltzmann equation \label{sec:new_scaling}}
    \subsubsection{Time step and forcing}
    A rescaling of the kinetic model in this section shall be maintained by introducing a {time step} $\delta t$.
    As a preliminary consideration, we evaluate the contribution of the force term over the time step. 
    To that end, as noted by \cite{kupershtokh_equations_2009}, for a generic force $\bm{F}$, we can write the action of the force on the distribution function as a full derivative in a frame that moves with the local fluid velocity,
    \begin{equation}
    	\frac{1}{\rho}\frac{\partial f^{\rm eq}}{\partial \bm{u}}\cdot\bm{F} = \frac{d f^{\rm eq}}{dt}.
    \end{equation}
    Introducing the \emph{velocity increment}, 
    \begin{equation}\label{eq:deltau}
    \delta\bm{u}=  \frac{\bm{F}}{\rho}\delta t,
    \end{equation}
    and integrating in time, leads to an approximation,
    \begin{equation}\label{eq:EDM}
    	\mathcal{F}=\int_{t}^{t+\delta t} \frac{1}{\rho}\frac{\partial f^{\rm eq}}{\partial \bm{u}}\cdot\bm{F} dt \approx f^{\rm eq}\left(\bm{u}+\delta\bm{u}\right) - f^{\rm eq}\left(\bm{u}\right).
    \end{equation}
    This so-called exact difference method (EDM) becomes accurate for a gravity force, $\bm{F}/\rho={\rm const}$, otherwise it often provides a reliable estimate for the force term and is widely used. 
    In what follows, the scaling to be applied assumes smallness of the velocity increment \eqref{eq:deltau} rather than smoothness of the spatial distribution of the force. Since the velocity increment is based on a time step, it is natural to proceed with a lattice Boltzmann realization of the kinetic equation.
    \subsubsection{Standard lattice and product-form}
    The lattice Boltzmann model shall be realized  with the standard discrete velocity set $D3Q27$, where $D=3$ stands for three dimensions and $Q=27$ is the number of discrete velocities,
    \begin{equation}\label{eq:d3q27vel}
    	\bm{c}_i=(c_{ix},c_{iy},c_{iz}),\ c_{i\alpha}\in\{-1,0,1\}.
    \end{equation}
    We first define a triplet of functions in two variables, $\xi_{\alpha}$ and $\zeta_{\alpha\alpha}$, 
    \begin{align}
    	&	\Psi_{0}(\xi_{\alpha},\zeta_{\alpha\alpha}) = 1 - \zeta_{\alpha\alpha}, 
    	\label{eqn:phi0}
    	\\
    	&	\Psi_{1}(\xi_{\alpha},\zeta_{\alpha\alpha}) = \frac{\xi_{\alpha} + \zeta_{\alpha\alpha}}{2},
    	\label{eqn:phiPlus}
    	\\
    	&	\Psi_{-1}(\xi_{\alpha},\zeta_{\alpha\alpha}) = \frac{-\xi_{\alpha} + \zeta_{\alpha\alpha}}{2},
    	\label{eqn:phis}
    \end{align}
    and consider a product-form associated with the discrete velocities $\bm{c}_i$ (\ref{eq:d3q27vel}),
    \begin{equation}\label{eq:prod}
    	\Psi_i= \Psi_{c_{ix}}(\xi_x,\zeta_{xx}) \Psi_{c_{iy}}(\xi_y,\zeta_{yy}) \Psi_{c_{iz}}(\xi_z,\zeta_{zz}).
    \end{equation}
    All pertinent populations below shall be determined by specifying the parameters $\xi_\alpha$ and $\zeta_{\alpha\alpha}$ in the product-form (\ref{eq:prod}). 
    A two-dimensional version of the model on the $D2Q9$ lattice is obtained by omitting the $z$-component in all formulas below.
    Finally, we use notation $\delta\bm{r}_{i}=\bm{c}_i\delta t$ for the {\it lattice links}, and denote the {\it grid spacing} in any direction $\alpha=x,y,z$ as $\delta r=\vert c_{i\alpha}\vert\delta t$, $c_{i\alpha}\ne 0$: For the $D3Q27$ discrete velocity set \eqref{eq:d3q27vel}, the lattice spacing is same in all Cartesian directions.
    \subsubsection{The lattice Boltzmann equation}
    The local density $\rho$ and  flow velocity $\bm{u}$ are defined using the populations $f_i$,
    \begin{align}
    &	\rho(\bm{r},t)=\sum_{i=0}^{Q-1}f_i(\bm{r},t),\\
    &	\rho\bm{u}(\bm{r},t)=\sum_{i=0}^{Q-1}\bm{c}_if_i(\bm{r},t).
    \end{align}
    With a reference pressure $P_0$, and by setting the parameters,
    \begin{align}
    &\xi_{\alpha}=u_{\alpha},\\
    &\zeta_{\alpha\alpha}=\frac{P_0}{\rho}+u_{\alpha}^2,
    \end{align} 
    the local equilibrium populations are represented with the product-form \eqref{eq:prod},
    \begin{equation}\label{eq:LBMeq}
        f_i^{\rm eq}=
        \rho\prod_{\alpha=x,y,z}\Psi_{c_{i\alpha}}\left(u_\alpha,\frac{P_0}{\rho}+u_{\alpha}^2\right).
    \end{equation}
    The LBGK equation, supplemented with a forcing term, is written,
    \begin{equation}\label{eq:LBGK}
	f_i\left(\bm{r}+\bm{c}_i \delta t, t+\delta t\right) - f_i\left(\bm{r}, t\right) = \omega\left(f_i^{\rm eq} - f_i\right)
	+ \left(f_i^*-f_i^{\rm eq}\right),
    \end{equation}
    where $\omega$ is a dimensionless relaxation parameter, the equilibrium populations are provided by \eqref{eq:LBMeq} while the extended equilibrium populations $f_i^*$ are defined by the product-form (\ref{eq:prod}) with the following assignment for the parameters,
	\begin{align}
	&\xi_{\alpha}^{*} = u_{\alpha}+\frac{F_{\alpha}\delta t}{\rho},\label{eq:xistar}	\\
	  &\zeta_{\alpha\alpha}^{*} = \frac{P_0}{\rho} +u_{\alpha}^2 + \frac{\Phi_{\alpha\alpha}}{\rho},\label{eq:zetastar}
	  \end{align}
	where $\Phi_{\alpha\alpha}/\rho$ is a correction term for the diagonals of the non-equilibrium momentum flux tensor,
	\begin{equation}\label{eq:correction}
	   \Phi_{\alpha\alpha} = \left(1-\frac{\omega}{2}\right) \delta t\partial_{\alpha}\left(\rho u_{\alpha} \left(u_{\alpha}^2 + \frac{3P_{0}}{\rho}-3\varsigma^2\right)\right),
	\end{equation}
    where $\varsigma= \delta r/\sqrt{3}\delta t$ is the so-called lattice speed of sound. Thus, the extended equilibrium is explicitly written as,
    \begin{equation}\label{eq:LBMstar}
	f_i^*=\rho\prod_{\alpha=x,y,z}\Psi_{c_{i\alpha}}\left(u_\alpha+\frac{F_{\alpha}\delta t}{\rho},\frac{P_0}{\rho}+u_{\alpha}^2+\frac{\Phi_{\alpha\alpha}}{\rho}\right).
    \end{equation}
    Comments are in order:
    \begin{itemize}
	\item If the correction term \eqref{eq:correction} is omitted in (\ref{eq:zetastar}), the population \eqref{eq:LBMstar} becomes the equilibrium with the increment $\delta\bm{u}$ \eqref{eq:deltau} due to force action added to the flow velocity $\bm{u}$. This corresponds to the EDM forcing maintained by the second bracket in the right hand side of the LBGK equation \eqref{eq:LBGK}.
	\item The correction term (\ref{eq:xistar}) is added following a proposal by \cite{saadat_extended_2021}. Its purpose is to compensate for the bias of the $D3Q27$ velocities \eqref{eq:d3q27vel}, $c_{i\alpha}^3=c_{i\alpha}$, and to restore the Galilean invariance of the normal components of the viscous stress tensor.
    \end{itemize}

    The LBGK model \eqref{eq:LBGK} is generic with respect to the choice of reference pressure $P_0$ and the force $\bm{F}$. 
    We now proceed with the special case of Korteweg's force in order to establish a representation thereof matched to the lattice Boltzmann system.
    \subsubsection{Pseudo-potential and capillarity}
    Following the representation (\ref{eq:Kforce}), Korteweg's force includes two distinct parts, the term supplying the nonideal gas equation of state and the capillarity term responsible for the surface tension. Introducing a pseudo-potential $\psi$,
    \begin{equation}
    \psi= 
    \begin{cases}
    \sqrt{ P-P_0}, & \text{if } P>P_0,\\
    \sqrt{P_0-P}, & \text{if } P \leq P_0,
    \end{cases}
    \end{equation}
    Korteweg's force is written,
    \begin{equation}\label{eq:force_form}
    \bm{F} = 
    \begin{cases}
    2\psi\bm{\nabla}\psi - \kappa\rho\bm{\nabla}\bm{\nabla}^2\rho, & \text{if } P>P_0,\\
    -2\psi\bm{\nabla}\psi - \kappa\rho\bm{\nabla}\bm{\nabla}^2\rho,      & \text{if } P \leq P_0,
    \end{cases}
    \end{equation}

    In the lattice Boltzmann setting, the pseudo-potential part is represented as a linear combination of the first- and second-neighbours contributions,
    \begin{align}\label{eq:PP_discrete}
	\delta t \psi(\bm{r})\bm{\nabla}\psi(\bm{r})= \psi(\bm{r})\sum_{i=0}^{Q-1} \frac{w_i}{\varsigma^2} \bm{c}_i
	\left[\mathcal{G}_1\psi(\bm{r}+\bm{c}_i \delta t) + \mathcal{G}_2\psi(\bm{r}+2\bm{c}_i \delta t)\right]  +  {O}\left([\delta r\bm{\nabla}]^5\right),
    \end{align}
    where the weights $w_i$ are defined by the product-form (\ref{eq:prod}) at $\xi_{\alpha}=0$, $\zeta_{\alpha\alpha}=\varsigma^2$, 
    \begin{equation}\label{eq:wi}
	w_i=\prod_{\alpha=x,y,z}\Psi_{c_{i\alpha}}\left(0,\varsigma^2\right),
    \end{equation}
    and where the parameters $\mathcal{G}_1$ and $\mathcal{G}_2$ satisfy the conditions,
    \begin{align}
	\mathcal{G}_1+2\mathcal{G}_2=1,\label{eq:PPcond1}\\
	\mathcal{G}_1+8\mathcal{G}_2=0.\label{eq:PPcond2}
    \end{align}
    Condition \eqref{eq:PPcond1} maintains the equation of state, while condition \eqref{eq:PPcond2} eliminates the third-order error. Non-compliance with the first and/or second condition would introduce respectively errors of order ${O}([\delta r\bm{\nabla}])$ and/or $ {O}([\delta r\bm{\nabla}]^3)$.

    The capillarity part of Korteweg's force in Eq. \eqref{eq:force_form} is represented in a similar way but using the density instead of the pseudo-potential,
    \begin{align}\label{eq:capillary_discrete}
	\delta t \tilde{\kappa}\rho(\bm{r})\bm{\nabla}\bm{\nabla}^2\rho(\bm{r}) = \rho(\bm{r})\sum_{i=0}^{Q-1} \frac{w_i}{\varsigma^2} \bm{c}_i \left[\mathcal{G}_3\rho(\bm{r}+\bm{c}_i \delta t) + \mathcal{G}_4 \rho(\bm{r}+2\bm{c}_i \delta t)\right] + {O}\left([\delta r\bm{\nabla}]^5\right),
    \end{align}
    where $\tilde{\kappa}=\kappa\delta r^2$ and the parameters $\mathcal{G}_3$ and $\mathcal{G}_4$ satisfy the conditions,
    \begin{align}
	&\mathcal{G}_3+2\mathcal{G}_4=0,\label{eq:Ccond1}\\
	&\mathcal{G}_3+8\mathcal{G}_4=6\kappa. \label{eq:Ccond2}
    \end{align}
    Condition \eqref{eq:Ccond1} cancels the first-order derivative, while condition \eqref{eq:Ccond2} maintains the capillarity contribution. 
    Combining both the pseudo-potential (\ref{eq:PP_discrete}) and the capillarity (\ref{eq:capillary_discrete}) contributions, we obtain the lattice Boltzmann form of Korteweg's force \eqref{eq:force_form} as,
    \begin{align}
	    \delta t\bm{F} = &\pm 2\psi(\bm{r})\sum_{i=0}^{Q-1} \frac{w_i}{\varsigma^2}  \bm{c}_i \left[\frac{4}{3}\psi(\bm{r}+\bm{c}_i\delta t) - \frac{1}{6} \psi(\bm{r}+2\bm{c}_i\delta t)\right]\nonumber\\
	    &+\tilde{\kappa}\rho(\bm{r})\sum_{i=0}^{Q-1} \frac{w_i}{\varsigma^2} \bm{c}_i \left[2\rho(\bm{r}+\bm{c}_i\delta t) - \rho(\bm{r}+2\bm{c}_i\delta t)\right] + {O}\left([\delta r\bm{\nabla}]^5\right),
	    \label{eq:Kforce_final}
	\end{align}
    where the sign convention follows (\ref{eq:force_form}).
	
	In the context of the lattice Boltzmann method, various pseudo-potential representations have long been in use, and comments are in order to make a distinction to the present formulation.
    
    \begin{itemize}
    \item By setting $\mathcal{G}_2=0$ in \eqref{eq:PP_discrete} and ignoring Korteweg's capillarity term \eqref{eq:capillary_discrete} by choosing $\mathcal{G}_3=\mathcal{G}_4=0$, one recovers a  model first proposed by \cite{shan_lattice_1993} for special equations of state (see Eqs.\ \eqref{eq:SCeos}, \eqref{eq:SC1}, \eqref{eq:SC2} in Appendix \ref{ap:Sound_speed})  and extended to a general equation of state by \cite{yuan_equations_2006}. Unlike the above condition \eqref{eq:PPcond2} which eliminates the third-order error in the pseudo-potential part of Korteweg's force (\ref{eq:Kforce}), the model of \cite{shan_lattice_1993} requires the third-order error to be retained in order that it mimics surface tension effects. 
    Consequently, the force in the models of \cite{shan_lattice_1993,yuan_equations_2006} becomes, 
    \begin{equation}\label{eq:SCforce}
     	\bm{F}_{\rm SC}=2\psi\bm{\nabla}\psi+\frac{1}{3}\psi\bm{\nabla}\bm{\nabla}^2\psi.
    \end{equation}
    The force \eqref{eq:SCforce} neither conforms with the van der Waals second-gradient theory and Korteweg's stress of section \ref{sec:vdW} (unless $\psi=A\rho$ which, however, does not lead to phase separation), nor can it be derived from the BBGKY equation with a central-force particles interaction of section \ref{sec:statmech}.
    Another, relatively minor issue is the fixed parameter $1/3$, which is mimicking the capillarity coefficient and is the result of the third-order error retained in the expansion.
     
     \item By imposing condition \eqref{eq:PPcond1} while still discarding Korteweg's capillarity contribution \eqref{eq:capillary_discrete}, one arrives at a dual-range force model of \cite{sbragaglia_generalized_2007}, 
     \begin{equation}\label{eq:DRforce}
     	\bm{F}_{\rm DR}=2\psi\bm{\nabla}\psi+\left(\frac{4}{3}-\mathcal{G}_1\right)\psi\bm{\nabla}\bm{\nabla}^2\psi.
     \end{equation}
     The model \eqref{eq:DRforce} is an improvement on the first-neighbour model \eqref{eq:SCforce} in that it allows for a variable capillarity-like coefficient. At the same time, it does not resolve the inconsistency with Korteweg's stress tensor.
     
     \item Various other modifications of the original models of \cite{shan_lattice_1993} were proposed to {improve} on the main inconsistency by introducing and tuning ad hoc numerical coefficients tailored to a selected equation of state~\citep{luo_unified_2021,li_lattice_2013,huang_thermodynamic_2019,kupershtokh_equations_2009}. 
     However, to the best of our knowledge, none of these proposals came to recognize that the problem lies in the fact that it is impossible to represent both parts of Korteweg's force, the non-ideal equation of state and the capillarity term while using a pseudo-potential alone. This fact follows both from the phenomenological thermodynamics reviewed in section \ref{sec:vdW}, 
     as well as from a more microscopic approach of section \ref{sec:statmech}. The pseudo-potential in both cases represents only the equation of state while the capillarity term requires the density field to be used, as featured by our Eq.\  (\ref{eq:capillary_discrete}).
     
    \item Pseudo-potential is a convenient {form} of representing the equation of state contribution to Korteweg's force, tailored to the lattice Boltzmann setting. If other numerical methods are used to evaluate the force \eqref{eq:force_form}, such as higher-order finite difference, the model is usually renamed to a free energy approach. 
     \end{itemize}
     
    With the generic LBGK model (\ref{eq:LBGK}) and Korteweg's force \eqref{eq:Kforce_final} both specified, we proceed to the analysis of the hydrodynamic limit under a suitable scaling.
    
    \subsubsection{Hydrodynamic limit under small velocity increment scaling}
    Chapman--Enskog analysis of the LBGK equation (\ref{eq:LBGK}) was performed under the following scaling: With the characteristic values of the flow velocity $\mathcal{U}$, the flow scale $\mathcal{L}$,  the density $\rho$, the force $\mathcal{F}$ and the velocity increment $\delta u$, the following conditions apply:
    \begin{align}
    &	\frac{\delta u}{\mathcal{U}}\sim \frac{\delta t \mathcal{F}}{\rho \mathcal{U}}\sim\varepsilon,\label{eq:smallu}\\
    & \frac{\delta r}{\mathcal{L}}\sim\varepsilon.\label{eq:resolution}
    \end{align}
    The first scaling condition (\ref{eq:smallu}) refers to a {\it smallness of velocity increment}, that is, to the smallness of the force action over time $\delta t$. The second scaling condition (\ref{eq:resolution}) is a {\it resolution requirement}. Both conditions are assumed to hold simultaneously. Details of the analysis are provided in Appendix \ref{ap:CE} while the result is summarized below.

    Let us introduce a transformed velocity $\bm{U}$ by shifting the local velocity $\bm{u}$ by half of the velocity increment $\delta\bm{u}$ (\ref{eq:deltau}),
    \begin{equation}
    	\bm{U}=\bm{u}+\varepsilon\frac{\delta t\bm{F}}{2\rho}.
    \end{equation}
    Then the following mass and momentum balance equations to second order in $\varepsilon$ are recovered when the force \eqref{eq:Kforce_final} is used under the scaling \eqref{eq:smallu} and \eqref{eq:resolution}, 
    \begin{align}
        &\partial_t \rho  + \varepsilon\bm{\nabla}\cdot\rho\bm{U} + {O}(\varepsilon^3) = 0,\\
        &\partial_t \rho \bm{U} + \varepsilon \bm{\nabla}\rho \bm{U}\otimes\bm{U} + \varepsilon\bm{F}_{\rm K} +  \varepsilon\bm{\nabla}\cdot\varepsilon\bm{T}_{\rm NS} + {O}(\varepsilon^3) = 0, \label{eq:momentum_LBM}
    \end{align}
    where the dynamic and the bulk viscosity in the Navier--Stokes stress tensor (\ref{eq:NS_stress}) are related to the relaxation parameter $\omega$ and the reference pressure $P_0$ as follows:
    \begin{align}
    &	\mu= \left(\frac{1}{\omega} - \frac{1}{2}\right)\delta t P_0,\\
    &\eta=	\left(\frac{5}{3} - \frac{\partial\ln P_0}{\partial\ln\rho}\right)\left(\frac{1}{\omega} - \frac{1}{2}\right)\delta t P_0. 
    \end{align}
    Unlike the previous result \eqref{eq:momentum_balance_force_weak}, the momentum balance \eqref{eq:momentum_LBM} includes not only the nonideal gas pressure but also the capillarity term, and is thus consistent with Korteweg's force in the momentum balance.
    It should be pointed out that the scaling \eqref{eq:smallu} refers to smallness of the increment of the flow velocity rather that to smallness of either the time step or of the force. Thus, rescaling the kinetic model \eqref{eq:gen_kinetic_model_ext} based on the smallness of flow velocity increments results in both the non-ideal gas equation of state and the capillarity revealed at the Euler level $O(\varepsilon)$ of the momentum balance \eqref{eq:momentum_LBM}. This is in a contrast to the conventional scaling, which is tight to the nonuniformity and surface tension would appear only at a Burnett level  $O(\epsilon^3)$. 

    In the remainder of this paper, and without loss of generality, we set the reference pressure $P_0=\varsigma^2\rho$ in order to minimize the correction term \eqref{eq:correction}. In the next section, the model is scrutinized by a set of numerical tests probing various aspects of thermo- and hydrodynamic consistency.
   
    \section{Thermodynamic consistency \label{sec:thermo_properties}}
    \subsection{Liquid-vapour  coexistence: The principle of corresponding states \label{sec:coexistence}}
    We begin with the validation of liquid-vapour coexistence. Two-dimensional flat interface simulations were performed on a grid $800\times10$, filled with the vapour phase of a fluid with a specified equation of state and periodic boundary conditions. A column of the liquid phase over 400 grid-points was placed at the centre of the domain. Simulation were ran until steady-state was reached. Steady-state was monitored via a $L_0$ norm convergence criterion based on the liquid density $\rho_l$ at the centre of the drop and the vapour density $\rho_v$ at a location outside the drop. A theoretical prediction for the coexistence density ratio $\rho_l/\rho_v$ is readily obtained via the  equilibrium condition leading to the Maxwell equal-area construction,
    \begin{equation}\label{eq:Maxwell_equal_area}
        \int_{\rho_v}^{\rho_l} \frac{P_{\rm sat}-P}{\rho^2}d\rho=0,
    \end{equation}
    where $P_{\rm sat}(T)$ is the saturation pressure at which the liquid and vapor phases coexist at a given temperature $T$ below the critical point.
    
    Initially four generally adopted equations of state (EoS) were considered: the van der Waals  EoS \citep{van_der_waals_over_1873},
    \begin{equation}\label{eq:vdWEoS}
        P = \frac{\rho R T}{1-b\rho} - a \rho^2,
    \end{equation}
    where parameters $a$ and $b$ are related to critical temperature $T_c$ and pressure $P_c$ as,
    \begin{equation}
        a=\frac{27}{64}\frac{R^2T_c^2}{P_c},\ b=\frac{1}{8}\frac{RT_c}{P_c};
    \end{equation}
    the Peng--Robinson EoS \citep{peng_new_1976},
    \begin{equation}\label{eq:PREoS}
        P = \frac{\rho R T}{1-b\rho} - \frac{a \alpha(T) \rho^2}{1+2\rho b - b^2 \rho^2},
    \end{equation}
    with
    \begin{equation}
        \alpha(T) = \left[1 + (0.37464 + 1.54226\omega' - 0.26992 \omega'^2) \left(1-\sqrt{T/T_c}\right)\right]^2,
    \end{equation}
    where  $\omega'$ the acentric factor ($\omega'=0.344$ for water), and
    \begin{equation}
        a=0.45724\frac{R^2T_c^2}{P_c},\ b=0.0778\frac{R T_c}{P_c};
    \end{equation}
    the Riedlich--Kwong--Soave EoS~\citep{redlich_thermodynamics_1949,soave_equilibrium_1972},
    \begin{eqnarray}\label{eq:RKSEoS}
        P = \frac{\rho R T}{1-b\rho} - \frac{a \alpha(T) \rho^2}{1+\rho b},
    \end{eqnarray}
    with
    \begin{equation}
        \alpha(T) = \left[1 + (0.480 + 1.574\omega' - 0.176 \omega'^2) \left(1-\sqrt{T/T_c}\right)\right]^2,
    \end{equation}
    and
    \begin{equation}
        a=0.42748\frac{R^2T_c^2}{P_c},\ b=0.08664\frac{R T_c}{P_c},
    \end{equation}
    and the Carnahan--Starling EoS~\citep{carnahan_equation_1969},
    \begin{equation}\label{eq:CSEos}
        P = \rho R T \frac{1+b\rho/4 + {(b\rho/4)}^2 - {(b\rho/4)}^3}{{(1-b\rho/4)}^3} - a\rho^2,
    \end{equation}
    with
    \begin{equation}
        a=0.4963\frac{R^2T_c^2}{P_c},\ b=0.18727\frac{R T_c}{P_c}.
    \end{equation}
    
    \begin{figure}
	    \centering
		    \includegraphics{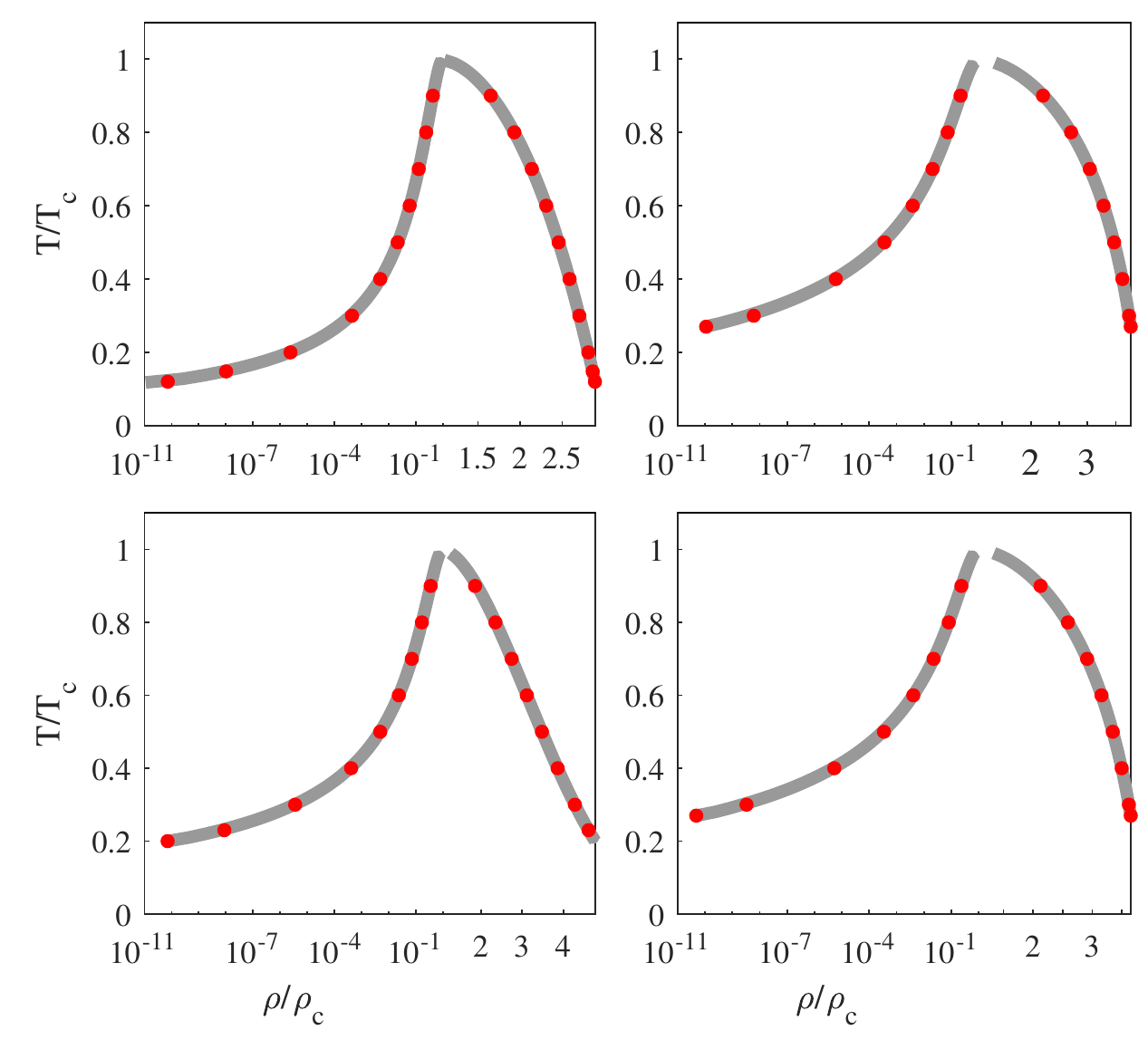}
	    \caption{Liquid-vapor coexistence for various equations of state. Gray lines: Maxwell's equal-area construction \eqref{eq:Maxwell_equal_area}; Red symbol:  Simulation.  Top left: van der Waals \eqref{eq:vdWEoS} ($a=0.000159$, $b=0.0952$);  Top right: Peng--Robinson \eqref{eq:PREoS} ($a=0.000159$, $b=0.0952$); Bottom left: Carnahan--Starling \eqref{eq:CSEos} ($a=0.000868$, $b=4$); Bottom right: Riedlich--Kwong--Soave \eqref{eq:RKSEoS} ($a=0.000159$, $b=0.0952$).
	    For all simulation $\tilde{\kappa}=0.02$.
	    }
	    \label{Fig:Coexistence}
    \end{figure}
    Fig.\ \ref{Fig:Coexistence} demonstrates that stationary density ratios $\rho_l/\rho_v$  obtained from the simulation are in excellent agreement with the theoretical coexisting liquid-vapour density ratios that are defined by Maxwell's equal-area rule \eqref{eq:Maxwell_equal_area}, for all four equations of state and for ratios as high as at least $\rho_l/\rho_v\sim 10^{11}$. It is noted that high coexisting density ratios were obtained without any tuning parameters, universally for all equations of state considered. Therefore, and without loss of generality, in the remainder of this article we only consider the van der Waals equation of state \eqref{eq:vdWEoS}.
    
\begin{figure}
	    \centering
		    \includegraphics{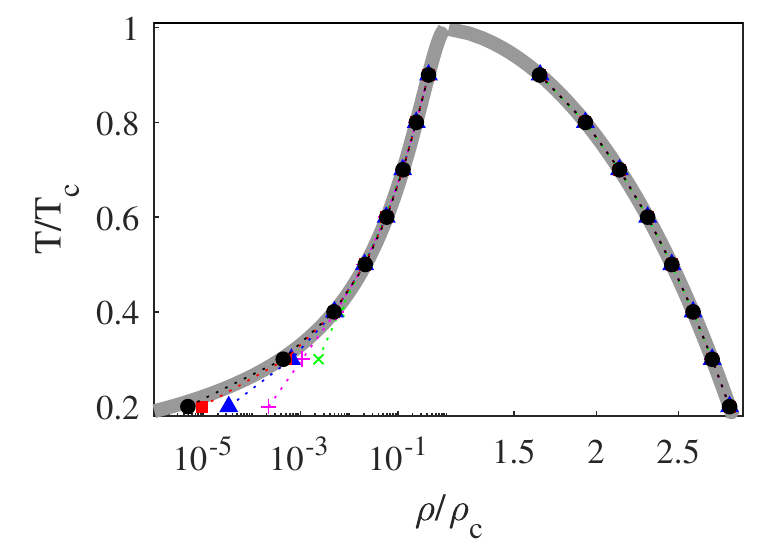}
	    \caption{Convergence to the principle of corresponding states. Coexistence densities as obtained from (grey lines) Maxwell's construction and (Markers) simulations with different choices of $a$: (green x) $a=0.0102$, (magenta +) $a=0.0051$, (blue triangles) $a=0.0026$, (red squares) $a=0.0013$ and (black circles) $a=0.00064$.}
	    \label{Fig:vdW_eqState}
    \end{figure}

    A discussion on the principle of corresponding states and the necessity of adherence to it in the simulation of realistic systems at large density ratios is in order. According to \cite{guggenheim_principle_1945}, 
    the principle of corresponding states is ``the most useful byproduct of van der Waals' equation of state''. 
    The principle maintains that all properties that depend on inter-molecular forces are related to the critical properties of the substance in a universal way, regardless of the molecular compound of interest. For the equation of state, the principle of corresponding states implies that the reduced pressure $P_r=P/P_c$ is a universal function of the reduced temperature $T_r=T/T_c$ and of the reduced density $\rho_r=\rho/\rho_c$,
    \begin{equation}
        \frac{P}{P_c} = f\left(\frac{T}{T_c}, \frac{\rho}{\rho_c}\right).\label{eq:correspondent_states}
    \end{equation}
    The universality of the reduced pressure \eqref{eq:correspondent_states} can be used to write Maxwell's equal-area rule in reduced form,
    \begin{equation}
        \int_{\rho_{r,v}}^{\rho_{r,v}} \frac{P_{r,\rm sat}(T_r) - P_r}{\rho_r^2} d\rho_r=0,
    \end{equation}
     The coexistence density ratio $\rho_l/\rho_v$ at a given reduced temperature $T_r$ is therefore also universal. 
     In order to probe the consistency with the principle of corresponding states in our model, we performed simulations for different values of the weak attraction parameter $a$ of the van der Waals equation of state \eqref{eq:vdWEoS}. Reduced coexistence densities are shown in Fig.\ \ref{Fig:vdW_eqState} for four different values of $a$. 
    It is observed that, down from the critical point to $T_r\approx 0.4$, coexistence densities essentially overlap for all the four values of $a$, in agreement with the principle of corresponding states. 
    Deviations from the principle of corresponding states are most pronounced on the vapour side, as observed in Fig.\ \ref{Fig:vdW_eqState}. This can be explained as follows: The local value of the scaling parameter $\varepsilon$ \eqref{eq:smallu} is estimated as $\varepsilon_{l,v}\sim \delta t \mathcal{F}/\rho_{l,v}\mathcal{U}$ on the liquid and on the vapour sides of the interface, respectively. Hence, their relative magnitude scales as the inverse of the density ratio, $\varepsilon_{v}/\varepsilon_l\sim\rho_l/\rho_v$. Thus, even if the scaling condition \eqref{eq:smallu} is satisfied on the liquid side, $\varepsilon_l\ll 1$, it is still prone to violation on the vapour side, if the density ratio $\rho_l/\rho_v$ becomes sufficiently large.

    Furthermore, due to the scaling relation between interface width and temperature detailed in section \ref{sec:interface_W}, the interface gets thinner at lower temperatures. This in turn means that, for a given $\delta r$, fewer grid-points resolve the interface with decreasing temperature. Whenever the parameters ${\delta r}/{W}$ and ${\delta t F}/{\rho \mathcal{U}}$  increase,
    contributions of higher orders in $\varepsilon$ become significant and lead to deviations from the analytical predictions. The characteristic interface width scales as $W \propto 1/\sqrt{a}$ \citep{jamet_second_2001}. As such lower values of $a$ lead to larger $W$, in parallel with smaller force increments over $\delta t$, which restores dominance of order $\varepsilon$ terms over Burnett and super-Burnett level contributions, and therefore the corresponding states principle. This last point is demonstrated by the convergence of the coexistence density ratio to the analytical predictions with decreasing $a$. For well-resolved simulations, as shown in Fig.\ \ref{Fig:Coexistence}, the model correctly recovers the coexistence densities and thus complies with the principle of corresponding states. 
    Below, we refer to the convergence of the scheme to the principle of corresponding states as the thermodynamic convergence.
    \subsection{Temperature dependence of the surface tension near the critical point \label{sec:surface_temperature}}
    Surface tension at liquid-vapour interface decreases with increasing temperature and vanishes at the critical point \citep{guggenheim_principle_1945}. 
    For the van der Waals equation of state,  the surface tension coefficient $\sigma$ follows a scaling law as $T_r\to 1$  \citep{van_der_waals_thermodynamische_1894,blokhuis_thermodynamic_2006},
    \begin{equation}\label{eq:vdw_surface_tension}
        \sigma = \frac{16a}{27b^2}\sqrt{\frac{\kappa}{a}} {\left(1-T_r\right)}^{3/2}.
    \end{equation}
    In order to probe the consistency of the proposed lattice Boltzmann model, the temperature dependence of the surface tension was numerically measured  in two different configurations, the flat interface and the circular drop, in a temperature interval $T_r\in[0.85,\, 1]$.
    
    In the first configuration, the surface tension coefficient was evaluated using its definition for the flat interface  \citep{kirkwood_statistical_1949},
    \begin{equation}\label{eq:surface_tension_int}
        \sigma = \int_{-\infty}^{+\infty} \left(P_{xx} - P_{yy}\right)dx,
    \end{equation}
    where the interface is  normal to the $x$-axis in a two-dimensional simulation setup. The normal $P_{xx}$ and the tangential $P_{yy}$ components of the discrete pressure tensor were computed 
    using a formalism developed in Appendix \ref{App:Discrete_Pressure}, following a proposal by \cite{shan_pressure_2008}.
    
    In the second configuration, simulations of  circular liquid drops surrounded with vapour at the center of a square domain were conducted. At each temperature, four different initial drop radii were considered, $R_0\in\{45,55,65,75\}$, chosen in such a way that the interface width is sufficiently small compared to initial drop radius. In the simulation, we used $W/R_0 \le 0.1$,
    where $W=(\rho_l-\rho_v)/\max\lvert\bm{\nabla}\rho\lvert$ is the interface width, in order to minimize curvature-dependence of surface tension. 
    The corresponding surface tension coefficient was evaluated using the Laplace law ($D=2$) in a form, 
    \begin{equation}\label{eq:Laplace_equimolar}
        \Delta P = \frac{(D-1)\sigma}{R_e},
    \end{equation} 
    where the radius $R_e$ corresponds to the \emph{equimolar} dividing surface \citep{gibbs_equilibrium_1874}. 
    
\begin{figure}
	    \centering
		    \includegraphics{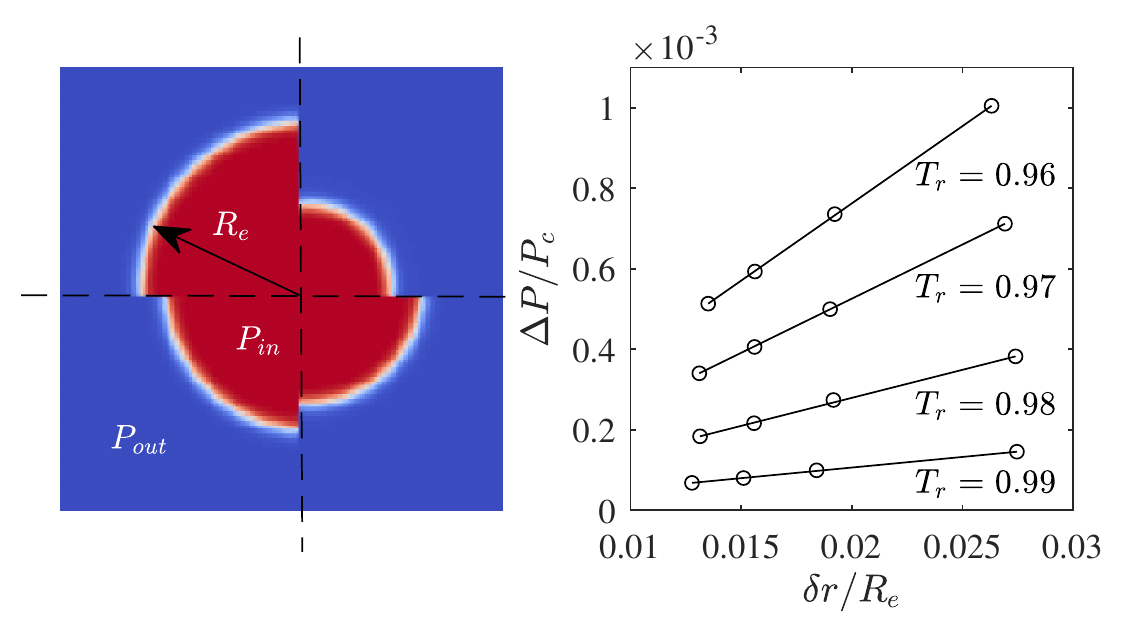}
	    \caption{Left: Circular $D=2$ drop configurations; Right: 
	    Pressure difference scaling with drop radius for $T_r=0.99,0.98,0.97$ and $0.96$.
	    The pressure difference is defined as $\Delta P=P_{\rm in}-P_{\rm out}$. The slope of the straight line is the surface tension coefficient.}
	    \label{Fig:SurfTenTempSlopes}
    \end{figure}

    A brief reminder of Gibbs' theory of dividing surfaces is in order. The total mass in both the diffuse and sharp interface pictures can be written as:
    \begin{equation}\label{eq:mass_sharp}
        \int_{V} \rho dV = \rho_l V_l + \rho_v V_v + \Gamma,
    \end{equation}
    where $\rho_l V_l$ and $\rho_v V_v$ are the masses in the bulk liquid and vapor phases in the sharp interface picture, while $\Gamma$ is the  excess mass on a dividing surface $\Sigma$, or mass \emph{adsorbance} \citep{gibbs_equilibrium_1874}.
    By requiring that no mass be stored on the dividing surface we get the definition of the \emph{equimolar} surface:
    \begin{equation}\label{eq:equimolar_adsorbance} 
        \Gamma = 0.
    \end{equation}
     The family of dividing surfaces in the case of drop or bubble  are concentric spheres ($D=3$) or concentric circles ($D=2$) parameterized by their radius $R$.
    In particular, for a two-dimensional drop, the mass adsorbance 
     can be written as a function of the radius of the dividing circle,
    \begin{equation}
       \Gamma(R)= \int_0^{2\pi}\int_0^{\infty}(\rho(r) - \rho_v)r drd\varphi - \int_0^{2\pi}\int_0^{R}(\rho_l - \rho_v)r drd\varphi,
    \end{equation}
    while the zero-adsorbtion condition \eqref{eq:equimolar_adsorbance}, $\Gamma(R_e)=0$, implies the {equimolar} radius $R_e$,
    \begin{equation}\label{eq:equmol2Ddrop}
        R_e = \sqrt{\frac{\int_0^{\infty}(\rho(r) - \rho_v)r dr}{\left(\rho_l - \rho_v\right)}}.
    \end{equation}
     The drop configuration along with the scaling of the pressure difference across the interface with drop radius are shown in Fig.\ \ref{Fig:SurfTenTempSlopes}.

\begin{figure}
	    \centering
		    \includegraphics{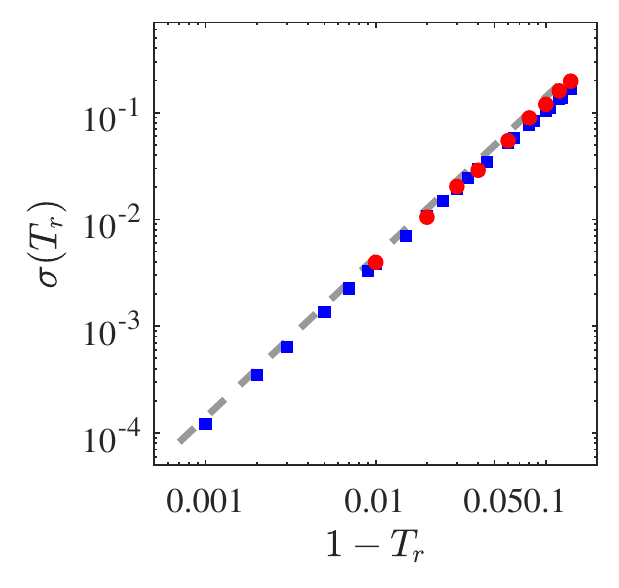}
	    \caption{Temperature dependence of the surface tension coefficient near the critical point. Dashed grey line: Theory, Eq.\ \eqref{eq:vdw_surface_tension}; Red circles:  Simulation results using Laplace's law \eqref{eq:Laplace_equimolar}; Blue squares: surface tension coefficient computed via Eq.\ \eqref{eq:surface_tension_int}.}
	    \label{Fig:SurfTenTemp}
    \end{figure}

    The results as obtained from both the flat interface and the drop configurations are shown in Fig.\ \ref{Fig:SurfTenTemp}.
    It is clearly observed that the surface tensions as obtained from the proposed formulation (using either one of the considered configurations) agree very well with Eq.\ \eqref{eq:vdw_surface_tension}, provided that $W \ll R_{e}$. Discussion of curvature-dependence of surface tension shall be continued in sec.\ \ref{sec:Tolman}.
    \subsection{Control of surface tension\label{sec:tunability}}
    The present formulation allows us to select the surface tension in the simulation via the capillarity parameter $\kappa$, independently from the density ratio and temperature. In order to demonstrate this feature, flat interface simulations were performed at three  reduced temperatures,  $T_r=\{0.99, 0.98, 0.97\}$, for different values of $\tilde{\kappa}$. The surface tension was evaluated using Eq.\ \eqref{eq:surface_tension_int}.   
    \begin{figure}
	    \centering
		    \includegraphics{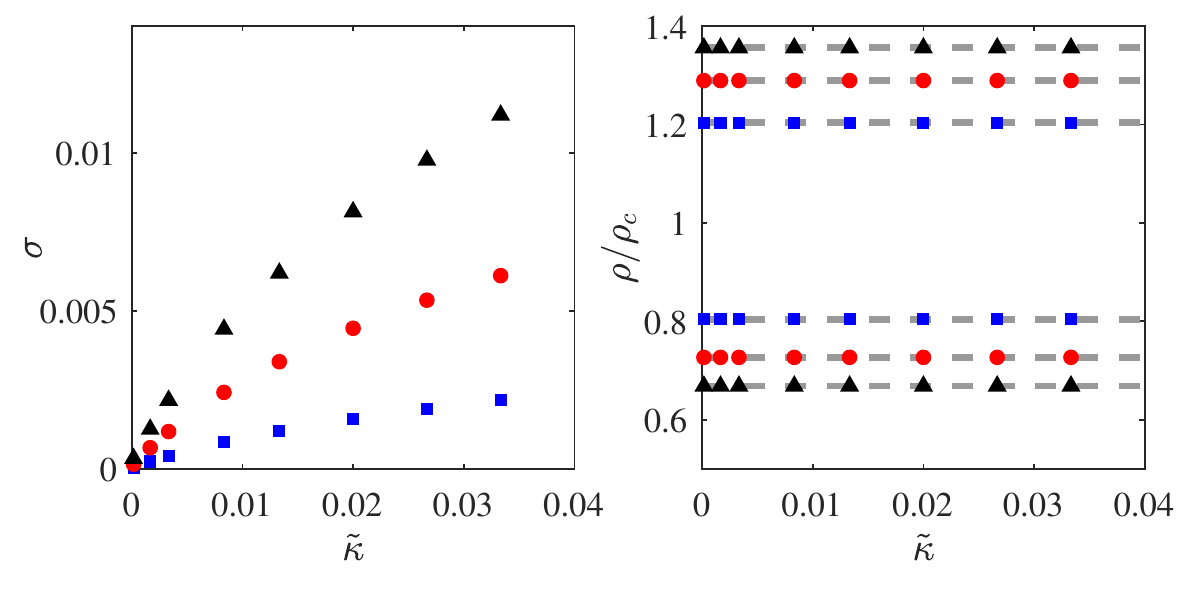}
	    \caption{Left: Surface tension as a function of capillarity parameter $\tilde{\kappa}$; Right: liquid/vapor densities as a function of $\tilde{\kappa}$. Results of simulation shown at three different reduced temperatures: Blue square: $T_r=0.99$;
	    Red circles: $T_r=0.98$; Black triangles: $T_r=0.97$. Dashed grey lines:  Theoretical coexistence densities from the Maxwell construction.
	    }
	    \label{Fig:SurfTenKappa}
    \end{figure}
    Results in Fig.\ \ref{Fig:SurfTenKappa} demonstrate that the surface tension can be effectively tuned using $\tilde{\kappa}$ and that changes in surface tension do not affect the equilibrium density ratios. This is further detailed in Table\ \ref{tab:capillary_coeff} where the values of surface tension and deviations of the vapor and liquid densities from theoretical values are given as a function of $\tilde{\kappa}$ for $T_r=0.97$.
    \begin{table}
    \begin{center}
    \def~{\hphantom{0}}
    \begin{tabular}{lcccccccc}
       $\tilde{\kappa}\times10^{3}$~& 0.2 & 1.7 & 3.3 & 8.3 & 13.3 & 20 & 26.7 & 33.3\\[3pt]
       $\sigma\times10^{2}$~& 0.03 & 0.13 & 0.22 & 0.44 & 0.62 & 0.81 & 0.98 & 1.12\\
       $\frac{\lvert\Delta\rho_v\lvert}{\rho_v}\times10^{3}$~& 0.41 & 0.33 & 0.27 & 0.15 & 0.1 & 0.06 & 0.04 & 0.03\\
       $\frac{\lvert\Delta\rho_l\lvert}{\rho_l}\times10^{3}$~& 0.11 & 0.09 & 0.08 & 0.04 & 0.03 & 0.02 & 0.01 & 0.009\\
    \end{tabular}
    \caption{Effect of the choice of $\tilde{\kappa}$ on surface tension and deviations in equilibrium vapor and liquid phases densities for $T_r=0.97$}
    \label{tab:capillary_coeff}
    \end{center}
    \end{table}
    It is clearly observed that while the surface tension covers two orders of magnitude
    the deviation from the predicted density of vapour is at most  $0.041$ percent.
    Furthermore, in the limit of vanishing $\kappa$ the surface tension also vanishes, as expected from the theory, Eq.\ \eqref{eq:vdw_surface_tension}.
    \subsection{Effect of curvature on surface tension: Tolman length \label{sec:Tolman}}
    
    The drop simulation in section \ref{sec:surface_temperature} made use of the Laplace law, relying on the equimolar dividing surface of radius $R_e$ \eqref{eq:equmol2Ddrop}. Further discussion on the non-uniqueness of the choice of dividing surface and curvature-dependence of surface tension is in order. Following \cite{gibbs_equilibrium_1874}, the free energy of 
     a  drop or bubble separated from the surrounding vapour or liquid by a dividing circle ($D=2$) or sphere ($D=3$) of length or area $\Sigma$ is, $A=U-TS+\sigma \Sigma$, where $U$ and $S$ are the internal energy and entropy of bulk phases while the last term is the adsorbance of free energy.
     The equilibrium condition requires vanishing of the variation $\delta A$; for the isothermal case we have,
    \begin{equation}
        \delta A=-P_{l,v}\delta V_{l,v}-P_{v,l}\delta V_{v,l} +\Sigma\delta \sigma+\sigma\delta \Sigma=0,
    \end{equation}
    where $P_{l,v}$ and $P_{v,l}$ are the pressures inside and outside the liquid drop or vapour bubble.
    Using {$\delta V_{l,v}=-\delta V_{v,l}=2(D-1)\pi R^{D-1} \delta R$} and {$\delta \Sigma=2(D-1)^2\pi R^{D-2} \delta R$} leads to a generalized Laplace law,
    \begin{equation}\label{eq:Laplace_gen}
        \Delta P = \frac{(D-1)\sigma(R)}{R} + \frac{d \sigma(R)}{dR}.
    \end{equation}
    The derivative of surface tension $d\sigma/dR$ is termed a \emph{notional} derivative by some authors \citep{blokhuis_pressure_1992} in order to stress that it refers to arbitrariness of the dividing surface.
    Apart from the equimolar surface \eqref{eq:equmol2Ddrop},
    the \emph{surface of tension} is another possible choice to lift the ambiguity of the dividing surface. 
    The notional derivative vanishes at the surface of tension,
    \begin{equation}\label{eq:condition_Rs}
    \frac{d\sigma}{dR}\bigg|_{R=R_s}=0,
    \end{equation}
    thereby reducing the generalized Laplace law \eqref{eq:Laplace_gen} to a standard form,
    \begin{equation}\label{eq:Laplace_surface_of_tension}
        \Delta P = \frac{(D-1)\sigma(R_s)}{R_s}.
    \end{equation}
    Integrating \eqref{eq:Laplace_gen} from $R_s$ to $R$, and eliminating $\Delta P$ using  \eqref{eq:Laplace_surface_of_tension}, one obtains  analytic expression for the notional surface tension $\sigma(R)$ relative to its minimum $\sigma_s$ at the surface of tension $R_s$,
    \begin{equation}\label{eq:surface_of_tension_drop}
       \frac{\sigma(R)}{\sigma_s} = \frac{1}{D}{\left(\frac{R_s}{R}\right)}^{D-1} + \frac{D-1}{D}\left(\frac{R}{R_s}\right).
    \end{equation}
    
\begin{figure}
	    \centering
		    \includegraphics{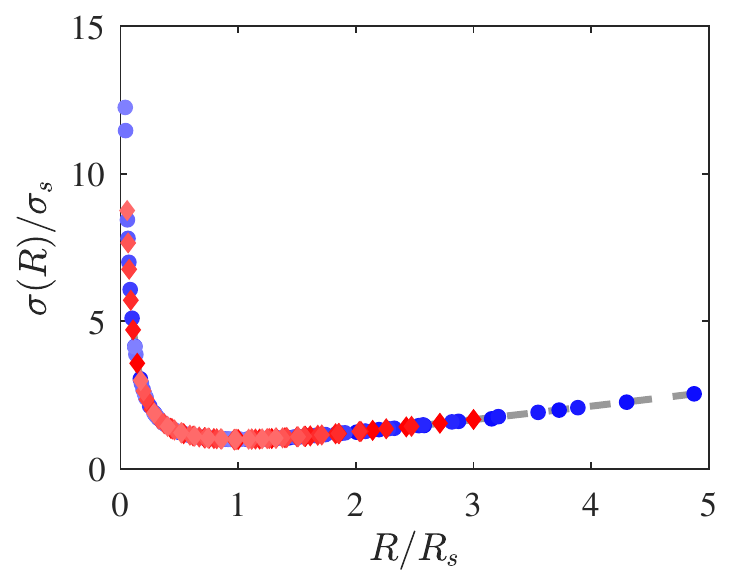}
	    \caption{Variation of surface tension $\sigma$ with the radius of the dividing circle $R$ at $T_r=0.98$. Gray dashed line: Theory,  Eq.\ \eqref{eq:surface_of_tension_drop}; Symbol: Simulations with van der Waals equation of state. Blue circles: Drop simulations; Red diamonds: Bubble simulations. Shading from dark to light: Drops and bubbles of increasing sizes are shown. 
	    }
	    \label{Fig:SigmasRs}
    \end{figure}

    Eq.\ \eqref{eq:surface_of_tension_drop} provides for a simple way of identifying the surface of tension and the corresponding surface tension.
    Two-dimensional simulations at $T_r=0.98$ were conducted with different initial drop and bubble sizes, $R_0\in\{30,40,50,60,70,80,100,120,140\}$.
     For each dividing surface of radius $R$, the corresponding surface tension  was evaluated as \citep{blokhuis_pressure_1992},
    \begin{equation}\label{eq:sigmaR}
        \sigma(R) = \int_{0}^{\infty} {\left(\frac{r}{R}\right)}^{D-1}\left[P_{\perp}(r,R) - P_{\parallel}(r)\right]dr,
    \end{equation}
    where $P_{\parallel}(r)$ is the tangential component of the pressure tensor,  computed via the discrete pressure tensor detailed in Appendix\ \ref{App:Discrete_Pressure}, 
    and 
    \begin{equation}\label{eq:PJ_eq}
        P_{\perp}(r,R) = P_{\rm in}  - (P_{\rm out} - P_{\rm in}) H(r-R),
    \end{equation}
    is the normal pressure component in the sharp interface system, where $H$ is the Heaviside step function. 
    For each drop or bubble, surface tension $\sigma(R)$ \eqref{eq:sigmaR} was probed at seventeen equidistant dividing circles between $R_{\min}=5\delta r$ and $R_{\max}=165\delta r$. The discrete values $\sigma(R)$ obtained in these simulations were fitted with Eq.\ \eqref{eq:surface_of_tension_drop}, with $\sigma_s$ and $R_s$ as the free fitting parameters. Fig.\ \ref{Fig:SigmasRs}
    shows that the data for all drops and bubbles collapsed on a single master curve, in excellent agreement with the theory \eqref{eq:surface_of_tension_drop}.
    Thus, the proposed model correctly identifies the surface of tension for the van der Waals fluid.
    
    While the notional derivative $d\sigma(R)/dR$ vanishes at the surface of tension, the same does not hold for the surface tension \emph{at} the surface of tension, $d\sigma_s/dR_s\neq0$. In other words,  surface tension $\sigma_s$ depends on the curvature of the surface of tension. In a seminal paper, \cite{tolman_effect_1949} characterized  the curvature-dependence of surface tension by the Tolman length: For sufficiently large $R_s$, the leading-order curvature-dependence of the surface tension may be written \citep{tolman_effect_1949},
    \begin{equation}\label{eq:tolman_length}
        \sigma(R_s) \approx \sigma_0\left(1 \mp \frac{(D-1)\delta_T}{R_s}\right),
    \end{equation}
    where $\delta_T$ is the Tolman length and $\sigma_0$ is the flat interface surface tension coefficient. Here, the negative (positive) sign corresponds  to drops (bubbles), respectively.
   
\begin{figure}
	    \centering
		    \includegraphics{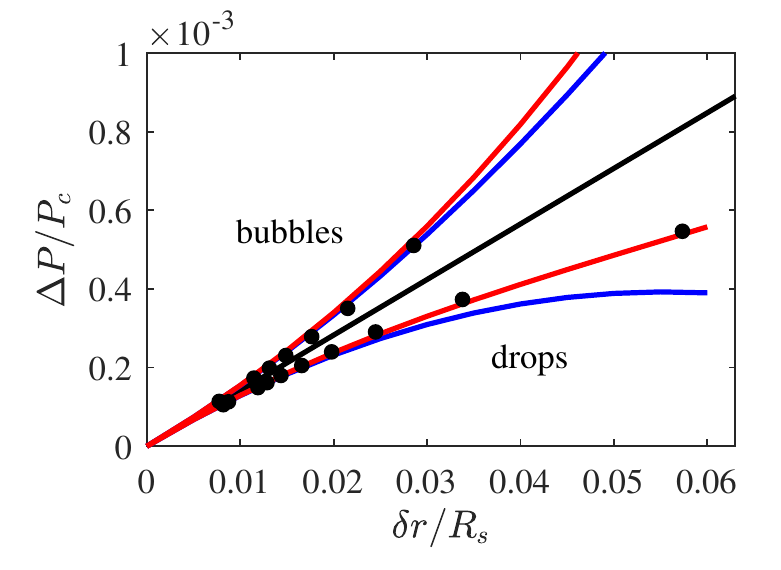}
	    \caption{Pressure difference scaling with surface of tension radius $R_s$ for liquid drops and vapour bubbles. Symbol: Simulation data for van der Waals equation of state with $a=0.18$ and $b=0.095$. 
	    Black line: Laplace law with $\sigma=\sigma_0$; Blue lines: Best fit with Eq.\ \eqref{eq:tolman_length} used to compute Tolman length $\delta_T$; Red lines: Best fit with the second-order Helfrich expansion \eqref{eq:Helfrich}.
	    }
	    \label{Fig:TolmanLaplace}
    \end{figure}

    We first compare the leading-order Tolman model \eqref{eq:tolman_length} with the simulation. The values of $R_s$ obtained in the previously described drops and bubbles simulations of Fig.\ \ref{Fig:SigmasRs} are plotted against the pressure difference for different drop and bubble sizes in Fig.\ \ref{Fig:TolmanLaplace}. It is clear that, for smaller drops and bubbles, the pressure difference deviates from the Laplace law with constant $\sigma_s=\sigma_0$, indicating a curvature-dependent surface tension.
    Fitting the data points with Eq.\ \eqref{eq:tolman_length}, the Tolman length can be extracted from the simulation, here $\delta_T=9\delta r$ for both drops and bubbles, at the reduced temperature $T_r=0.98$. 
    
    While the leading-order Tolman correction \eqref{eq:tolman_length}  improves agreement with data at moderate $R_s$, deviation persists for smaller drops and bubbles at $\delta r/R_s> 0.03$.
    Higher-order terms in the inverse powers of $R_s$, important for droplets or bubbles of small radius, are neglected by \eqref{eq:tolman_length} and were addressed by \cite{helfrich_steric_1978}:
   \begin{equation}\label{eq:Helfrich}
       \sigma = \sigma_0 \mp \sigma_0\frac{(D-1)\delta_T}{R_s} + \frac{k{(D-1)}^2}{2R_s^2} + \frac{\bar{k}(D-2)}{R_s^2} + \dots,
   \end{equation}
   where $k$ and $\bar{k}$ are the bending and Gaussian rigidities; note that the latter vanishes for $D=2$.
   Taking the second-order term \eqref{eq:Helfrich} into account, the best fit in Fig.\ \ref{Fig:TolmanLaplace} results in bending rigidity $k=1.049\times10^{5}\sigma_0\delta r^2$.

\begin{figure}
	    \centering
		    \includegraphics{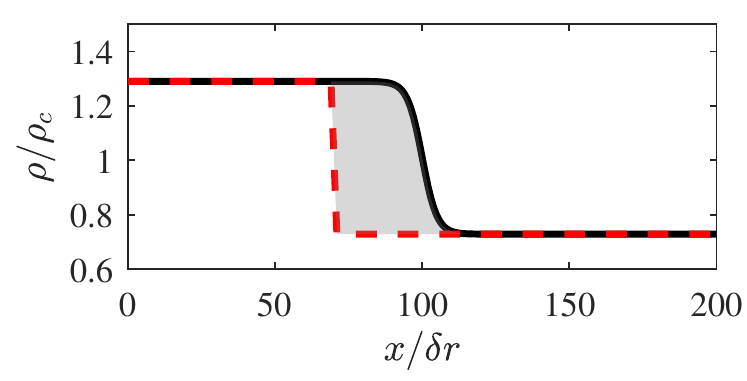}
	   \caption{Example of mass adsorbance for a flat interface. Black continuous line: density profile at $T_r=0.98$;
	   Red dashed line: Sharp interface profile with the dividing surface at $X/\delta r$=70. Grey area represents the mass   adsorbance $\Gamma(X)$ \eqref{eq:Gamma_flat}.}
	   \label{Fig:DivSurfIllustration}
    \end{figure}

    Finally we consider the limit of flat interface where the Tolman length can be derived independently from the above considerations.
    In this case, location of the surface of tension $X_s$ can be found as the normalized first-order moment of the {normal stress difference}
    \citep{rao_location_1979},
    \begin{equation}\label{eq:surface_of_tension_flat}
        X_s = \frac{\int_{-\infty}^{\infty} x(P_{xx} - P_{yy}) dx}{\int_{-\infty}^{\infty} (P_{xx} - P_{yy}) dx}.
    \end{equation}
    With the dividing surface as a vertical straight line at $X$ in two dimensions, the mass adsorbance $\Gamma(X)$ is defined as (see example in Fig.\ \ref{Fig:DivSurfIllustration}),
    \begin{equation}\label{eq:Gamma_flat}
       \Gamma(X)= \int_{-\infty}^{\infty} \left[\rho(x) - \rho_v - (\rho_l-\rho_v) H(x-X)\right] dx.
    \end{equation}
    Similar to the case of cylindrical symmetry considered in sec.\ \ref{sec:surface_temperature}, the equimolar surface is found by annihilating the mass adsorbance, $\Gamma(X_e)=0$.
    
\begin{figure}
	    \centering
		    \includegraphics{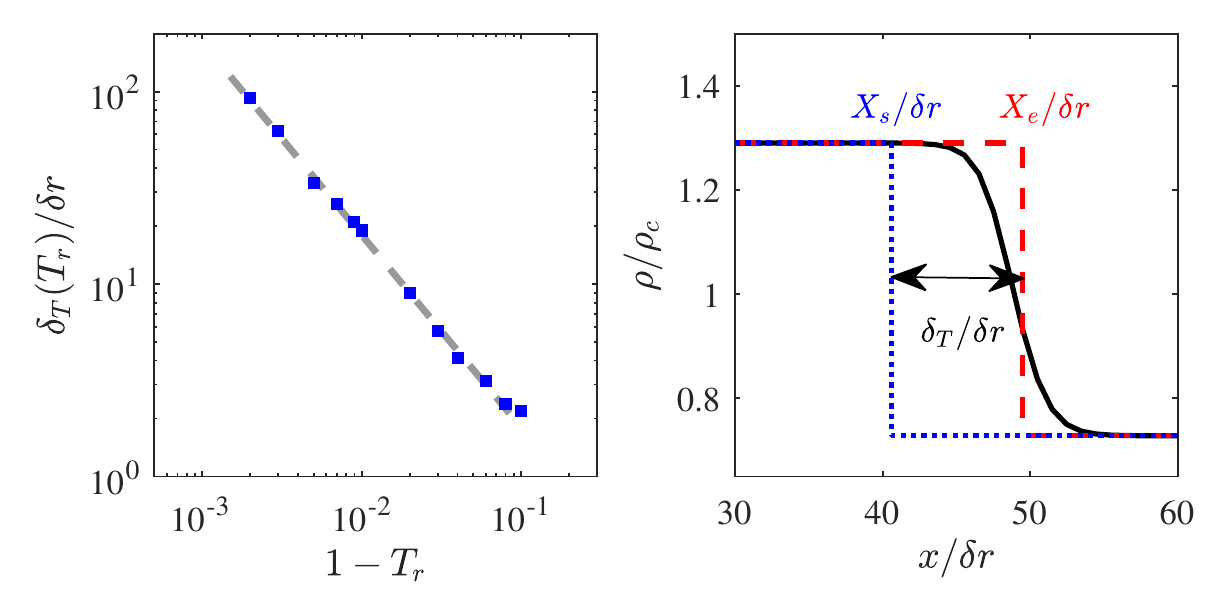}
	    \caption{(Left) Temperature dependence of  Tolman length $\delta_T$.
	    Results from the flat interface simulation for van der Waals fluid are shown with blue squares while the grey dashed line represents  theoretical scaling \eqref{eq:TolmanT}. (Right) Surface of tension, equimolar surface and Tolman length for flat interface. Continuous black line: Density profile at $T_r=0.98$; Red dashed line:  Sharp approximation with the equimolar surface as the dividing surface; Blue dotted line:
	    Sharp approximation with the surface of tension as the dividing surface. Distance between the surface of tension and the equimolar surface: Tolman length. In all simulations $a=0.18$ and $b=0.095$.}
	    \label{Fig:TolmanTempScale}
    \end{figure}

    The Tolman length in the limit of flat interface is the distance between the surface of tension and the equimolar surface~\citep{blokhuis_pressure_1992,blokhuis_thermodynamic_2006},
    \begin{equation}\label{eq:Tolman_flat}
        \delta_T=X_e - X_s.
    \end{equation}
    An example from the simulation is presented in Fig.\ \ref{Fig:TolmanTempScale}. 
    As shown by \cite{blokhuis_pressure_1992}, the Tolman length scales with the reduced temperature as,
    \begin{equation}\label{eq:TolmanT}
        \delta_T\propto{(1-T_r)}^{-1}.
    \end{equation}
     In order to validate the scaling \eqref{eq:TolmanT} in our model, flat interface simulations were conducted over a range of temperatures in the vicinity of the critical state. The Tolman length \eqref{eq:Tolman_flat} for various temperatures is shown in Fig.\ \ref{Fig:TolmanTempScale}. 
    The results obtained from simulations are in excellent agreement with the theoretical scaling \eqref{eq:TolmanT}. A similar study using the Shan-Chen equations of states and the pseudo-potential model was presented in \citep{lulli2021mesoscale}. Furthermore,  the flat interface simulations lead a to $\delta_T=9.2\delta r$ at $T_r=0.98$,
    in agreement with the value obtained from drop and bubble simulations, $\delta_T=9\delta r$ at same temperature. The relatively small discrepancy can be attributed to higher-order terms in the curvature, neglected in the fitting process.
    \subsection{Interface width: Temperature scaling and control\label{sec:interface_W}}
In the present work we use a definition for the interface width bearing numerical information as to how well the stiff gradients are resolved on a given mesh, making it directly related to the velocity increment per time-step:
    \begin{equation}\label{eq:interface_W}
        W = \frac{\rho_l-\rho_v}{\max\lvert\bm{\nabla}\rho\lvert},
    \end{equation}
    where $\rho_l$ and $\rho_v$ are densities of saturated liquid and vapor, respectively. It can readily be observed that in the limit of a sharp interface, i.e. resolved with $\delta r$, $\delta r/W\to 1$. As previously noted for the co-existence densities and surface tension, the model recovers thermodynamical properties/scalings of the second-gradient fluid in the limit $\varepsilon\to 0$, akin to $\delta r/W\to 0$. As such in the limit of $\delta r/W \to 1$ one expects numerical effects to dominate and observe deviations from thermodynamics of the van der Waals fluid.
    
    Surface tension vanishes as the temperature approaches the critical, cf.\ section \ref{sec:surface_temperature}, Eq.\ \eqref{eq:vdw_surface_tension} and Fig.\ \ref{Fig:SurfTenTemp}, while the interface  diverges as $T\to T_c$.  As noted by \cite{widom_surface_1965}, the van der Waals theory predicts the temperature scaling  of the interface width as,
    \begin{align}\label{eq:Wscaling}
    	W(T_r)\propto (1-T_r)^{-1/2}.
    \end{align} 
    In order to validate consistency of the proposed model, simulations of flat interfaces were carried out in a range of reduced temperatures $T_r$ near the critical point, and corresponding interface widths $W(T_r)$ \eqref{eq:interface_W} were measured. Given the effect of the choice of $a$ on interface thickness simulations were also carried out with different $a$.
    \begin{figure}
	    \centering
		    \includegraphics{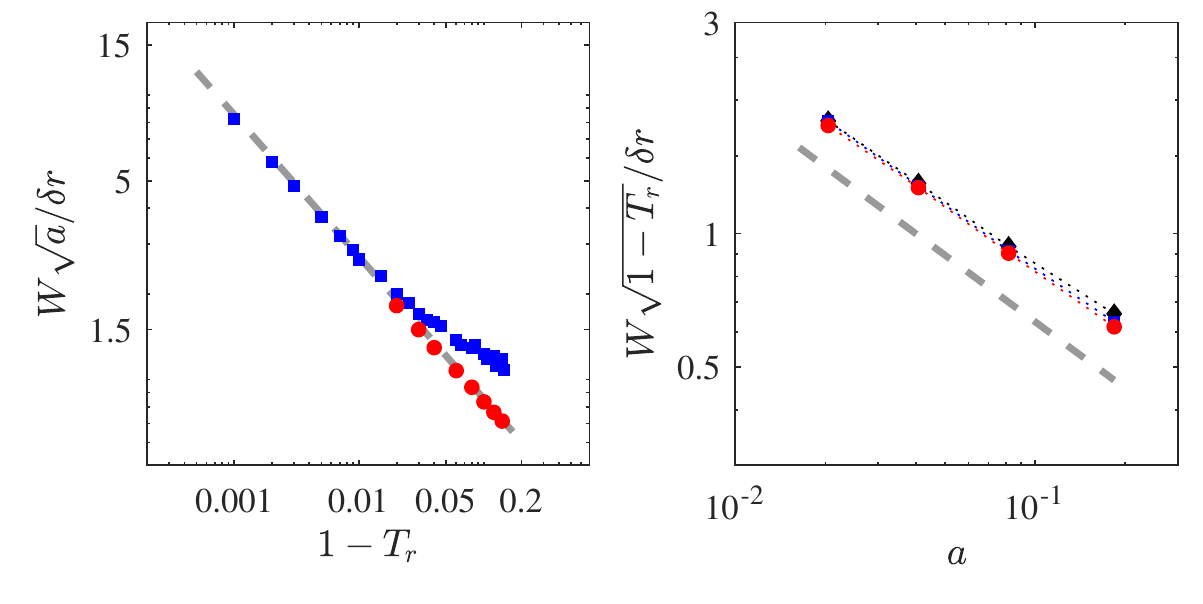}
	    \caption{(Left) Interface width as a function of temperature. Blue square:  Simulation with $a=0.184$;  Red circle: Simulation with $a=0.02$;  Grey dashed line: Theoretical scaling \eqref{eq:Wscaling}.  
	    (Right) Effect of the choice of $a$ on the interface width. Diamond, square and circle: Simulation for $T_r= 0.98, 0.99, 0.995$, respectively.}
	    \label{Fig:WidthTemp}
    \end{figure}
    The results obtained from simulations near the critical point are shown in Fig.\ \ref{Fig:WidthTemp}. As noted by many authors in the literature \citep{jamet_second_2001}, the parameter $a$ can be used to control the interface thickness, as done in the present work, at a given density ratio, leaving the ratios and Maxwell construction unaffected. In agreement with the equivalent states theory, the right hand side plot in Fig.\ \ref{Fig:WidthTemp} points to the universality of the scaling of the interface width near the critical point regardless of the choice of $a$. In addition it is interesting to note that for a fixed grid-size $\delta r$, as $(1-T_r)\to1$, $\delta r/W\to 1$ (equivalent to the scaling parameter $\varepsilon$ introduced in the multi-scale analysis) indicating deviation from the thermodymically converged state. This is illustrated by the deviation of the numerical interface thickness, starting at $T_r\approx0.98$ from the theoretical predictions. Lowering the value of $a$, i.e. rescaling the interface by a factor $1/\sqrt{a}$ and therefor lowering $\varepsilon$, it is observed that interface is again well-resolved and the scaling \eqref{eq:Wscaling} restored. In agreement with previous sections, in the limit of $\varepsilon\to 0$, the model is shown to be thermodynamically converged and recovers the properties of the second-gradient fluid.
    \section{Hydrodynamic consistency \label{sec:hydro_consistent}}
    \subsection{Shear stress: Layered Poiseuille flow \label{sec:Poiseuille}}
    The setup consists of a rectangular domain filled with the liquid phase at the bottom and the vapor phase on top. The flow is driven by a body force. Top and bottom are subject to no-slip boundary conditions while the inlet and outlet are fixed by periodicity.
    The momentum balance at steady state reduces to a well-posed system of ordinary differential equations,
    \begin{subequations}
	\begin{align}
		\partial_y \mu_l \partial_y u + \rho_l g = 0, &\forall y: 0\leq y \leq h_l, \\
		\partial_y \mu_v \partial_y u + \rho_v g = 0, &\forall y: h_l\leq y \leq H, 
		\end{align}
	\label{Eq:poiseuille_odes}
    \end{subequations}
    closed by the following boundary conditions:
    \begin{subequations}
	\begin{align}
		u\lvert_{y=0} &= 0, \\
		u\lvert_{y=H} &= 0, \\
		u\lvert_{y=h_l^{-}} &= u\lvert_{y=h_l^{+}}, \\
		\mu_l\partial_y u\lvert_{y=h_l^{-}} &= \mu_v \partial_y u\lvert_{y=h_l^{+}}.
		\end{align}
	\label{Eq:poiseuille_odes_BD}
    \end{subequations}
    Here $H$ is the height of the channel, $h_l$ the height of the section filled with the liquid phase, $\mu_l$ and $\mu_v$ are the dynamic viscosities of the liquid and vapor, respectively, and $g$ is the acceleration due to the body force.
    Analytical solution is given in Appendix \ref{App:Poiseuille_analytical}.
    This configuration is of particular interest as it probes ability of a two-phase model to capture jump conditions at the interface which happens only if deviatoric components of the viscous stress tensor \eqref{eq:viscous_stress}
    are correctly recovered. 
    
\begin{figure}
	    \centering
		    \includegraphics{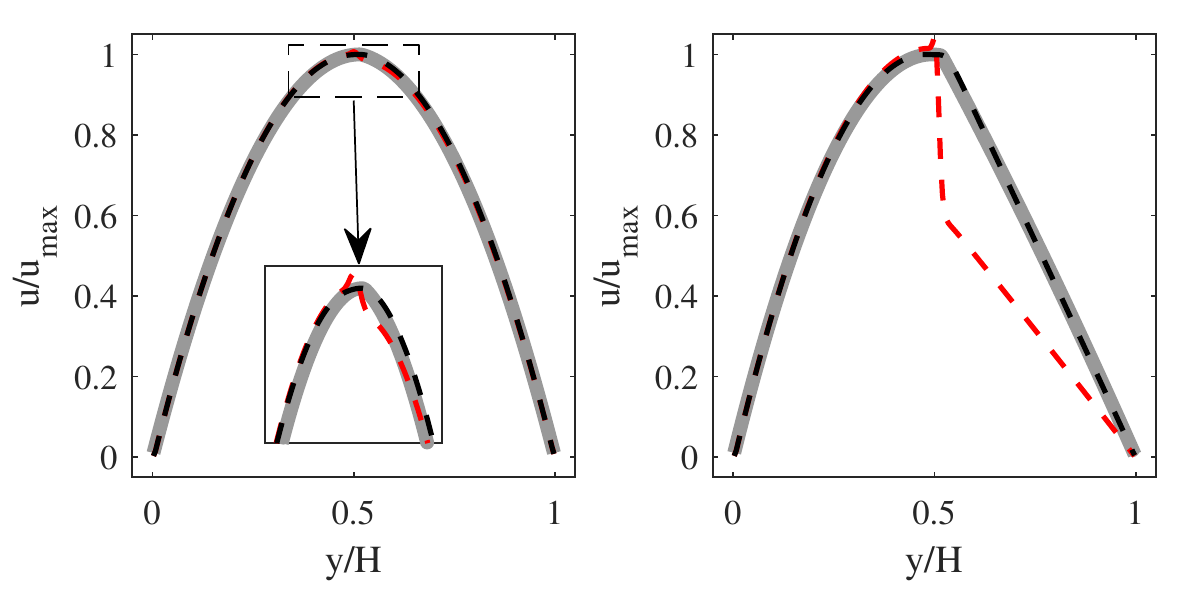}
	    \caption{Steady-state velocity profiles for the layered Poiseuille flow. Left: Configuration (a), $T_r=0.77$, $\rho_l/\rho_v=10.1$, $\mu_l/\mu_v=10.1$ ; Right: Configuration (b), and (right) $T_r=0.36$,  $\rho_l/\rho_v=1030$ and $\mu_l/\mu_v=11.3$. Grey plain line: analytical solution; Black dashed line: LBM with product-form equilibrium; Red dashed line: LBM with conventional second-order equilibrium.}
	    \label{Fig:Poiseuille_R10_R1000}
    \end{figure}

     Two sets of parameters were considered: (a) $\mu_l/\mu_v=10.1$, $\rho_l/\rho_v=10.1$ ($T_r=0.77$) and (b) $\mu_l/\mu_v=11.3$, $\rho_l/\rho_v=1030$ ($T_r=0.36$). 
    Simulations were conducted on a grid of size $5\times200$, with a constant acceleration  
    $g=10^{-8}\delta r/\delta t^2$.
    Furthermore, the same simulations were performed with the conventional second-order equilibrium \citep{succi_lattice_2002} instead of the product-form \eqref{eq:LBMeq}. 
    The steady-state results are compared to analytical solution  in Fig.\ \ref{Fig:Poiseuille_R10_R1000}.
    Contrary to the conventional second-order equilibrium, the use of the product form \eqref{eq:LBMeq} in our model recovers the correct jump conditions and therefore results in continuous velocity profiles, also for a large density ratio, in excellent agreement with analytical solution.
    \subsection{Normal stress: Dissipation of acoustic waves \label{sec:acoustics}}
\begin{figure}
	    \centering
		    \includegraphics{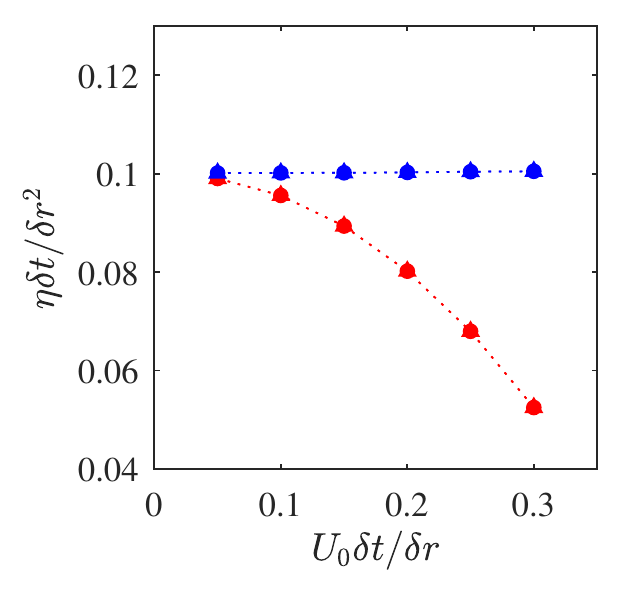}
	    \caption{Dissipation rate of acoustic mode at $T_r=0.36$. Circle: Saturated liquid; Triangle: Saturated vapour. Blue: Simulation of the model with the correction term; Red: Simulation of the model without correction term.}
	    \label{Fig:AcDiss_vdW_R1000}
    \end{figure}

    To assess Galilean invariance of the diagonal components of the viscous stress tensor and the effect of the correction term,  dissipation of acoustic waves was measured numerically. Acoustic waves were initialized as a small perturbation of density with an amplitude $\delta \rho$ and subject to uniform background velocity $U_0$ representing a moving reference frame,
    \begin{equation}
        \rho(x,0) = \rho_0 + \delta \rho \sin\left(2\pi/\lambda\right),
    \end{equation}
    where $\rho_0$ is the density of saturated liquid or vapor at a given temperature and $\lambda$ is the wavelength of the perturbation. The maximum velocity in the domain was monitored over time and fitted to an exponential function of the form, $\max(u) - U_0 = \exp\left(-\Lambda t\right)$, where the coefficient $\Lambda$ is tied to the dissipation rate of the normal modes as,
    \begin{equation}\label{eq:viscosity_acoustic}
        \eta = \frac{\Lambda}{{(2\pi/\lambda)}^2}.
        \end{equation}
    Simulations were performed on a periodic domain of the size $L=128\delta r\times \delta r$ with $\lambda=64\delta r$ at the temperature $T_r=0.36$, corresponding to density ratio  $\rho_l/\rho_v\approx 10^3$. Perturbation amplitude was set to $\delta\rho = 10^{-4}\rho_0$ while the reference frame velocity varied in the range $U_0\delta t/\delta r\in[0, 0.3]$.
    The dissipation rate \eqref{eq:viscosity_acoustic} with and without the correction term are shown in Fig.\ \ref{Fig:AcDiss_vdW_R1000}. It is observed that the correction term for the third-order diagonal moments of the equilibrium populations \eqref{eq:correction} restores the Galilean invariance (independence from the reference frame velocity) of the dissipation rate of normal modes.
    \subsection{Viscosity-capillarity coupling: Oscillating drop \label{sec:Rayleigh}}
\begin{figure}
	\centering
	\includegraphics{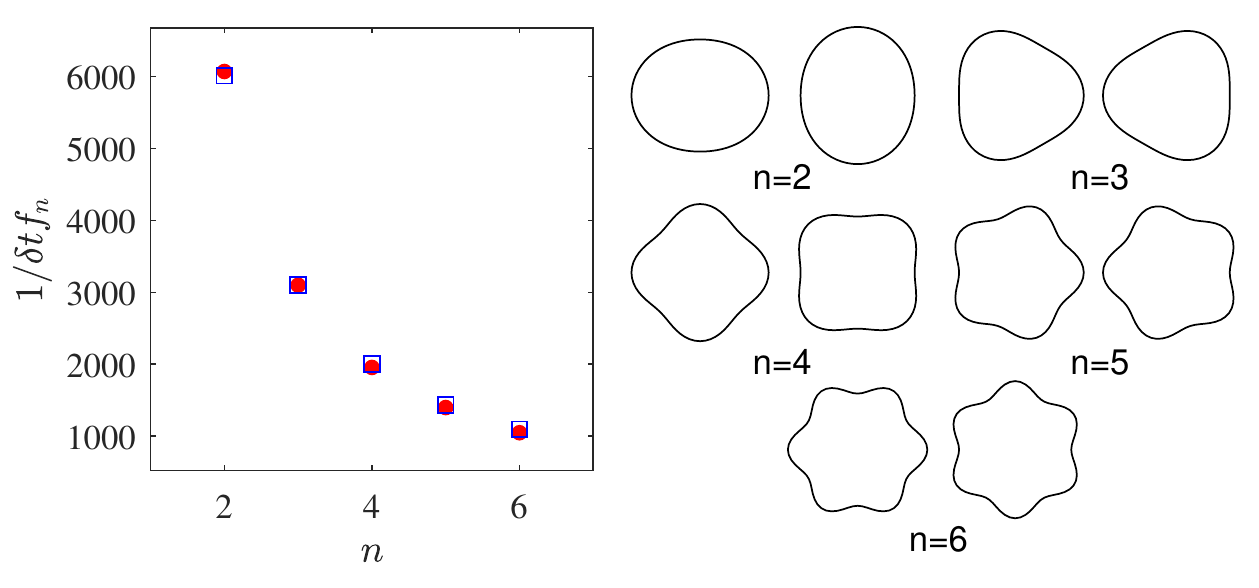}
	\caption{
		Left: Drop oscillation period for modes $n=2,3,4,5,6$. Red circle: Simulation; Blue square:
		Rayleigh's modes, Eq.\ \eqref{eq:Rayleigh_frequency};
		Right:
		Corresponding shapes of the drop at $t=1/2\delta t f_n$ and $t=1/\delta t f_n$.}
	\label{Fig:Rayleigh_modes_n}
    \end{figure}

    Rayleigh's classical study of a freely oscillating drop of non-viscous liquid under small deformations laid ground for the understanding of capillary waves \citep{rayleigh_capillary_1879}. Assuming purely axisymmetric oscillation modes,  Rayleigh's oscillation frequency of 
      $n$th mode, $n=2,3,\dots$, for large density ratios, $\rho_l/\rho_v\gg 1$, is given by,
    \begin{equation}\label{eq:Rayleigh_frequency}
        f_n = \frac{1}{2\pi}\sqrt{\frac{\sigma n (n-1)(n+2)}{\rho_l R_0^3}}.
    \end{equation}
    Two-dimensional simulations were performed for a drop of initial radius $R_0=40\delta r$, for a relatively low kinematic viscosity, $\nu_l\delta t/\delta r^2=0.01$. Modes of order $n=2,3,4,5,6$ were initiated with a monochromatic perturbation, 
    $R(\theta,0)=R_0\left(1 + a_n \cos(n\theta) - \frac{1}{4}na_n^2\right)$, with the amplitude $a_n=0.1$. Reduced temperature of van der Waals fluid was set to $T_r=0.36$, corresponding to density ratio $\rho_l/\rho_v\approx 10^3$.
    The initial and mid-cycle shapes of Rayleigh's  modes are shown in  Fig.\ \ref{Fig:Rayleigh_modes_n}.
    Simulations were performed over eight oscillation cycles and corresponding oscillation periods were identified. Fig.\  \ref{Fig:Rayleigh_modes_n} demonstrates excellent comparison of oscillation frequencies measured in the simulation with corresponding theoretical values \eqref{eq:Rayleigh_frequency}, even for higher-order modes.  
    While Rayleigh's analysis was restricted to non-viscous liquids, 
    a number of analytical solutions for viscous oscillating drops have been obtained over the years. In particular, \cite{aalilija_analytical_2020} derived a time-dependent solution of the following form,
    \begin{equation}\label{eq:oscillation_analytical}
        R(\theta,t) = R_0 \left( 1 + \epsilon_n \cos(n\theta) - \frac{1}{4}n\epsilon_n^2\right),
    \end{equation}
    where
        $\epsilon_n = a_n \exp(-\lambda_n t) \cos\left(2\pi f_n t\right)$,
    while  $\lambda_n$ is the damping rate of the $n$th mode,
    \begin{equation}\label{Eq:oscillating_drop_diss_rate}
        \lambda_n = 2n(n-1){\nu_l}{R_0^{-2}}.
    \end{equation}
   
\begin{figure}[h]
	    \centering
		    \includegraphics{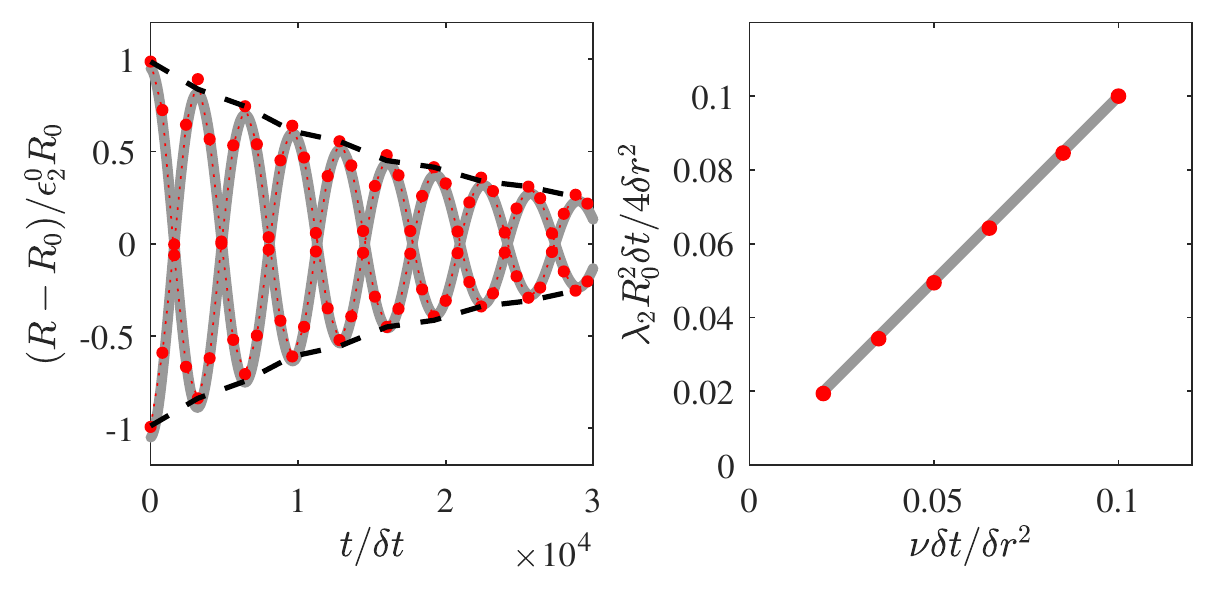}
    \caption{Left: Time evolution of the drop radius in directions $\theta=0$ and $\theta=\pi/2$ for kinematic viscosity $\nu\delta t/\delta r^2=0.02$. Gray line: Analytical solution \eqref{eq:oscillation_analytical}; Symbol: Simulation; Dashed black line: The $\exp(-\Lambda t)$ fit used to compute the dissipation rate in Eq.\ \eqref{Eq:oscillating_drop_diss_rate};
    Right: Apparent viscosity as computed via Eq.\ \eqref{Eq:oscillating_drop_diss_rate}. Gray line: Analytical solution \eqref{Eq:oscillating_drop_diss_rate}; Symbol: Simulation; All results correspond to the oscillation mode $n=2$.}
    \label{Fig:Rayleigh_mode2}
    \end{figure}

    In order to validate the damping rate of capillary waves, the previously described two-dimensional drop was simulated in  a range of kinematic viscosities, $\nu_l\delta t/\delta r^2\in[0.01, 0.1]$,  subject to a monochromatic perturbation with the lowest mode $n=2$ and 
    the initial perturbation amplitude was set to $a_2=0.2$. 
    The envelope of oscillations was fitted using the exponential function $\exp (-\Lambda t)$, see Fig.\ \ref{Fig:Rayleigh_mode2}. The effective damping coefficient $\Lambda$ was used along with Eq.\ \eqref{Eq:oscillating_drop_diss_rate} to evaluate the apparent kinematic viscosity. 
    As shown in Fig.\ \ref{Fig:Rayleigh_mode2}, the dissipation rates obtained by simulation agree well with the analytical solution \eqref{Eq:oscillating_drop_diss_rate}. 
    \subsection{Isothermal speed of sound \label{sec:SoS}}
\begin{figure}
	    \centering
		    \includegraphics{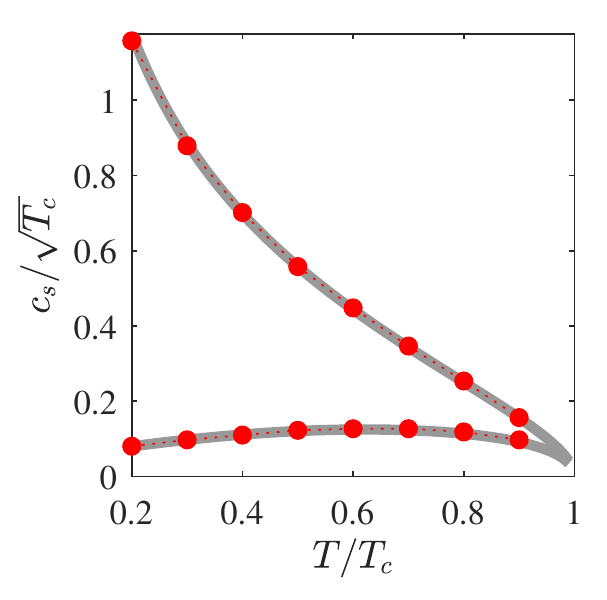}
	    \caption{Isothermal sound speed for van der Waals fluid at various temperatures.  Line: theory, Eq.\ \eqref{eq:sound_vdW}; Symbol: simulation. Upper branch: saturated liquid; Lower branch: saturated vapour.
	    }
	    \label{Fig:SoundSpeed_VdW}
    \end{figure}

    Finally, we validate the speed of sound at different temperatures in both the liquid and vapor phases. For the van der Waals fluid, the isothermal sound speed is,
    \begin{equation}\label{eq:sound_vdW}
        c_s = \sqrt{\frac{\partial P}{\partial \rho}\bigg|_{T}} = \sqrt{\frac{RT}{{(b\rho-1)}^2} - 2a\rho}.
    \end{equation}
    The sound speed was measured by monitoring the position of a pressure front over time in a quasi-one-dimensional simulation at different temperatures. The obtained results are compared to theory \eqref{eq:sound_vdW} in Fig.\ \ref{Fig:SoundSpeed_VdW}. 
    It is observed that the simulations accurately capture the speed of sound in both liquid and vapor phases, for entire range of density ratios on the coexistence diagram Fig.\ \ref{Fig:Coexistence}, up to at least $\rho_l/\rho_v\sim 10^{11}$. Results for other equations of state are compared in Appendix\ \ref{ap:Sound_speed}.
    \subsection{Interaction with solid boundaries: Equilibrium contact angle\label{sec:contact}}
    A comprehensive review of various implementations of  the contact angle on solid boundaries in the lattice Boltzmann setting can be found in \cite{li_contact_2014}. 
    Here we follow a proposal by \cite{benzi_mesoscopic_2006} where a virtual density, and therefor corresponding virtual pseudo-potential, is attributed to the solid nodes. The calculations of the force \eqref{eq:Kforce_final} is then carried at all fluid nodes. The no-slip condition is imposed via the modified bounce-back scheme of \cite{bouzidi_momentum_2001} for curved boundaries. The virtual density attributed to the solid nodes is the free parameters controlling the contact angle.
    The ability of this approach to correctly reproduce the dynamics of the contact line in a high-density ratio regime will be validated below. 
    \subsubsection{Young--Laplace equation}
    As a first validation of the solid/fluid interaction the case of a two-dimensional channel of height $H$ and length $L$ is considered. Initially a rectangular column of liquid of length $L/3$ and height $H$ surrounded by vapor was placed at the center of the channel. Simulations were performed until the system reaches steady-state, and the angle between the liquid/vapor interface and the solid wall at the triple-point was measured. The static contact angle measured at the triple-point should verify the Young--Laplace law for this configuration:
    \begin{equation}
        \Delta P = \frac{2\sigma \sin \theta}{H},
    \end{equation}
    where $\theta$ is the equilibrium contact angle and $\sigma$ the liquid/vapor surface tension.
    
\begin{figure}
	    \centering
		    \includegraphics{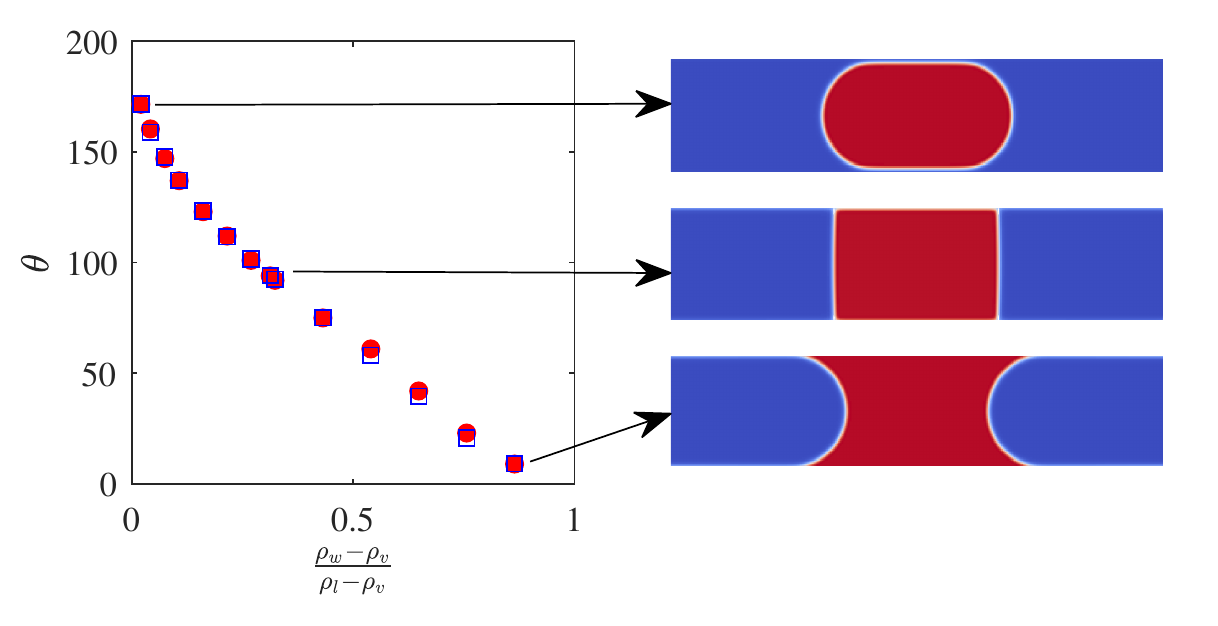}
	    \caption{
	    Static contact angles as (blue squares) obtained from the Young--Laplace equation and (red circles) measured directly from the simulations. 
	    }
	    \label{Fig:CAResults}
    \end{figure}

    Here to validate the static angle, the reduced temperature of van der Waals fluid was set to $T_r=0.36$ which corresponds to density ratio  $\rho_l/\rho_v=10^3$. The pressure difference $\Delta P$ is computed as the difference in pressure between two monitoring points located respectively inside and outside the liquid column. The convergence criterion used to assess steady-state is based on the time-variations of pressure at these two monitoring points. The contact angles were measured using the low-bond axisymmetric drop shape analysis module~\citep{stalder_low-bond_2010} in ImageJ. The results obtained for the entire contact angle range are shown in Fig.~\ref{Fig:CAResults}.
    The measured contact angles show excellent agreement with the Young--Laplace equation. Furthermore, the results show that, for $\rho_w\to\rho_v$, the contact angle tends to $\theta\to 180^{\circ}$, while as  $\rho_w\to\rho_l$, the contact angles tends to $\theta\to 0^{\circ}$ therefore covering the entire range of contact angles.
    \subsubsection{Wettability-driven drop motion in a channel}
\begin{figure}
	    \centering
		    \includegraphics{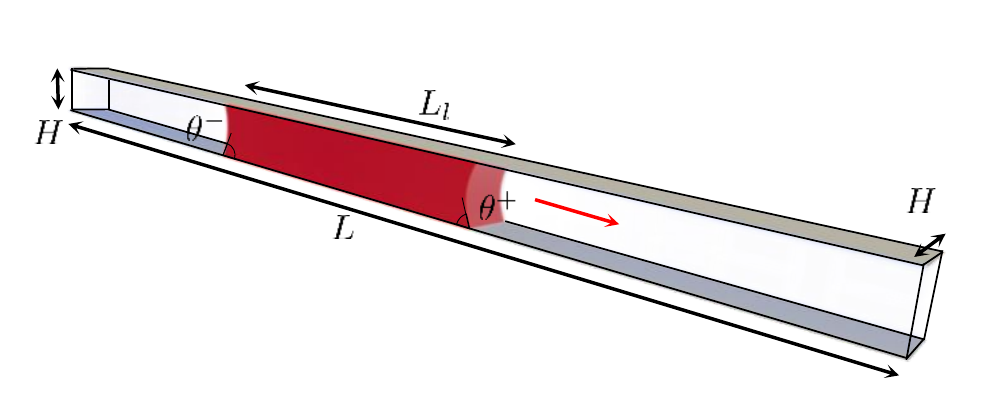}
	    \caption{Schematics of the wettability-driven liquid column. Arrow indicates the direction of motion of liquid column when the contact angle $\theta_{+}$ (right) exceeds the contact angle $\theta_{-}$ (left).}
	    \label{Fig:liquid_column_geo}
    \end{figure}

\begin{figure}
	\centering
	\includegraphics{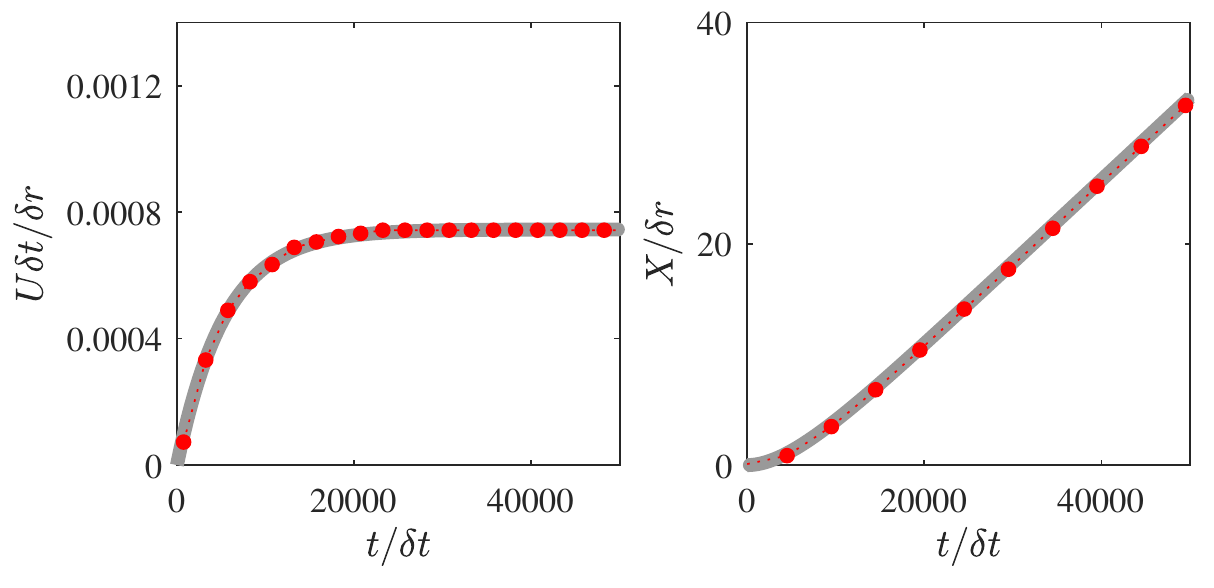}
	\caption{(left) Centroid velocity and (right) center-of-mass position as obtained from (grey line) analytical solution and (red markers) LB simulations for the wettability-driven motion case.}
	\label{Fig:CapFillResults}
    \end{figure}

    Difference in contact angle on opposite sides of a liquid drop placed on a surface with a non-uniform wettability may initiate a motion of the drop. 
    As a further validation of  the solid/liquid interaction of the proposed model, 
    the case of a
    liquid column placed in a channel and subjected to a wettability step function is considered. 
    The configuration, illustrated in Fig.\ \ref{Fig:liquid_column_geo}, consists of a channel of length $L$ and width/height of $H$. Initially a liquid column of length $L_l$ is placed at the channel center.
    Once the liquid column has reached equilibrium, the wettability on the right-half of the channel is changed, which results in unequal contact angles on the different sides of the liquid column. An approximate analytical solution for the center-of-mass position $X$ and centroid velocity $U$ was found by \cite{esmaili_dynamics_2012},
    \begin{align}
        &X(t) = \tau_{\infty} U_{\infty} \left( e^{-{t}/{\tau_{\infty}}} + (t/{\tau_{\infty}}) - 1 \right),\\
         &   U(t) = U_{\infty} \left( 1 - e^{-{t}/{\tau_{\infty}}} \right),
    \end{align}  
    where the saturated centroid velocity $U_{\infty}$ and transition time $\tau_{\infty}$ are computed respectively as:
    \begin{align}
        &U_{\infty} = \frac{\sigma H \left(\cos \theta^{-} - \cos \theta^{+} \right)}{6\left[ \rho_l \nu_l L_l + \rho_v \nu_v \left(L - L_l\right)\right]},\\
        &\tau_{\infty} = \frac{H^2\left[ \rho_l L_l + \rho_v \left(L - L_l\right)\right]}{12\left[ \rho_l \nu_l L_l + \rho_v \nu_v \left(L - L_l\right)\right]}.
    \end{align}
    Here $H=70\delta r$  while $L=1260\delta r$, $L_l=420\delta r$, $\nu_l=\nu_v=0.075\delta r^2/\delta t$ and density ratio is set to $10^3$ as in previous cases.
    Initially the contact angles on both sides of the water column were set to $\theta^{-}=59.3^{\circ}$. Once a steady-state was reached, the contact angle on the right-hand side of the column was changed to $\theta^{+}=63.4^{\circ}$, which resulted in a net force acting on the liquid. The liquid column centroid velocity and center-of-mass position are compared to analytical solutions in Fig.\ \ref{Fig:CapFillResults}. Both quantities closely match analytical predictions.
    \section{Application to dynamic cases \label{sec:droplets}}
    \subsection{Thermodynamic convergence and resolution requirements}
    Two-phase lattice Boltzmann models are routinely used to simulate impact of liquid drops on solids. While experiments are typically conducted with water and other liquids in air, the applicability of the two-phase LBM is justified whenever the dynamic effects are concerned. However, majority of LB models do not reach the experimentally relevant density ratios of, say, water/air density. The proposed LBM was demonstrated to be valid and thermodynamically well-posed at high density ratios.
    However, thermodynamic convergence, especially at higher density ratios comes at a cost in terms of interface resolution requirements. For instance, at $T_r=0.36$ resulting in a density ratio of $10^3$, to guarantee deviation below one percent from theory in the vapor phase one must have $W>8\delta r$. In large-scale hydrodynamic-dominated cases targeted in the present section full thermodynamic convergence is not necessary. The only manifestations of deviation from the thermodynamically converged state that must be kept under control are spurious currents. Here, for density ratios $10^3$ the interface thickness, \eqref{eq:interface_W}, is fixed at $W=5\delta r$ resulting in a seven percent deviation of the vapor phase density from theory and spurious currents below $10^{-3}\delta r/\delta t$. The Tolman length for this choice of parameter is $\delta_T/\delta r=2.4$ which for the resolutions considered below ($R_0=75\delta r$) results in $\delta_T/R_0=0.032$ therefor minimizing the impact of curvature on effective surface tension.
    Below, we consider benchmark simulations of dynamic effects at realistic density ratios of increasing complexity to probe the accuracy and numerical stability of our model in the dynamic setting.
    \subsection{Contact time on flat non-wetting surfaces}
\begin{figure}
	    \centering
		    \includegraphics[width=0.7\linewidth]{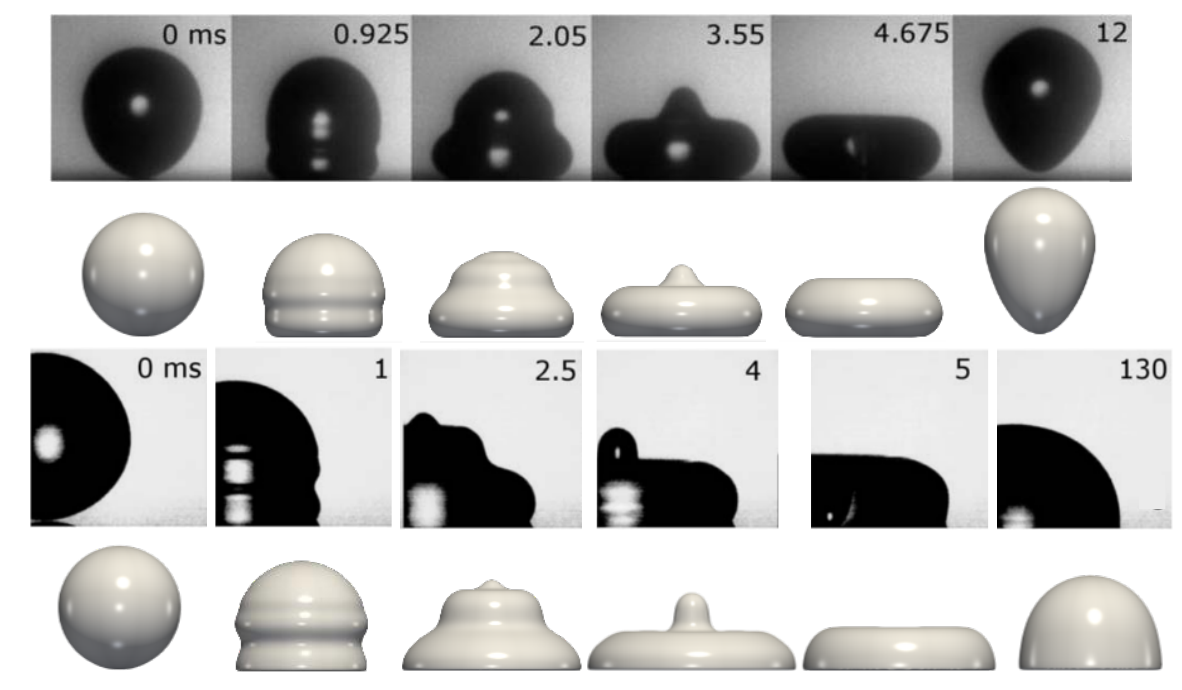}
	    \caption{Drop impacting flat solid surface with different contact angles (first and second row) $\theta$=180 and (third and fourth row) $\theta$=90. The Weber and Reynolds number are respectively 3.5 and 750. Experimental data from~\cite{vadillo_dynamic_2009} are shown in the first and third rows. Density ratios in both experiments and simulations are $10^3$ and kinematic viscosity ratios are set to 15.}
	    \label{Fig:We3o5Impact}
    \end{figure}

\begin{figure}
	    \centering
		    \includegraphics{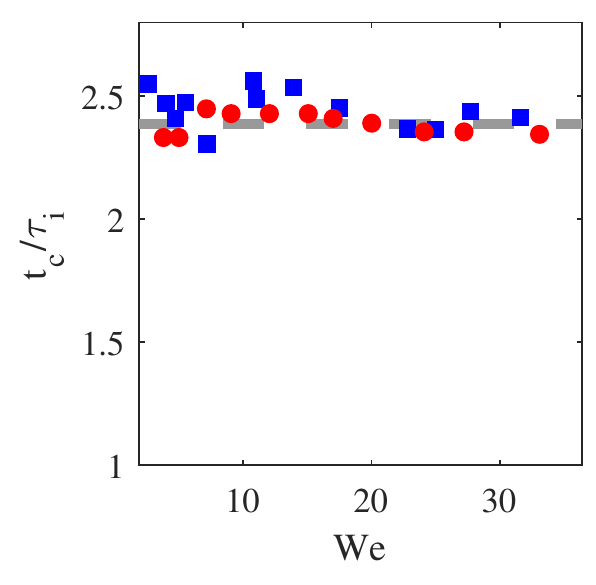}
	    \caption{Drop contact times on flat non-wetting surface (contact angle $\theta$=165 for different Weber numbers as obtained from simulations and experiments. Simulations results are shown with red circular markers while experimental data reported by~\cite{gauthier_water_2015} are illustrated with blue square markers. The dashed grey line represents the average contact time as obtained from simulations, $\bar{t_c}/\tau_i=2.4$.}
	    \label{Fig:contact_times}
    \end{figure}

    Extensive studies of drop impact dynamics have shown that the contact time on so-called super-hydrophobic surfaces is independent of the Weber number, ${\rm We}=\rho_l D_0 U_0^2/\sigma$, and only scales with the inertio-capillary time, $\tau_i=\sqrt{\rho_l D_0^3/8\sigma}$, i.e. for a given drop initial diameter $D_0$ the contact time is not affected by impact velocity~\citep{gauthier_water_2015}. 
    The different stages of the impact process are illustrated in Fig.~\ref{Fig:We3o5Impact} through water drops impacting two surfaces with different contact angles. Simulations carried out under the same conditions are shown and already point to very good agreement between simulations and experiments.
    To quantify the ability of the proposed model to capture the dynamics of drop impact and to investigate the Weber-independence of the contact time on non-wetting surfaces, simulations were performed for a wide range of Weber numbers We $\in [1,40]$, resulting in Reynolds numbers in the range of Re $\in[400, 1400]$, with Re=$U_0 D_0/\nu_l$. 
    Simulations have been performed in boxes of size $4D_0\times4D_0\times4D_0$ with $D_0=150\delta r$. The temperature was set to $T_r=0.36$ resulting in a density ratio of $10^3$. The obtained data show very good agreement with experimental contact times measured by~\cite{gauthier_water_2015} and confirm the Weber-independence of the contact times. Both numerical and experimental data are shown in Fig.~\ref{Fig:contact_times}.
    \subsection{Reducing the contact time: Pancake bouncing}
\begin{figure}
	    \centering
		    \includegraphics[scale=0.75]{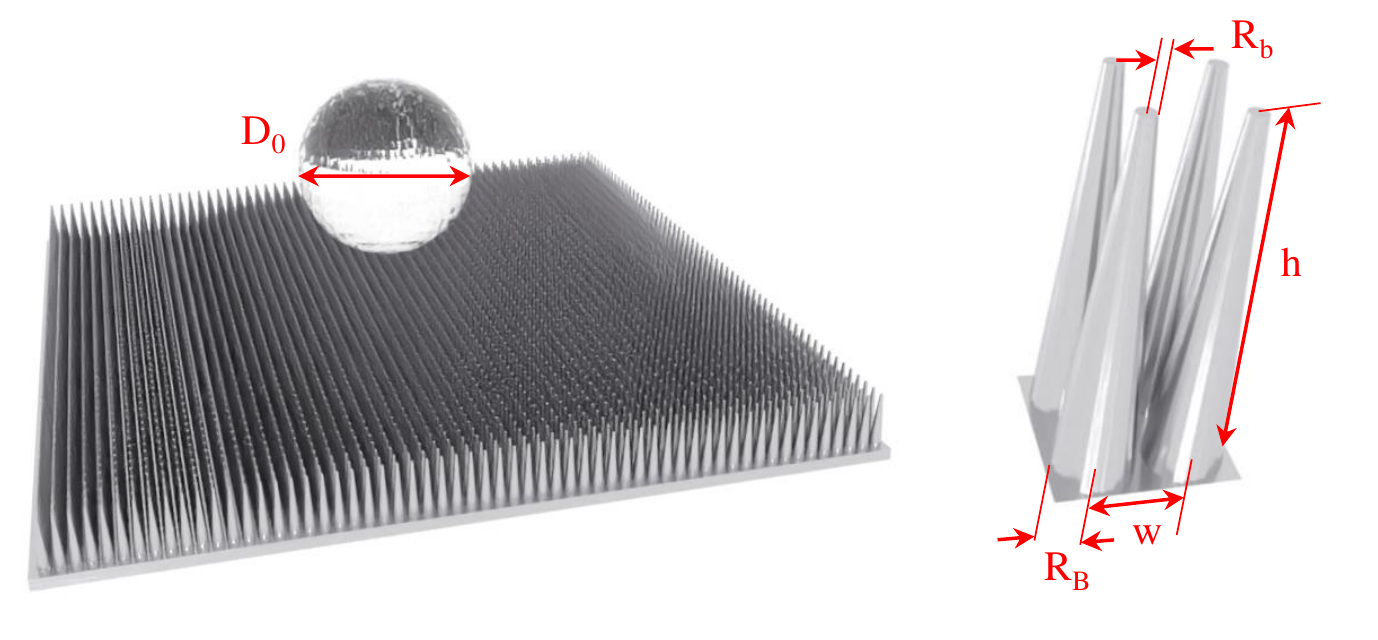}
	    \caption{Illustration of the geometry of tapered posts. Configuration follows the experiment setup  by \cite{liu_pancake_2014}.}
	    \label{Fig:posts_geometry}
    \end{figure}

\begin{figure}
	    \centering
		    \includegraphics{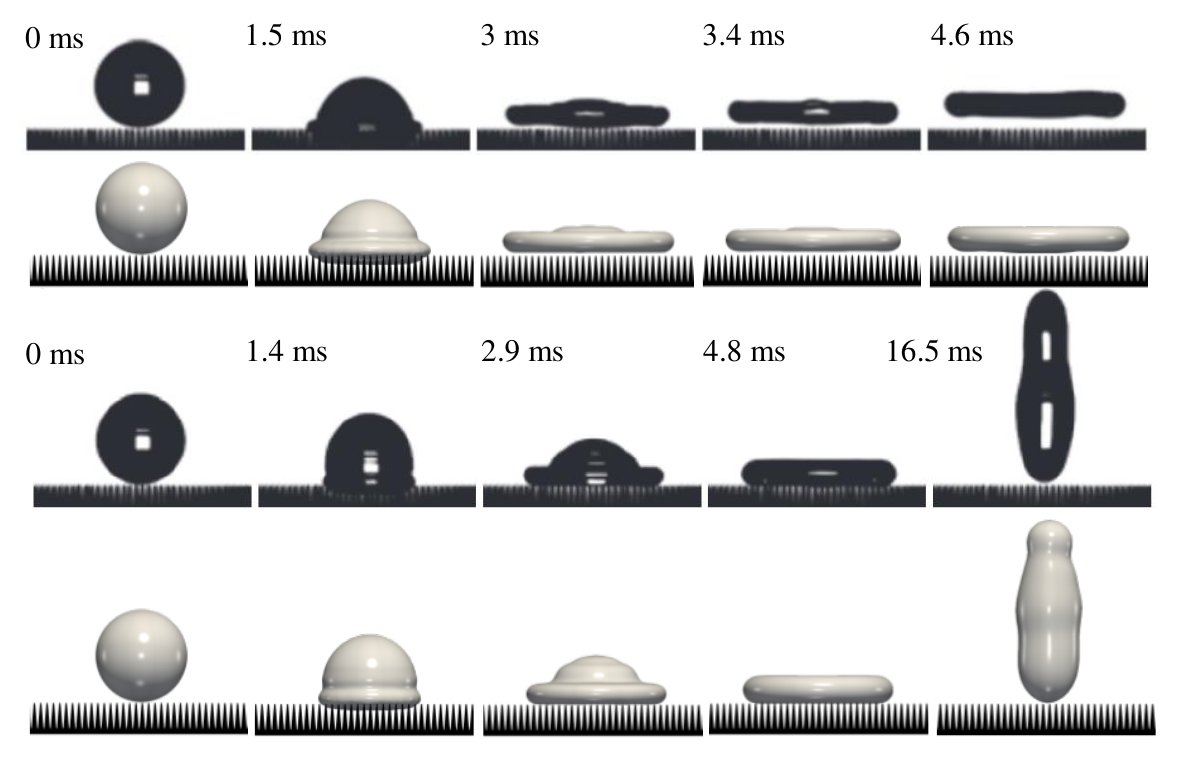}
	    \caption{Drop impacting tapered posts at different Weber numbers (first and second rows) We=28.2 with pancake bouncing and (third and fourth rows) We=14.2. The first and third rows are experiments from~\citep{liu_pancake_2014} while the second and fourth rows are from simulations.}
	    \label{Fig:PostImpactShots}
    \end{figure}

\begin{figure}
	    \centering
		    \includegraphics{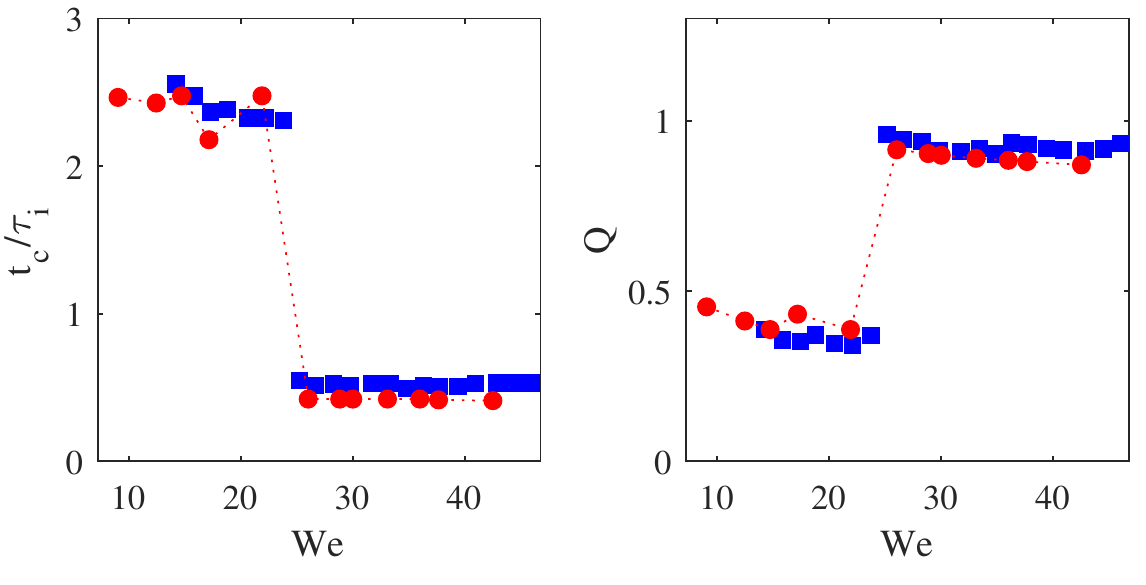}
	    \caption{(left) Drop contact times and (right) pancake quality at rebound on tapered posts for different Weber numbers as obtained from simulations and experiments. Simulations results are shown with red circular markers while experimental data reported by~\cite{liu_pancake_2014} are illustrated with blue square markers.}
	    \label{Fig:contact_times_posts}
    \end{figure}

    To further reduce the drop contact times, a number of different strategies have been devised during the past decades. 
    Recently, \cite{liu_pancake_2014} proposed to use macroscopic structures, in the form of tapered posts, to reduce the contact time. It has been shown that above a certain threshold Weber number these structures can decrease the contact time by approximately $75$ percent. 
    This mechanism is also known as pancake bouncing, due to the pancake-like shape of the drop at take off. A detailed numerical study of pancake bouncing using LBM has been is presented in~\citep{mazloomi_moqaddam_drops_2017}, for a limited range of densities. Here, to further demonstrate the versatility of our model we consider a realistic density ratio of $10^3$.
    The geometry consists of a flat substrate populated by tapered posts of base and tip radii $R_B$ and $R_b$ and height $h$ with a center-to-center distance $w$ in a simple square arrangement. The simulations were run following the the geometrical configurations considered in experiments~\citep{liu_pancake_2014}. 
    Curved wall boundaries were implemented via the modified bounce-back scheme introduced by \cite{bouzidi_momentum_2001}.The geometry is illustrated in Fig.~\ref{Fig:posts_geometry}.
    As for the flat substrate simulations, the domain size was set to $4D_0\times4D_0\times4D_0$ with $D_0=150\delta r$. Snapshots from the different stages of impact for two different Weber numbers, one below the pancake bouncing threshold and one above, from both simulations and experiments are shown in Fig.~\ref{Fig:PostImpactShots}. The results show excellent agreement between simulations and experiments.
    The contact times for different Weber numbers as obtained from simulations are compared to experimental results from ~\citep{liu_pancake_2014} in Fig.~\ref{Fig:contact_times_posts}. It is shown that the numerical simulations not only accurately capture the contact time reduction due to pancake bouncing and the shape of the drop at take off, they also correctly predict the onset of pancake bouncing. The shape of the drop at take off is assessed via the \emph{pancake quality} parameter $Q$ defined as the ratio of the drop radius at take off to its radius at maximum spreading~\citep{liu_pancake_2014}.
    \subsection{Extreme density ratios: Inertia-dominated coalescence of mercury drops}
\begin{figure}
	    \centering
		    \includegraphics{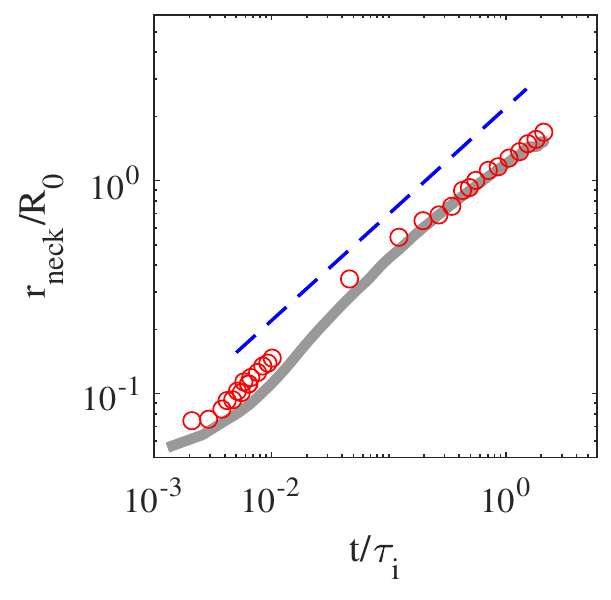}
	    \caption{Time evolution of the neck radii from both (grey line) simulations and (red circular markers) experiments as reported by  \cite{menchaca-rocha_coalescence_2001}. The dashed blue line represents the $\sqrt{t/\tau_i}$ scaling with $\tau_i=\sqrt{\rho R_0^3/\sigma}$.}
	    \label{Fig:colescence_validation}
    \end{figure}

    The sudden and pronounced topological changes involved in the coalescence of drops, especially the formation and subsequent evolution of the neck between them has been of interest and subject to study for many years. The dynamics of the dimensionless neck radius $r_{\rm neck}/R_0$ (with $R_0$ the drops initial radius and $r_{{\rm neck}}$ the neck radius) have been shown to belong to one of two regimes~\citep{eggers_coalescence_1999} characterized by the Ohnesorge number, ${\rm Oh}=\mu/\sqrt{2\rho_l\sigma R_0}$: highly viscous Stokes or inertial. The neck radius evolution over time has been shown to scale either as $r_{\rm neck}/R_0\propto t/\tau_i$ (in the viscous regime) or $r_{\rm neck}/R_0\propto \sqrt{t/\tau_i}$ (in the inertial regime).
    Here, to better illustrate the ability of the proposed scheme to model flows with extreme density ratios, we will focus on the first regime, more specifically the coalescence of mercury droplets, with Ohnesorge numbers of the order of $10^{-4}$ (for 1~g drops). We consider a case following the experimental configuration of~\cite{menchaca-rocha_coalescence_2001} with two 1~g mercury droplets. To match the proper density ratio of a mercury/air system ($\rho_{\ce{Hg}}=13600~\hbox{kg}/\hbox{m}^3$) the temperature is set to $T_r=0.267$, resulting in a density ratio of $\rho_l/\rho_v\sim 11500$. The two drops, resolved with 150 points on their radii are placed in a rectangular domain of size $400\times400\times800$ with a center-to-center distance of 304 points. The drops are connected initially via a neck of 4 points radius.
    \begin{figure}
	    \centering
		    \includegraphics[width=0.8\linewidth]{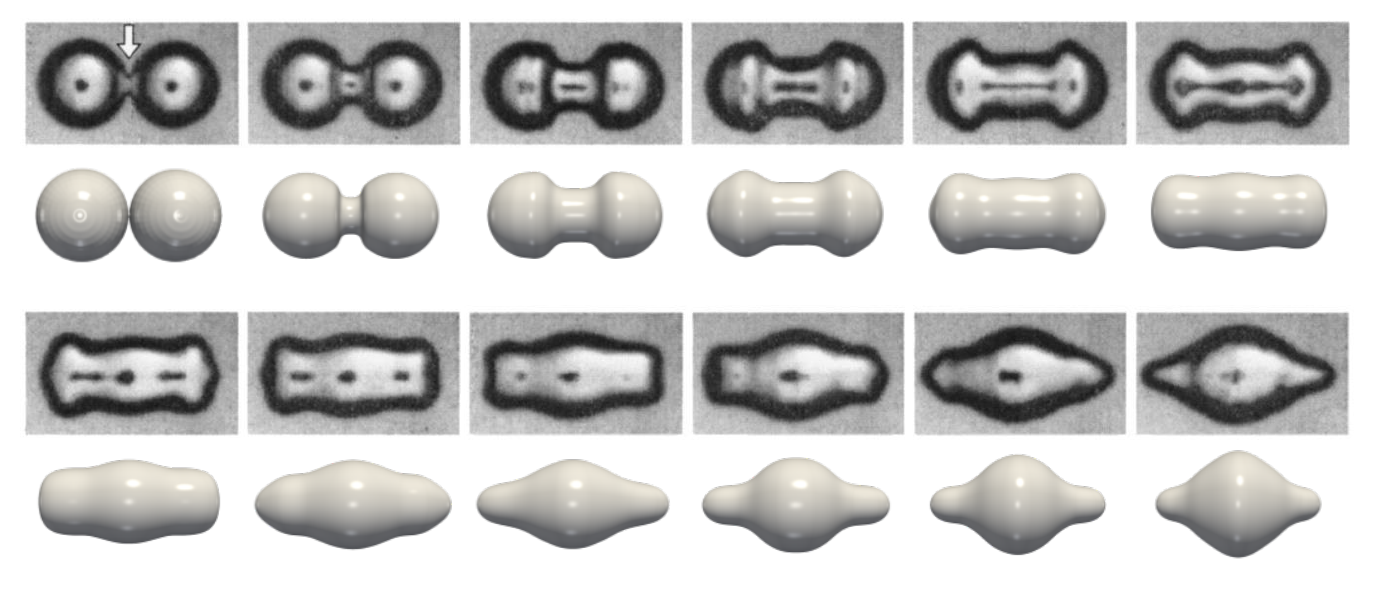}
	    \caption{Sequential images from different stages of the mercury drops coalescence and sub-sequential capillary waves propagation. First and third-row images are from experiments of \cite{menchaca-rocha_coalescence_2001} while second and fourth rows are from LB simulations. Snapshots are taken at $\Delta t=3.5$~ms intervals.}
	    \label{Fig:mercury_shapes}
    \end{figure}
    The evolution of the drops shape over time are compared to experiments from~\cite{menchaca-rocha_coalescence_2001} in Fig.~\ref{Fig:mercury_shapes}. To further illustrate the agreement of numerical simulations with experiments, the evolution of the neck radii $r_{\rm neck}$ over time are shown in Fig.~\ref{Fig:colescence_validation}.
    Both qualitative comparison of the drops shape and quantitative comparison of the neck radius show that the presented solver is able to correctly model multi-phase flow dynamics with extreme density ratio. To the authors knowledge this is the first lattice Boltzmann simulation at such high density ratios, regardless of the formulation, i.e. pseudo-potential, free energy, phase-field etc.
    \section{Conclusion \label{sec:conclusion}}
    Multi-phase flows are a well-established area in kinetic theory and more specifically in the context of the LBM. While dynamic 2-D/3-D simulations are routinely carried out with the different LB-based formulations, which are subject to continuous numerical \emph{improvements}, a clear and concise kinetic framework along with a characterisation of the resulting fluid thermodynamic properties is lacking.
    
    The aim of the present work was to develop a general framework for iso-thermal single-component multi-phase flow simulations consistent with the capillary fluid thermodynamics. Starting from the first-order BBGKY equation and using a projector operator in phase-space, a flexible model, in terms of pressure contribution partition, is proposed. A specific realization of the partition minimizing deviations from the first-neighbor stencil reference (optimal) state was then discretized using the LBM. The discrete solver was shown to recover the full Navier-Stokes-Korteweg under the proposed scaling. The resulting discrete model was then probed for thermodynamic consistency and shown to abide by all the scaling laws of the corresponding meanfield, i.e. surface tension, interface thickness and Tolman length. Finally the model was shown to allow for not only static, but dynamic large density ratios simulations with complex geometries, illustrated best by the extreme case of mercury drops coalescence, therefor effectively removing the well-known limitations on maximum density ratio.
    
    The detailed study of the properties of the proposed model and accompanying theoretical analyses lead us to believe that the present framework fills important gaps in the multi-phase literature, more specifically in the context of discrete kinetic models such as the LBM. It provides a consistent framework for the simulation of challenging physics as demonstrated by the curvature-dependence of the surface tension. Furthermore, it paves the way for extension to fully compressible non-ideal fluids, which will be topic of future publications.
    \section*{Acknowledgement}
    This work was supported by European Research Council (ERC) Advanced Grant no. 834763-PonD (S.A.H, B.D. and I.K.) and the Swiss National Science Foundation (SNSF) grant No. 200021-172640 (S.A.H.). Computational resources at the Swiss National Super Computing Center CSCS were provided under grant no. s1066. S.A.H. thanks M. Lulli for discussions on the Tolman length.
    \section*{Declaration of interests}
    The authors report no conflict of interest.
    \appendix
    \section{Short-range interaction \label{ap:ShortRange}}
    The non-local contributions to particle's interaction in Eq.\ \eqref{eq:Boltzmann_eq} consist of short- and long-range interactions and are classically treated by splitting the integral over physical space into two domains,  $\lvert\bm{r}_1-\bm{r}\lvert\leq d$ and $\lvert\bm{r}_1-\bm{r}\lvert > d$, where $d$ is the hard-sphere  diameter. 
    To derive the pressure-form of the short-range interaction, we start with the Enskog hard-sphere collision integral \citep{enskog_warmeleitung_1921,chapman_mathematical_1939},
    \begin{multline}\label{eq:enskog_integral}
	\mathcal{J}_{\rm E} = d^2\int \int \left[\chi\left(\bm{r}+\frac{d}{2}\bm{k}\right)f\left(\bm{r},\bm{v}'\right)f\left(\bm{r}+d\bm{k},\bm{v}_1'\right) \right. \\
	\left. - \chi\left(\bm{r}-\frac{d}{2}\bm{k}\right)f\left(\bm{r},\bm{v}\right)f\left(\bm{r}-d\bm{k},\bm{v}_1\right) \right] \bm{g}\cdot\bm{k}d\bm{k}d\bm{v}_1,
    \end{multline}
    where $\bm{k} = (\bm{r}_1-\bm{r})/\lvert\bm{r}_1-\bm{r}\lvert$, $\bm{g}=\bm{v}_1-\bm{v}$, $\bm{v}'=\bm{v}+\bm{k}(\bm{g}\cdot\bm{k})$, $\bm{v}_1'=\bm{v}_1-\bm{k}(\bm{g}\cdot\bm{k})$ and $\chi$ is equilibrium pair correlation function, evaluated at local density taking into account the effect of volume of particles in the collision probability \citep{chapman_mathematical_1939}. 
    Using a Taylor expansion around $\bm{r}$,
    \begin{eqnarray}
	\chi\left(\bm{r}\pm \frac{d}{2}\bm{k}\right) &=& \chi\left(\bm{r}\right) \pm \frac{d}{2}\bm{k}\cdot\bm{\nabla}\chi\left(\bm{r}\right) + {O}(d^2),\\
	f\left(\bm{r}\pm d\bm{k},\bm{w}\right) &=& f\left(\bm{r},\bm{w}\right) \pm d\bm{k}\cdot\bm{\nabla} f\left(\bm{r},\bm{w}\right) + {O}(d^2),
    \end{eqnarray}
    and neglecting terms of order ${O}(d^4)$, the resulting approximation of the Enskog collision integral becomes,
    \begin{equation}
	\mathcal{J}_{\rm E} = \chi\mathcal{J}_{\rm B} + \mathcal{J}_{\rm E}^{ (1)},
    \end{equation}
    where $\mathcal{J}$ is the Boltzmann collision integral for hard-spheres,
    \begin{equation}
	\mathcal{J}_{\rm B} = d^2  \int \int \left[f\left(\bm{r},\bm{v}'\right)f\left(\bm{r},\bm{v}_1'\right)  - f\left(\bm{r},\bm{v}\right)f\left(\bm{r},\bm{v}_1\right) \right] \bm{g}\cdot\bm{k}d\bm{k}d\bm{v}_1,
    \end{equation}
    while $\mathcal{J}_{\rm E}^{(1)}$ is the non-local contribution to the lowest order,
    \begin{align}
	\mathcal{J}_{\rm E}^{(1)} =& d^3 \chi\left(\bm{r}\right) \int \int \bm{k}\cdot\left[f\left(\bm{r},\bm{v}'\right)\bm{\nabla}f\left(\bm{r},\bm{v}'_1\right)  + f\left(\bm{r},\bm{v}\right)\bm{\nabla}f\left(\bm{r},\bm{v}_1\right) \right] \bm{g}\cdot\bm{k}d\bm{k}d\bm{v}_1 \nonumber\\
    &	+ \frac{d^3}{2} \int \int \bm{k}\cdot\bm{\nabla}\chi\left(\bm{r}\right)\left[f\left(\bm{r},\bm{v}'\right)f\left(\bm{r},\bm{v}'_1\right) + f\left(\bm{r},\bm{v}\right)f\left(\bm{r},\bm{v}_1\right) \right] \bm{g}\cdot\bm{k}d\bm{k}d\bm{v}_1,
    \end{align}
    Since the Boltzmann collision integral conserves the mass and momentum locally, it is annulled by the projector,
    \begin{equation}
	\mathcal{K}\mathcal{J}_{\rm B} = 0.
    \end{equation}
    Furthermore,  $\mathcal{J}_{\rm SR}^{\rm 1}$ is evaluated at the local equilibrium $f^{\rm eq}$ \eqref{eq:LM} to get \citep{chapman_mathematical_1939},
    \begin{align}
	\mathcal{J}_{\rm E}^{(1)} = d^3 \int \int f^{\rm eq}\left(\bm{r},\bm{v}\right)f^{\rm eq}\left(\bm{r},\bm{v}_1\right) \bm{k}\cdot\bm{\nabla}\ln\left[\chi\left(\bm{r}\right)f^{\rm eq}\left(\bm{r},\bm{v}\right)f^{\rm eq}\left(\bm{r},\bm{v}_1\right)\right] \bm{g}\cdot\bm{k}d\bm{k}d\bm{v}_1,
    \end{align}
    which, after integration in $\bm{v}_1$ and $\bm{k}$, for the isothermal flow results in,
    \begin{align} 
	\mathcal{J}_{\rm E}^{(1)} =& -b\rho\chi f^{\rm eq}\left[(\bm{v}-\bm{u})\cdot\bm{\nabla}\ln\rho^2\chi T\right] \nonumber\\  
	&-b\rho\chi f^{\rm eq}\left[ \frac{2}{5RT}(\bm{v}-\bm{u})(\bm{v}-\bm{u}):\bm{\nabla}\bm{u} + \left(\frac{1}{5RT}{\lvert\bm{v}-\bm{u}\lvert}^2-1\right)\bm{\nabla}\cdot\bm{u}\right],
	\label{eq:E_approximation}
    \end{align}
    where $b=2\pi d^3/3m$. Finally, applying the iso-thermal projector $\mathcal{K}$, we obtain
    \begin{align}
	\mathcal{K}\mathcal{J}_{\rm E}^{(1)}
	=&\frac{1}{\rho}\frac{\partial f^{\rm eq}}{\partial\bm{u}}\cdot\int\bm{v}\mathcal{J}^{(1)}_{\rm E}d\bm{v}\nonumber\\
	= &-\frac{1}{\rho}\frac{\partial f^{\rm eq}}{\partial\bm{u}}\cdot\int\bm{v}\left(\bm{v}-\bm{u}\right)f^{\rm eq}\left(\bm{r},\bm{v}\right)d\bm{v}\cdot\frac{b}{\rho T}\bm{\nabla}\rho^2\chi T\nonumber\\
	=&-\frac{1}{\rho}\frac{\partial f^{\rm eq}}{\partial\bm{u}}\cdot\bm{\nabla}b\rho^2\chi RT.
    \end{align}
    While the phenomenological Enskog's collision integral \citep{enskog_warmeleitung_1921} was used above, the lowest-order approximation \eqref{eq:E_approximation} is identical in other versions of hard-sphere kinetic equations such as the revised Enskog theory (RET) \citep{van_beijeren_modified_1973} or kinetic variational theory \citep{karkheck_kinetic_1981}. 
    \section{Long-range interaction \label{ap:LongRangeMeanField}}
    Vlasov's mean-field approximation for a long-range interaction is reviewed next. Assuming absence of correlations, the two-particle distribution function is approximated as
    \begin{equation}
        f_{2}(\bm{r},\bm{v}, \bm{r}_1, \bm{v}_1) \approx f(\bm{r},\bm{v}) f(\bm{r}_1, \bm{v}_1),
    \end{equation}
    while the long-range interaction integral can be simplified,
    \begin{equation}\label{eq:LR_o}
        \mathcal{J}_{\rm V} = \frac{\partial f(\bm{r},\bm{v}) }{\partial\bm{v}}\cdot\bm{\nabla}\left[\int_{\lvert\bm{r}_1-\bm{r}\lvert > d} \rho(\bm{r}_1)V(\lvert \bm{r}_1-\bm{r}\lvert) d\bm{r}_1\right].
    \end{equation}
    With  a Taylor expansion around $\bm{r}$, 
    \begin{equation}\label{eq:TmcL_density_longrange}
        \rho(\bm{r}_1) = \rho(\bm{r}) + (\bm{r}_1-\bm{r})\cdot\bm{\nabla}\rho(\bm{r}) + \frac{1}{2}(\bm{r}_1-\bm{r})\otimes(\bm{r}_1-\bm{r}):\bm{\nabla}\otimes\bm{\nabla}\rho(\bm{r}) + \mathcal{O}(\bm{\nabla}^3\rho),
    \end{equation}
    and neglecting higher-order terms, Eq.\ \eqref{eq:LR_o} leads to,
    \begin{equation}
        \mathcal{J}_{\rm V} = -\bm{\nabla}\left[2 a \rho(\bm{r}) +\kappa \bm{\nabla}^2 \rho(\bm{r})\right]\cdot\frac{\partial}{\partial\bm{v}}f(\bm{r},\bm{v}),
    \end{equation}
    where parameters $a$ and $\kappa$ are, after integration over a unit sphere,
    \begin{align}
	a &= -2\pi\int_{d}^{\infty} r^2V(r) dr,\\
	\kappa &= -\frac{2\pi}{3}\int_{d}^{\infty} r^4V(r) dr.
    \end{align}
    Applying the projector, we obtain,
    \begin{align}
	\mathcal{K}\mathcal{J}_{\rm V}
	=&\frac{1}{\rho}\frac{\partial f^{\rm eq}}{\partial\bm{u}}\cdot\int\bm{v}\mathcal{J}_{\rm V}d\bm{v}\nonumber\\
	 = &-\frac{1}{\rho}\frac{\partial f^{\rm eq}}{\partial\bm{u}}\cdot\int\bm{v}\frac{\partial}{\partial\bm{v}}f(\bm{r},\bm{v},t)d\bm{v}\cdot\bm{\nabla}\left[2 a \rho +\kappa \bm{\nabla}^2 \rho\right]\nonumber\\
	=&\frac{\partial f^{\rm eq}}{\partial\bm{u}}\cdot\bm{\nabla}\left[2 a \rho +\kappa \bm{\nabla}^2 \rho\right].
    \end{align}
    Finally, we can estimate the relative magnitude of the parameters $a$ and $\kappa$ by introducing a range of the attraction potential $\delta$. Assuming $d\ll \delta$, we have $a\sim \bar{V}\delta^3$ and $\kappa\sim \bar{V}\delta^5$, where $\bar{V}$ is a characteristic value of the potential, thus   
	\begin{equation}\label{eq:longestimate}
	\sqrt{\kappa/a}\sim\delta.
	\end{equation}.
    \section{Hydrodynamic limit of the Enskog--Vlasov--BGK kinetic model\label{ap:CE_cont}}
    \subsection{Rescaled kinetic equation}
    For the Enskog--Vlasov--BGK kinetic model,
    \begin{equation}\label{eq:EV_model_dim}
		\partial_{t} f + \bm{v}\cdot\bm{\nabla} f = -\frac{1}{\tau}\left(f - {f^{\rm eq}}\right) - \frac{1}{\rho}\frac{\partial {f^{\rm eq}}}{\partial \bm{u}}\cdot \left[\bm{\nabla}\left(b\rho^2\chi RT-a\rho^2 \right)-\kappa\rho\bm{\nabla}\bm{\nabla}^2\rho\right],
    \end{equation}
    let us introduce the following parameters:
    characteristic flow velocity $\mathcal{U}$, 
    characteristic flow scale $\mathcal{L}$,
    characteristic flow time $\mathcal{T}=\mathcal{L}/\mathcal{U}$,
    characteristic density $\bar{\rho}$,
    isothermal speed of sound of ideal gas $c_s=\sqrt{RT}$, and
    kinematic viscosity of the BGK model of ideal gas $\nu=\tau c_s^2$.
    With the above, the variables are reduced as follows (primes denote non-dimensional variables):
	time $t=\mathcal{T}t'$,
	space $\bm{r}=\mathcal{L}\bm{r}'$,
	flow velocity $\bm{u}=\mathcal{U}\bm{u}'$,
	particle velocity $\bm{v}=c_s\bm{v}'$,
	density $\rho$=$\bar{\rho}\rho'$,
	distribution function $f=\bar{\rho}c_s^{-3}f'$.
    Furthermore, the following non-dimensional groups are introduced:
    Viscosity-based Knudsen number
	${\rm Kn}={\tau c_s}/{\mathcal{L}}$, 
    Mach number 
    $	{\rm Ma}={\mathcal{U}}/{c_s}$,
    Enskog number 
	${\rm En}=
	b\bar{\rho}{{\rm Kn}}/{{\rm Ma}}$ and
    Vlasov number 
	${\rm Vs}={a}/{bRT}$.
    With this, the Enskog--Vlasov--BGK kinetic model is rescaled as follows:
    \begin{multline}\label{eq:EV_model_scaled}
		{\rm Ma}\,{\rm Kn}\,\partial_{t}' f' + \bm{v}'\cdot{\rm Kn}\bm{\nabla}' f' = -\left(f' - {f^{\rm eq}}'\right) \\
		- \frac{1}{\rho'}\frac{\partial {f^{\rm eq}}'}{\partial \bm{u}'}\cdot {\rm En}\left[\bm{\nabla}'\left(\chi \left(\rho'\right)^2-{\rm Vs}\,\left(\rho'\right)^2\right){-\left(\frac{\delta}{\mathcal{L}}\right)^2{\rm Vs}\, \left(\rho'\bm{\nabla}'\bm{\nabla}^{'2}\rho'\right)}\right],
    \end{multline}
    where $\delta$ is the range of the attraction potential, estimated according to Eq.\ \eqref{eq:longestimate}.
    The following scaling assumptions are applied: 
	Acoustic scaling, ${\rm Ma}\sim 1$;
	Hydrodynamic scaling,  ${\rm Kn}\sim {\rm En}\sim \delta/\mathcal{L}\sim\epsilon$;
	Enskog--Vlasov parity,  ${\rm Vs}\sim 1$. In other words, the conventional hydrodynamic limit treats all non-dimensional groups that are inversely proportional to the flow scale $\mathcal{L}$ (${\rm Kn}$, ${\rm En}$ and $\delta/\mathcal{L}$) as a small parameter while the Enskog--Vlasov parity ensures that both the short- and long-range contributions to the pressure are treated on equal footing.
    Returning to dimensional variables, we may write, 
    \begin{equation}\label{eq:overall_eq}
	\epsilon\partial_{t} f + \bm{v}\cdot\epsilon\bm{\nabla} f = -\left(f - {f^{\rm eq}}\right) - \frac{1}{\rho}\frac{\partial {f^{\rm eq}}}{\partial \bm{u}}\cdot {\left(\epsilon \bm{F}^{(1)}+\epsilon^3\bm{F}^{(3)}\right)},
    \end{equation}
    where,
    \begin{align}
    &	\bm{F}^{(1)}=\bm{\nabla}\left(b\rho^2\chi RT-a\rho^2\right),\\
    &	\bm{F}^{(3)}=-\kappa\rho\bm{\nabla}\bm{\nabla}^2\rho.\label{eq:F3}
    \end{align}
    Finally, taking into account the reference equilibrium \eqref{eq:eq_ref} and the reference pressure $P_0$, the rescaled kinetic equation \eqref{eq:gen_kinetic_model_ext} takes the form \eqref{eq:overall_eq} with 
    \begin{equation}
	\bm{F}^{(1)}=\bm{\nabla}(P-P_0).\label{eq:F1gen}
    \end{equation}
    Chapman--Enskog analysis of the rescaled kinetic equation \eqref{eq:gen_kinetic_model_ext} is presented in the next section.

    \subsection{Chapman--Enskog analysis}
    Expanding the distribution function as:
    \begin{equation}
    f = f^{(0)} + \epsilon f^{(1)} + \epsilon^2 f^{(2)} + {O}(\epsilon^3),
    \end{equation}
    introducing it back into \eqref{eq:overall_eq} and separating terms with different orders in $\epsilon$, at order zero one recovers:
    \begin{equation}
    f^{(0)} = f^{\rm eq}.
    \end{equation}
    This latter implies the solvability conditions,
    \begin{align}
    & \int f^{(k)} d\bm{v} = 0,\ \forall k\neq 0,\\
    &\int \bm{v} f^{(k)} d\bm{v} = 0,\ \forall k\neq 0.
    \end{align}
    At order $\epsilon$:
    \begin{equation}
    \partial_t^{(1)}f^{(0)}+\bm{v}\cdot\bm{\nabla} f^{(0)} = - \frac{1}{\tau} f^{(1)} - \frac{1}{\rho}\frac{\partial f^{\rm eq}}{\partial \bm{u}}\cdot\bm{F}^{(1)},
    \end{equation}
    which, upon integration in $\bm{v}$, leads to
    \begin{align}
    &  \partial_t^{(1)}\rho+\bm{\nabla}\cdot\rho \bm{u} = 0,\\
    &  \partial_t^{(1)}\rho \bm{u}+\bm{\nabla}\rho \bm{u}\otimes \bm{u} + \bm{\nabla}\cdot P_0\bm{I} + \bm{F}^{(1)} = 0.
    \end{align}
    At order $\epsilon^2$:
    \begin{equation}
    \partial_t^{(2)}f^{(0)} + \partial_t^{(1)}f^{(1)}+ \bm{v}\cdot\bm{\nabla}f^{(1)} = -\frac{1}{\tau} f^{(2)},
    \end{equation}
    which leads to the following equation for mass conservation:
    \begin{equation}
    \partial_t^{(2)}\rho = 0,
    \end{equation}
    while for momentum:
    \begin{equation}
    \partial_t^{(2)}\rho \bm{u} + \bm{\nabla}\cdot\left[\int \bm{v}\otimes\bm{v} f^{(1)}d\bm{v}\right] = 0.
    \end{equation}
    The last term on the left hand side can be evaluated using the previous order in $\epsilon$ as:
    \begin{multline}
    \int \bm{v}\otimes\bm{v} f^{(1)}d\bm{v} = -\tau\left[\partial_t^{(1)}\int \bm{v}\otimes\bm{v} f^{(0)}d\bm{v} + \bm{\nabla}\cdot\int \bm{v}\otimes\bm{v}\otimes\bm{v} f^{(0)}d\bm{v} \right. \\ \left. +\frac{1}{\rho} \int \bm{v}\otimes\bm{v} \frac{\partial f^{\rm eq}}{\partial \bm{u}}\cdot\bm{F}^{(1)} d\bm{v}\right],
    \end{multline}
    where:
    \begin{equation}
    \partial_t^{(1)}\int \bm{v}\otimes\bm{v} f^{(0)}d\bm{v} = -\bm{\nabla}\cdot \rho \bm{u}\otimes\bm{u}\otimes\bm{u} - \left[\bm{u}\otimes(\bm{F}^{(1)}+\bm{\nabla}P_0) +  (\bm{F}^{(1)}+\bm{\nabla}P_0)\otimes\bm{u}\right] + \partial_t^{(1)} P_0\bm{I},
    \end{equation}
    and:
    \begin{equation}
    \bm{\nabla}\cdot\int \bm{v}\otimes\bm{v}\otimes\bm{v} f^{(1)}d\bm{v} = \bm{\nabla}\cdot \rho \bm{u}\otimes\bm{u}\otimes\bm{u} +
    \left[\bm{\nabla}P_0\bm{u} + {\bm{\nabla} P_0\bm{u}}^{\dagger}\right] + \bm{I}\bm{\nabla}\cdot P_0 \bm{u},
    \end{equation}
    \begin{equation}
    \frac{1}{\rho} \int \bm{v}\otimes \bm{v} \frac{\partial f^{\rm eq}}{\partial \bm{u}}\cdot\bm{F}^{(1)} d\bm{v}= \bm{u} \otimes\bm{F}^{(1)} + {\bm{F}^{(1)}\otimes\bm{u}},
    \end{equation}
    which leads to:
    \begin{equation}
    \int \bm{v}\otimes\bm{v} f^{(1)}d\bm{v} = -\tau\left[ P_0\left(\bm{\nabla}\bm{u}+{\bm{\nabla}\bm{u}}^{\dagger}\right)+ \left(\partial_t^{(1)}P_0 + \bm{\nabla}\cdot P_0 \bm{u}\right)\bm{I}\right],
    \end{equation}
    where the last two terms can be re-written as:
    \begin{align}
	\partial_t^{(1)}P_0 + \bm{\nabla}\cdot P_0 \bm{u}
	= &\frac{\partial P_0}{\partial \rho}(\partial_t^{(1)}\rho + \bm{\nabla}\cdot\rho \bm{u}) + \left(P_0 -\rho\frac{\partial P_0}{\partial \rho}\right)\bm{\nabla}\cdot \bm{u}\nonumber\\
	 = & P_0\left(1 - \frac{\partial \ln P_0}{\partial \ln \rho}\right)\bm{\nabla}\cdot\bm{u},
    \end{align}
    in turn recovering the Navier-Stokes-level momentum equation:
    \begin{equation}
    \partial_t^{(2)}\rho \bm{u} - \bm{\nabla}\cdot\mu\left(\bm{\nabla}\bm{u} + {\bm{\nabla}\bm{u}}^{\dagger} - \frac{2}{3}\bm{\nabla}\cdot\bm{u}\bm{I}\right) - \bm{\nabla}\cdot\left(\eta \bm{\nabla}\cdot\bm{u}\bm{I}\right) = 0,
    \end{equation}
    where:
    \begin{align}
    &\mu=\tau P_0,\\
    &\eta=\tau P_0\left(\frac{5}{3} - \frac{\partial \ln P_0}{\partial\ln\rho}\right).
    \end{align}
    \section{Chapman--Enskog analysis of the lattice Boltzmann model\label{ap:CE}}
    Using a Taylor expansion around $(\bm{r},t)$, 
    \begin{equation}
        f_i\left(\bm{r}+\bm{c}_i\delta t,t+\delta t\right) - f_i\left(\bm{r},t\right) = \left[\delta t \mathcal{D}_t + \frac{\delta t^2}{2} \mathcal{D}^2_t\right] f\left(\bm{r},t\right) + {O}(\delta t^3)
    \end{equation}
    the discrete time-evolution equation is re-written as:
	\begin{equation}
	    \delta t \mathcal{D}_t f_i + \frac{{\delta t}^2}{2}{\mathcal{D}_t}^2 f_i + {O}({\delta t}^3)= \omega\left(f_i^{\rm eq} - f_i\right) + \left( f^*_i - f_i^{\rm eq}\right),
	\end{equation}
	where we have only retained terms up to order two. Then introducing characteristic flow size $\mathcal{L}$ and velocity $\mathcal{U}$ the equation is made non-dimensional as:
	\begin{equation}
	    \left(\frac{\delta r}{\mathcal{L}}\right) \mathcal{D}_t' f_i + \frac{1}{2}{\left(\frac{\delta r}{\mathcal{L}}\right)}^2{\mathcal{D}_t'}^2 f_i = \omega\left(f_i^{\rm eq} - f_i\right) + \left( f^*_i(\bm{u}'+\frac{\delta u}{ \mathcal{U}}\delta \bm{u}') - f_i^{\rm eq}(\bm{u}')\right),
	\end{equation}
	where primed variables denote non-dimensional form and
	\begin{equation}
	    \mathcal{D}_t' = \frac{\mathcal{U}}{c}\partial_t' + \bm{c}_i'\cdot\bm{\nabla}',
	\end{equation}
	where $c=\delta r/\delta t$. Assuming acoustic, i.e. $\frac{\mathcal{U}}{c}\sim 1$ and hydrodynamic, i.e. $\frac{\delta r}{\mathcal{L}}\sim\frac{\delta u}{\mathcal{U}}\sim\varepsilon$, scaling and dropping the primes for the sake of readability:
	\begin{equation}
	    \varepsilon \mathcal{D}_t f_i + \frac{1}{2}\varepsilon^2{\mathcal{D}_t}^2 f_i + {O}(\varepsilon^3)= \omega\left(f_i^{\rm eq} - f_i\right) + \left( f^*_i(\bm{u}+\varepsilon\delta \bm{u}) - f_i^{\rm eq}(\bm{u})\right).
	\end{equation}
	Then introducing multi-scale expansions:
	\begin{eqnarray}
	    f_i &=& f_i^{(0)} + \varepsilon f_i^{(1)} + \varepsilon^2 f_i^{(2)} + O(\varepsilon^3),\\
	    {f^*_i} &=& {f^*_i}^{(0)} + \varepsilon {f^*_i}^{(1)} + \varepsilon^2 {f^*_i}^{(2)} + O(\varepsilon^3),
	\end{eqnarray}
	the following equations are recovered at scales $\varepsilon$ and $\varepsilon^2$:
	\begin{subequations}
	\begin{align}
	\varepsilon &: \mathcal{D}_{t}^{(1)} f_i^{(0)} = -\omega f_i^{(1)} + {f^*}_i^{(1)},\\
	\varepsilon^2 &: \partial_t^{(2)}f_i^{(0)} + \mathcal{D}_{t}^{(1)} \left(1-\frac{\omega}{2}\right)f_i^{(1)} = -\omega f_i^{(2)} + {f^*}_i^{(2)} - \frac{1}{2}\mathcal{D}_{t}^{(1)}{f^*}_i^{(1)},
	\end{align}
    \label{Eq:CE_Eq_orders}
    \end{subequations}
    with $f_i^{(0)}={f_i^*}^{(0)}=f_i^{\rm eq}$. The moments of the non-local contributions (including both non-ideal contributions to the thermodynamic pressure, surface tension and the correction for the diagonals of the third-order moments tensor) are:
    \begin{subequations}
	    \begin{align}
		\sum_i {f^*}_i^{(k)} &= 0, \forall k>0,\\
		\sum_i \bm{c}_{i} {f^*_i}^{(1)} &= \bm{F},\\
		\sum_i \bm{c}_{i}\otimes\bm{c}_{i} {f^*_i}^{(1)} & = (\bm{u}\otimes\bm{F} + {\bm{F}\otimes\bm{u}}) + \Phi\\
        \sum_i \bm{c}_{i}\otimes\bm{c}_{i} {f^*_i}^{(2)} & = \frac{1}{\rho}\bm{F}\otimes\bm{F}.
		\end{align}
	    \label{Eq:CE_solvability}
    \end{subequations}
	Taking the moments of the Chapman-Enskog-expanded equation at order $\varepsilon$:
	\begin{eqnarray}
	    \partial_t^{(1)}\rho + \bm{\nabla}\cdot\rho \bm{u} &=& 0,\label{eq:approach2_continuity1}\\
	    \partial_t^{(1)}\rho \bm{u} + \bm{\nabla}\cdot\rho \bm{u}\otimes\bm{u} + \bm{\nabla}\cdot P_0\bm{I} + \bm{F} &=& 0,\label{eq:approach2_NS1}
	\end{eqnarray}
	while at order $\varepsilon^2$ the continuity equation is:
	\begin{equation}
	    \partial_t^{(2)}\rho + \bm{\nabla}\cdot\frac{\bm{F}}{2} = 0.\label{eq:approach2_continuity2}
	\end{equation}
	Summing up Eqs.~\ref{eq:approach2_continuity1} and \ref{eq:approach2_continuity2} we recover the continuity equation as:
	\begin{equation}
	    \partial_t \rho + \bm{\nabla}\cdot\rho \bm{U} = 0,
	\end{equation}
	where $\bm{U} = \bm{u} + \frac{\delta t}{2\rho}\bm{F}$. For the momentum equations we have:
	\begin{multline}\label{eq:eps2_mom1_1}
	    \partial_t^{(2)}\rho \bm{u} + \frac{1}{2} \partial_t^{(1)} \bm{F} + \frac{1}{2}\bm{\nabla}\cdot\left(\bm{u}\otimes\bm{F} + {\bm{F}\otimes\bm{u}} \right)
	    + \bm{\nabla}\cdot\left(\frac{1}{2}-\frac{1}{\omega}\right)\left[\partial_t^{(1)}\Pi_{2}^{(0)}+\bm{\nabla}\cdot\Pi_{3}^{(0)}\right]
	     \\ - \bm{\nabla}\cdot\left(\frac{1}{2}-\frac{1}{\omega}\right) \left(\bm{u}\otimes\bm{F} + {\bm{F}\otimes\bm{u}}\right) + \bm{\nabla}\cdot\frac{1}{\omega}\Phi = 0,
	\end{multline}
	where $\Pi_{2}^{(0)}$ and $\Pi_{3}^{(0)}$ are the second- and third-order moments of $f_i^{(0)}$ defined as:
    \begin{eqnarray}
        \Pi_{2}^{(0)} &=& \rho \bm{u}\otimes\bm{u} + P_0\bm{I},\\
        \Pi_{3}^{(0)} &=& \Pi_{3}^{\rm MB} - \rho\bm{u}\otimes \bm{u} \otimes\bm{u}\circ\bm{J} - 3(P_0 - \rho \varsigma^2)\bm{J}
    \end{eqnarray}
    where $\Pi_{\alpha\beta\gamma}^{\rm MB}=\rho u_\alpha u_\beta u_\gamma + P_0{\rm perm}(u_\alpha \delta_{\beta\gamma})$ is the third-order moment of the Maxwell-Boltzmann distribution, and for the sake of simplicity we have introduced the diagonal rank three tensor $\bm{J}$, with $J_{\alpha\beta\gamma}=\delta_{\alpha\beta}\delta_{\alpha\gamma}\delta_{\beta\gamma}$ and $\circ$ is the Hadamard product.
	The contributions in the fourth term on the left hand side can be expanded as:
    \begin{align}
	\partial_t^{(1)} \Pi_{2}^{(0)}=&
	 \partial_t^{(1)}\rho \bm{u}\otimes\bm{u} + \partial_t^{(1)} P_0\bm{I}\nonumber\\
	 = & \bm{u}\otimes\partial_t^{(1)}\rho \bm{u} + {(\partial_t^{(1)}\rho \bm{u})\otimes\bm{u}} - \bm{u}\otimes\bm{u} \partial_t^{(1)} \rho + \partial_t^{(1)} P_0 \bm{I}\nonumber\\
	=& -\bm{\nabla}\cdot\rho \bm{u}\otimes\bm{u}\otimes\bm{u} 
	    -\left[\bm{u}\otimes\left(\bm{\nabla} P_0 - \bm{F}\right) + {\left(\bm{\nabla} P_0 - \bm{F}\right)\otimes\bm{u}}
	     \right] + \partial_t^{(1)}P_0 \bm{I}
    \end{align}
	and:
	\begin{multline}
	    \bm{\nabla}\cdot\Pi_{3}^{(0)} = \bm{\nabla}\cdot\rho \bm{u}\otimes\bm{u}\otimes\bm{u} 
	    + \left(\bm{\nabla} P_0 \bm{u} + {\bm{\nabla} P_0 \bm{u}}^{\dagger}\right) + (\bm{\nabla}\cdot P_0\bm{u})\bm{I}\\
	    - \bm{\nabla}\cdot\left[\rho\bm{u}\otimes \bm{u} \otimes\bm{u}\circ\bm{J} + 3(P_0 - \rho \varsigma^2)\bm{J}\right],
	\end{multline}
	resulting in:
	\begin{multline}
	    \partial_t^{(1)} \Pi_{2}^{(0)} + \bm{\nabla}\cdot\Pi_{3}^{(0)} = P_0\left(\bm{\nabla}\bm{u} + {\bm{\nabla}\bm{u}}^{\dagger} \right) + \left(\bm{u}\otimes\bm{F} + {\bm{u}\otimes\bm{F}}^{\dagger}\right)\\
	    + \left(\bm{\nabla}\cdot P_0 \bm{u} + \partial_t^{(1)}P_0\right)\bm{I} - \bm{\nabla}\cdot\left[\rho\bm{u}\otimes \bm{u} \otimes\bm{u}\circ\bm{J} + 3(P_0 - \rho \varsigma^2)\bm{J}\right].
	\end{multline}
	Plugging this last equation back into Eq.\ \eqref{eq:eps2_mom1_1}:
	\begin{multline}
	    \partial_t^{(2)}\rho \bm{u} + \partial_t^{(1)}\frac{\bm{F}}{2} + \frac{1}{2}\bm{\nabla}\cdot(\bm{u}\otimes\bm{F} + {\bm{F}\otimes\bm{u}}) + \bm{\nabla}\cdot\left(\frac{1}{2}-\frac{1}{\omega}\right)P_0\left(\bm{\nabla}\bm{u} + {\bm{\nabla}\bm{u}}^{\dagger}\right)\\
	    + \bm{\nabla}\left(\frac{1}{2}-\frac{1}{\omega}\right) \left(\partial_t^{(1)}P_0+\bm{\nabla}\cdot P_0 \bm{u}\right) \\+ \bm{\nabla}\cdot\left[\left(\frac{1}{2}-\frac{1}{\omega}\right)\bm{\nabla}\cdot\left(\rho\bm{u}\otimes \bm{u} \otimes\bm{u}\circ\bm{J} + 3(P_0 - \rho \varsigma^2)\bm{J}\right) + \frac{1}{\omega}\Phi\right] = 0.
	\end{multline}
	where the last term cancels out by setting:
	\begin{equation}
	    \Phi = \left(1-\frac{\omega}{2}\right) \bm{\nabla}\cdot\left(\rho\bm{u}\otimes \bm{u} \otimes\bm{u}\circ\bm{J} + 3(P_0 - \rho \varsigma^2)\bm{J}\right),
	\end{equation}
	and the fourth and fifth terms reduce to the viscous stress tensor by defining $\mu/P_0 = \left(\frac{1}{\omega} - \frac{1}{2}\right)$ and:
	\begin{equation}
        P_0\left(\frac{2+D}{D} - \frac{\partial\ln P_0}{\partial\ln\rho}\right)\left(\frac{1}{\omega} - \frac{1}{2}\right) = \eta.
    \end{equation}
	Furthermore, using $\bm{U} = \bm{u} + \frac{\delta t}{2\rho}\bm{F}$ and:
	\begin{equation}
	    \rho \bm{U}\otimes\bm{U} = \rho \bm{u}\otimes\bm{u} + \frac{\delta t}{2}(\bm{u}\otimes\bm{F} + {\bm{F} \otimes\bm{u}}) + \frac{\delta t^2\bm{F}\otimes\bm{F}}{4\rho},
	\end{equation}
	in combination with the Euler-level equation, and keeping in mind that errors of the form $\bm{\nabla}\cdot\frac{\delta t^2\bm{F}\otimes\bm{F}}{4\rho}$ in the convective term and $\delta t\bm{\nabla}\mu\left(\bm{\nabla}\frac{\bm{F}}{\rho}+{\bm{\nabla}\frac{\bm{F}}{\rho}}^{\dagger}\right)$ in the viscous stress are of order $\varepsilon^3$ one recovers:
	\begin{align}
	    \partial_t\rho \bm{U} + \bm{\nabla}\cdot \rho \bm{U}\otimes\bm{U} - \bm{\nabla}\cdot\mu\left( \bm{\nabla}\bm{U} + {\bm{\nabla}\bm{U}}^{\dagger} - \frac{2}{D}\bm{\nabla}\cdot\bm{U} \bm{I}\right) - \bm{\nabla}\cdot\left(\eta \bm{\nabla}\cdot\bm{U}\right) + {O}(\varepsilon^3) = 0.
	\end{align}
    \section{Discrete non-local contributions to pressure tensor\label{App:Discrete_Pressure}}
\begin{figure}
	    \centering
		    \includegraphics{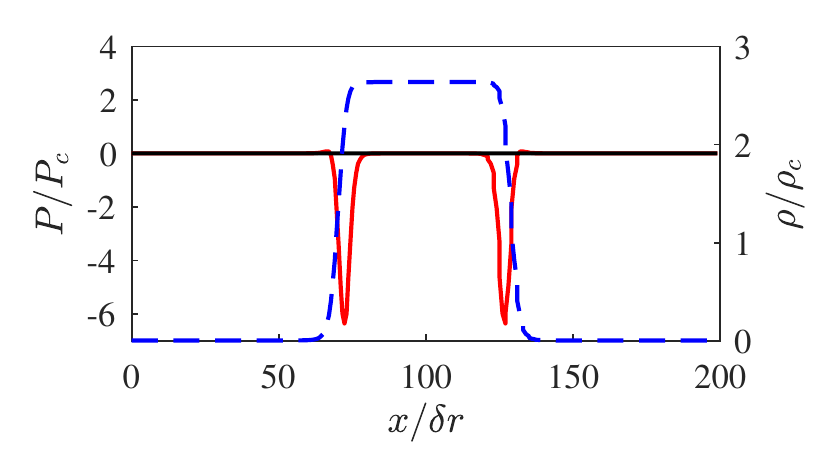}
	    \caption{Pressure distribution from a simulation at $T_r=0.36$, corresponding to $P_r=0.0022$ and $\rho_l/\rho_v=10^3$. Black line: Evaluation using the discrete pressure tensor; Red line: Evaluation using continuous pressure tensor. Dashed blue line: Density profile. 
	    }
	    \label{Fig:discrete_pressure}
    \end{figure}

    Following the analysis presented by \cite{shan_pressure_2008}, we write the force contribution for the present model as,
    \begin{align}
        \bm{F} &= \bm{F}^{A} + \bm{F}^{B} + \bm{F}^{C} + \bm{F}^{D},\\
            \bm{F}^{A} &= \pm\frac{8}{3} \psi(\bm{r}) \sum_{i=0}^{Q-1} \frac{w_i}{\varsigma^2} \bm{c}_{i} \psi(\bm{r}+\bm{c}_i\delta t),\\
        \bm{F}^{B} &= \mp \frac{1}{3} \psi(\bm{r}) \sum_{i=0}^{Q-1}  \frac{w_i}{\varsigma^2}\bm{c}_{i}\psi(\bm{r}+2\bm{c}_i\delta t),\\
        \bm{F}^{C} &= 2\tilde{\kappa} \rho(\bm{r}) \sum_{i=0}^{Q-1}  \frac{w_i}{\varsigma^2}\bm{c}_{i}\rho(\bm{r}+\bm{c}_i\delta t),\\
        \bm{F}^{D} &= -\tilde{\kappa} \rho(\bm{r}) \sum_{i=0}^{Q-1}  \frac{w_i}{\varsigma^2}\bm{c}_{i}\rho(\bm{r}+2\bm{c}_i\delta t).
    \end{align}
    The pressure tensor contributions from forces $\bm{F}^{A}$ and $\bm{F}^{C}$ can be readily written as:
    \begin{eqnarray}\label{eq:discrete_pressure_psi1}
        \bm{P}^{A} &=& \mp\frac{8}{6} \psi(\bm{r}) \sum_{i=0}^{Q-1} \frac{w_i}{\varsigma^2} \bm{c}_{i}\otimes\bm{c}_{i} \psi(\bm{r}+\bm{c}_i\delta t),\\
        \bm{P}^{C} &=& -\tilde{\kappa} \rho(\bm{r}) \sum_{i=0}^{Q-1} \frac{w_i}{\varsigma^2} \bm{c}_{i}\otimes\bm{c}_{i} \rho(\bm{r}+\bm{c}_i\delta t),
    \end{eqnarray}
    while  $\bm{F}^{B}$ and $\bm{F}^{D}$ contribute to the pressure tensor as follows:
    \begin{align}\label{eq:discrete_pressure_psi2}
        \bm{P}^{B} =& \pm \frac{1}{6} \left[\psi(\bm{r}) \sum_{i=0}^{Q-1} \frac{w_i}{\varsigma^2} \bm{c}_{i}\otimes\bm{c}_{i} \psi(\bm{r}+2\bm{c}_i\delta t) \right.
         + \left. \sum_{i=0}^{Q-1} \frac{w_i}{\varsigma^2}\bm{c}_{i}\otimes\bm{c}_{i} \psi(\bm{r}-\bm{c}_i\delta t)\psi(\bm{r}+\bm{c}_i\delta t)\right],\\
        \bm{P}^{D} =& \frac{\tilde{\kappa}}{2} \left[\rho(\bm{r}) \sum_{i=0}^{Q-1} \frac{w_i}{\varsigma^2}\bm{c}_{i}\otimes\bm{c}_{i} \rho(\bm{r}+2\bm{c}_i\delta t) \right. 
         + \left. \sum_{i=0}^{Q-1} \frac{w_i}{\varsigma^2}\bm{c}_{i}\otimes\bm{c}_{i} \rho(\bm{r}-\bm{c}_i\delta t)\rho(\bm{r}+\bm{c}_i\delta t)\right].
    \end{align}
    These expressions allow to compute the discrete pressure tensor with high accuracy. Fig.\ \ref{Fig:discrete_pressure} shows the distribution of the normal pressure, $P_{xx}$, in a flat interface simulation as computed from both the discrete and continuous pressure tensors, 
    \begin{equation}\label{eq:pressure_cont}
        P_{xx}^{\rm cont} = P + \kappa\left(\partial_x^2 \rho - \frac{1}{2}{\lvert\partial_x \rho\lvert}^2\right),
    \end{equation}
     While the discrete evaluation method correctly results in a uniform pressure distribution throughout the domain, also across the interface, the continuous approximation 
    evaluated using a finite differences approximation fails to do so, indicating errors due to higher-order terms. This points to the necessity of using the discrete pressure tensor instead of Eq.\ \eqref{eq:pressure_cont} for evaluation of quantities such as  surface tension and Tolman length in sec.\ \ref{sec:surface_temperature} and \ref{sec:Tolman}.
    \section{Analytical solution of layered Poiseuille flow \label{App:Poiseuille_analytical}}
    Solving the system presented in Eqs.\ \eqref{Eq:poiseuille_odes} and \eqref{Eq:poiseuille_odes_BD} one gets:
    \begin{subequations}
	\begin{align}
		u(y) &= a_l y^2 + b_l y, &\forall y: 0\leq y \leq h_l,\\
		u(y) &= a_v y^2 + b_v y + c_v, &\forall y: h_l\leq y \leq H,
	\end{align}
	\label{Eq:poiseuille_analytical}
    \end{subequations}
    with $a_l=-\frac{\rho_l g}{2\mu_l}$, $a_v=-\frac{\rho_v g}{2\mu_v}$, and:
    \begin{subequations}
	\begin{align}
		b_l &= g\frac{ H^2 \mu_l \rho_v - 2 h_l^2 \mu_l \rho_l + h_l^2 \mu_l \rho_v + h_l^2 \mu_v \rho_l + 2 H h_l \mu_l \rho_l - 2 H h_l \mu_l \rho_v}{2 \mu_l (H\mu_l - h_l \mu_l + h_l \mu_v)},\\
		b_v &= g\frac{H^2 \mu_l \rho_v - h_l^2 \mu_l \rho_v + 2 h_l^2 \mu_v \rho_v}{2 \mu_v (H\mu_l - h_l \mu_l + h_l \mu_v)},\\
		c_v &= gH\frac{ h_l^2\mu_l\rho_v + h_l^2\mu_v\rho_l - 2h_l^2\mu_v\rho_v - H h_l \mu_l\rho_v + H h_l\mu_v\rho_v }{2 \mu_v (H\mu_l - h_l \mu_l + h_l \mu_v)}.
	\end{align}
	\label{Eq:LayPoiseuilleCoeffs}
    \end{subequations}
    \section{Isothermal sound speed for other equations of states \label{ap:Sound_speed}}
    Apart from the van der Waals equation of state in the main text, present LBM formulation was also used to simulate the speed of sound for the three other cubic  equations of state, Peng--Robinson \eqref{eq:PREoS}, Riedlich--Kwong--Soave \eqref{eq:RKSEoS} and Carnahan--Starling \eqref{eq:CSEos},  along with two equations of state proposed by \cite{shan_lattice_1993}, 
    \begin{equation}\label{eq:SCeos}
    P_{\rm SC} = P_0 + \frac{P_0 \mathcal{G}}{2}\psi_{\rm SC}^2,
    \end{equation}
    with:
    \begin{eqnarray}
    \psi_{\rm SC-I} &=& 1 - \exp\left({-\rho}\right),\label{eq:SC1}\\
    \psi_{\rm SC-II} &=& \exp\left({-1/\rho}\right),\label{eq:SC2}
    \end{eqnarray}
    while $P_0=\varsigma^2\rho$. Corresponding critical densities $\rho_c$ and {pseudo-temperatures} $\mathcal{G}_c$ are readily computed by solving the  conditions at critical point,
    \begin{equation}
    \frac{\partial P_{\rm SC}}{\partial \rho}\bigg|_{\mathcal{G}_c,\rho_c} = 0,\
    \frac{\partial^2 P_{\rm SC}}{\partial \rho^2}\bigg|_{\mathcal{G}_c,\rho_c} = 0,
    \end{equation}
    which results in $(\mathcal{G}_c=-4,\rho_c=0.6931)$ and $(\mathcal{G}_c=-7.389,\rho_c=1)$ for the pseudo-potentials \eqref{eq:SC1} and \eqref{eq:SC2}, respectively. 
    \begin{figure}
	    \centering
		    \includegraphics{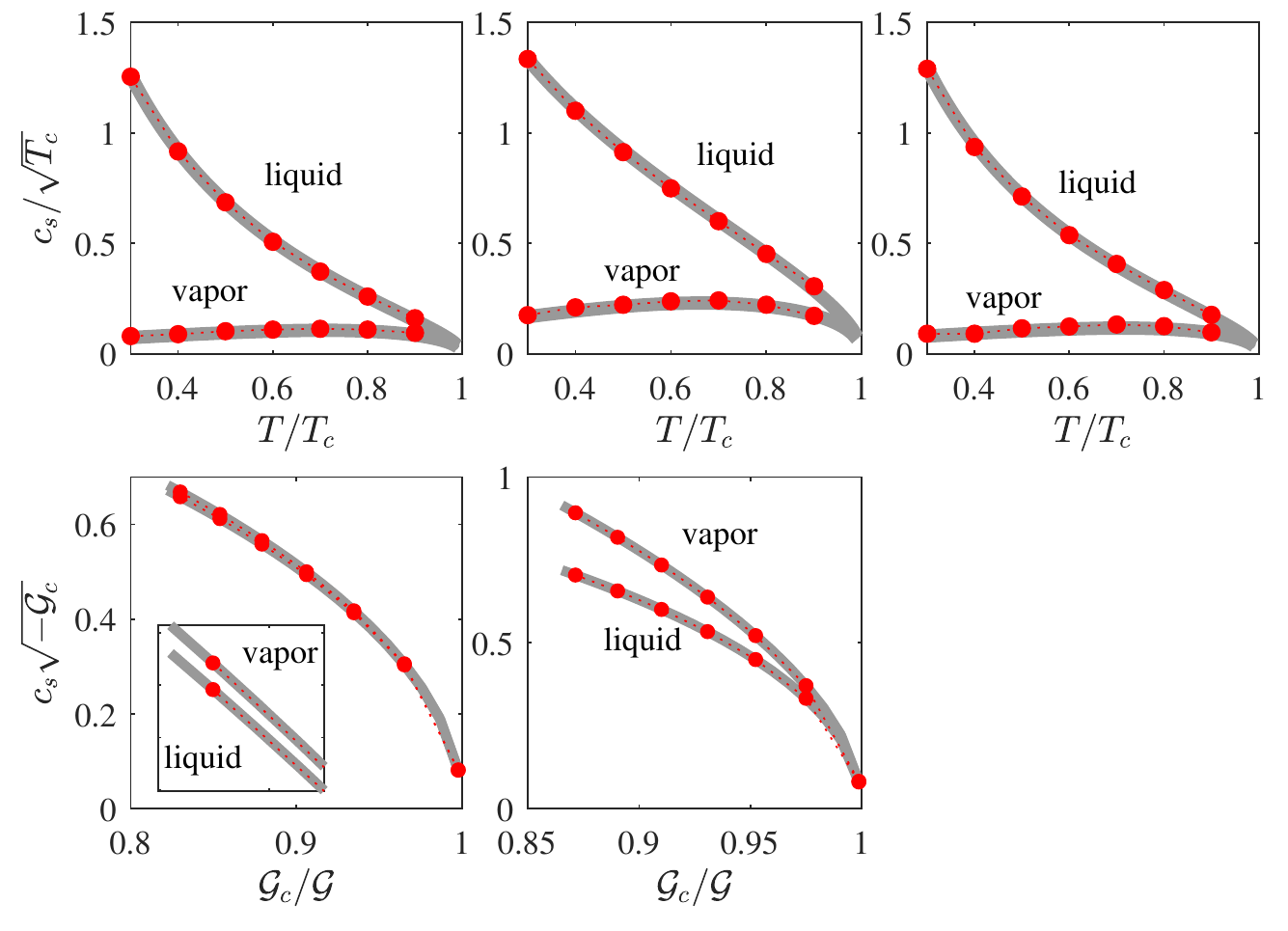}
	    \caption{Isothermal sound speed for various equations of state. Top row, from left to right:
	    Peng--Robinson \eqref{eq:PREoS}, 
	    Carnahan--Starling \eqref{eq:CSEos} and Riedlich--Kwong--Soave  \eqref{eq:RKSEoS};
	    Bottom row, from left to right: Shan--Chen \eqref{eq:SC1} and \eqref{eq:SC2}.
	    Grey plain lines: Theory; Symbol: Simulations with the present LB model. 
	    }
	    \label{Fig:sound_speed_all}
    \end{figure}
    Fig.\ \ref{Fig:sound_speed_all} shows excellent agreement between simulations using present model and theory for all equations of state.
    It is interesting to note that, contrary to the conventional equations of state of the van der Waals type,  equations of state \eqref{eq:SCeos}, \eqref{eq:SC1} and \eqref{eq:SC2} demonstrate higher speed of sound in the vapor phase than in the liquid. While this, by itself, does not contradict thermodynamics, it remains unclear which substances may feature such an unusual behaviour.
    \bibliographystyle{jfm}
    \bibliography{references}
\end{document}